\documentclass[12pt]{article}
\usepackage{slashed}
\usepackage{amsmath,amssymb,graphicx}
\usepackage{hyperref}
\usepackage{cite}
\usepackage{mathtools}
\usepackage{comment}
\usepackage{bm}
\usepackage{upgreek}
\newcommand{\CR}{\nonumber \\*}

\usepackage{mdframed}

\usepackage{multicol,color,longtable}

\definecolor{darkred}{rgb}{0.65,0.15,0}
\hypersetup{pdfborder={0 0 0},colorlinks=true,urlcolor=blue,citecolor=blue,linkcolor=darkred,linktocpage=true}

\usepackage{young}
\usepackage[vcentermath]{youngtab}

\usepackage{mathrsfs}

\newcommand{\be}{\begin{equation}}
\newcommand{\ee}{\end{equation}}

\newcommand{\bea}{\begin{eqnarray}}
\newcommand{\eea}{\end{eqnarray}}
\newcommand{\nn}{\nonumber}

\newcommand{\eq}[1]{\eqref{#1}}
\newcommand{\w}[1]{\\[0.#1cm]}

\newcommand{\ba}{\begin{array}}
\newcommand{\ea}{\end{array}}

\newcommand{\bn}{\begin{align}}
\newcommand{\en}{\end{align}}

\newcommand{\E}{{\mathcal E}}

\newcommand{\cL}{{\mathcal L}}

\newcommand{\cO}{{\mathcal O}}

\newcommand{\eb}{{\bar\epsilon}}

\newcommand{\eps}{{\bar\epsilon}}

\newcommand{\ag}{a}

\newcommand{\Tr}{\mathrm{Tr}}
\newcommand{\tr}{\mathrm{tr}}

\definecolor{darkgreen}{rgb}{0.42, 0.56, 0.14}

\DeclareFontFamily{U}{matha}{\hyphenchar\font45}
\DeclareFontShape{U}{matha}{m}{n}{
      <5> <6> <7> <8> <9> <10> gen * matha
      <10.95> matha10 <12> <14.4> <17.28> <20.74> <24.88> matha12
      }{}
\DeclareSymbolFont{matha}{U}{matha}{m}{n}

\DeclareMathSymbol{\oleft}{2}{matha}{"68}
\DeclareMathSymbol{\oright}{2}{matha}{"69}

\DeclareFontFamily{U}{mathx}{}
\DeclareFontShape{U}{mathx}{m}{n}{ <-> mathx10 }{}
\DeclareSymbolFont{mathx}{U}{mathx}{m}{n}
\DeclareFontSubstitution{U}{mathx}{m}{n}

\DeclareMathAccent{\widecheck}{0}{mathx}{"71}

\makeatletter

\@addtoreset{equation}{section}
\makeatother

\newcommand{\e}{\epsilon}

\newcommand{\m}{\mu}
\newcommand{\n}{\nu}

\newcommand{\ho}{{\widehat\omega}}

\newcommand{\rh}{\rho}

\newcommand{\vp}{\varphi}

\newcommand{\ve}{\varepsilon}

\usepackage{ifthen}
\reversemarginpar

\numberwithin{equation}{section}

\newcommand{\R}{{\mathcal R}}

\newcommand{\oh}{{\widehat\omega}}

\newcommand{\bchi}{{\bar\chi}}
\newcommand{\hrho}{{\widehat\rho}}

\newcommand{\ua}{\underline{a}}
\newcommand{\ub}{\underline{b}}
\newcommand{\uc}{\underline{c}}
\newcommand{\ud}{\underline{d}}
\newcommand{\ue}{\underline{e}}
\newcommand{\uf}{\underline{f}}

\begin{document}

\begin{flushright} CPHT-RR083.112024\\MI-HET-845 \end{flushright} 
 \vspace{8mm}

\begin{center}

{\large \bf Higher derivative couplings with multi-tensor multiplets \\[3mm]
in 6D supergravity, action and anomalies }\\[5mm]

\vspace{6mm}
\normalsize
{\large  Guillaume Bossard${}^{1}$, Axel Kleinschmidt${}^{2,3}$ and Ergin Sezgin${}^4$}

\vspace{10mm}
${}^1${\it Centre de Physique Th\'eorique, CNRS,  Institut Polytechnique de Paris\\
91128 Palaiseau cedex, France}
\vskip 1 em
${}^2${\it Max-Planck-Institut f\"{u}r Gravitationsphysik (Albert-Einstein-Institut)\\
Am M\"{u}hlenberg 1, DE-14476 Potsdam, Germany}
\vskip 1 em
${}^3${\it University of Vienna, Faculty of Physics, Boltzmanngasse 5, 1090 Vienna, Austria}
\vskip 1 em
${}^4${\it Mitchell Institute for Fundamental Physics and Astronomy\\ Texas A\&M University
College Station, TX 77843, USA}

\vspace{20mm}

\hrule

\vspace{5mm}

 \begin{tabular}{p{14cm}}
{\small
We revisit six-dimensional $(1,0)$ supergravity coupled to $n_T$ tensor multiplets and Yang--Mills fields for $n_T>1$ for which no covariant action exists. We construct the action in the Henneaux--Teitelboim approach and in the presence of a gauge anomaly. 
We moreover obtain the supersymmetric Green--Schwarz counterterm for the gravitational anomaly for  arbitrary matter content.}
\end{tabular}
\vspace{6mm}
\hrule
\end{center}

\thispagestyle{empty}

\newpage

\setcounter{page}{1}

\setcounter{tocdepth}{2}
\tableofcontents

\vspace{4mm}
\hrule

\section{Introduction}
Six dimensions is the highest dimension in which minimal supergravity couples to matter multiplets other than  vector multiplets. They are the so-called $(1,0)$ supergravity theories with eight left-handed supersymmetries. Because they are chiral, they suffer from local and global anomalies. When there is more than one tensor multiplet, the cancellation of anomalies involves a generalisation of the Green--Schwarz mechanism \cite{Sagnotti:1992qw}. The associated Green--Schwarz--Sagnotti type Lagrangian cannot be written with manifest diffeomorphism invariance, but the two-derivative equations of motion and pseudo-Lagrangian were worked out using supersymmetry in  \cite{Nishino:1984gk,Nishino:1997ff,Ferrara:1997gh,Riccioni:2001bg}. 

The Yang--Mills coupling constants turn out to diverge at regular values of the scalars when there is a gauge anomaly with a negative coefficient \cite{Sagnotti:1992qw}. These singular loci define walls separating different phases of the theory where non-critical strings living in six dimensions become tensionless and gravity decouples \cite{Ganor:1996mu,Seiberg:1996vs}. The quantum consistency of the theory implies the existence of strings with charges valued in a self-dual lattice \cite{Seiberg:2011dr}. In particular the coefficients defining the Green--Schwarz--Sagnotti Lagrangian are quantised for the theory to be free of global anomalies 
\cite{Monnier:2017oqd,Monnier:2018nfs}. In this way, six-dimensional $(1,0)$ supergravity theories with more than one tensor multiplet provide a fruitful landscape for exploring the Swampland program \cite{Kumar:2009us,Kumar:2009ae,Kumar:2009ac,Kumar:2010ru}, whose aims include finding apparently consistent theories which have no known string/M-theory origin \cite{Vafa:2005ui}. Explicit perturbative string theories with more than one tensor multiplet were first constructed as free field orientifolds  in \cite{Bianchi:1990tb,Dabholkar:1996zi,Gimon:1996ay,Dabholkar:1996ka,Angelantonj:1996mw}.

\vskip 2mm

One of the salient features of these theories with more than one tensor multiplet is the presence of chiral 2-forms. There is unfortunately no totally satisfactory way to write a Lagrangian for chiral $p$-forms in $2p{+}2$ dimensions. One may only write the equations of motion as in \cite{Nishino:1984gk,Nishino:1997ff}, or write a pseudo-Lagrangian, whose Euler--Lagrange equations must be supplemented by first order duality equations as in \cite{Ferrara:1997gh,Riccioni:2001bg}. While the computation of anomalies has been achieved without having to appeal to an action \cite{Alvarez-Gaume:1983ihn}, it is desirable to have an action principle which lends itself to a proper quantisation of the model. However, the perturbative quantisation of the theory calls for a proper Lagrangian with well-defined Ward identities. We believe the most legitimate way to do so is to give up manifest Lorentz covariance by choosing a timelike foliation~\cite{Henneaux:1987hz,Henneaux:1988gg,Schwarz:1993vs,Schwarz:1997mc,Hillmann:2009zf}, an approach known as the Henneaux--Teitelboim formulation. There are alternative formulations admitting a covariant Lagrangian, but they involve other complications. One may restore covariance by defining the foliation through the introduction of an auxiliary field for the time function as in \cite{Pasti:1996vs,DallAgata:1997gnw,DallAgata:1998ahf} that appears non-polynomially, or using more auxiliary fields to render the theory polynomial~\cite{Mkrtchyan:2019opf}. The quantisation of the time function field requires a gauge-fixing that is equivalent to choosing a timelike foliation or is expected to involve infinitely many fields \cite{Devecchi:1996cp}. Thus, these classically covariant approaches seem to lead to not manifestly covariant quantum theories.

Another option is to decouple the unwanted $p$-forms with the opposite chirality as in \cite{Witten:1996hc,Witten:1999vg,Belov:2006jd}. This formulation is very useful for understanding the global properties of the free chiral $p$-forms through the definition of a half level Chern--Simons theory in $2p{+}3$ dimensions and was used to determine the global anomalies \cite{Monnier:2017oqd,Monnier:2018nfs}. The coupling to other fields was proposed in \cite{Sen:2015nph} in connection to string field theory \cite{Sen:2015uaa}, but it is not clear to us how Ward identities could enforce the decoupling of the wrong chirality gauge fields in perturbation theory. There are also other proposals involving infinitely many auxiliary fields \cite{McClain:1990sx,Wotzasek:1990zr,Martin:1994np,Devecchi:1996cp,Bengtsson:1996fm,Berkovits:1996em} that lead to other difficulties in the quantisation. To our knowledge, the proper perturbative quantisation of chiral gauge fields in these covariant formulations has not been addressed in the paradigm of quantum field theory. Only in the Henneaux--Teitelboim approach~\cite{Henneaux:1987hz,Henneaux:1988gg} one knows how to define local Ward identities to impose the stability of the bare action through the master equation, see e.g. \cite{Bossard:2010dq}. For these reasons, this is the approach we adopt.

\vskip 2mm

In this paper we wish to clarify the structure of the supergravity effective action in six-dimensional models with $n_T>1$ tensor multiplets.  First we define a proper Lagrangian in the Henneaux--Teitelboim formalism~\cite{Henneaux:1987hz,Henneaux:1988gg,Schwarz:1993vs} consistent with the duality equation and the pseudo-Lagrangian derived in \cite{Ferrara:1997gh,Riccioni:2001bg}. We will show in particular that the Henneaux--Teitelboim Lagrangian is simply related to the covariant pseudo-Lagrangian by an additional term quadratic in the duality equation along the chosen timelike direction.

Second, we shall construct the supersymmetric four-derivative Green--Schwarz counterterm associated to the gravitational anomaly. The construction is based on  superconformal tensor calculus and relies on the Bergshoeff--de Roo map from the Poincar\'e to the Yang--Mills multiplet, generalising a previous construction for a single tensor multiplet~\cite{Bergshoeff:2012ax}. This correction is obtained for any number of tensor, vector and hyper multiplets at leading order in $\alpha'$, up to terms associated to the mixed anomaly. The structure of the invariant is rather simple and consistent with its dimensional reduction on a circle to five dimensions \cite{Hanaki:2006pj}. To obtain this result we first derive a map from  (1,0) supergravity coupled to tensor multiplets to the off-shell Poincar\'e multiplet~\cite{Bergshoeff:1985mz}. We extend this map in the presence of hyper and vector multiplets when there is no mixed gravitational-gauge anomaly, and explain the nature of the modifications when there is a mixed anomaly. With this map one can simply use the results of \cite{Bergshoeff:2012ax} to obtain the Riemann squared invariant. The map is naturally defined in a ``string frame'' that generalises the ten-dimensional Einstein frame in type I string theory in the presence of an anomaly. For $n_T>9$ tensor multiplets, the $R^2$ coupling coefficient can in principle vanish at finite values of the scalar fields and one exhibits in this ``string frame'' that this implies a decoupling of gravity.  We discuss the relation of this singularity with the more standard Yang--Mills strong coupling limits in Section \ref{String4}. 

For simplicity we only consider semi-simple gauge groups. When the gauge group is reductive and includes abelian factors, one generically needs additional counterterms to cancel the mixed anomaly involving the abelian gauge fields. This mechanism requires the gauging of axion shift isometries of the hypermultiplet scalar fields with respect to these abelian gauge fields, such that the abelian vector multiplets and the associated hypermultiplets combine into massive vector multiplets  \cite{Berkooz:1996iz}, see~\eq{FtrF3} below for an example.\footnote{Note that while we will concentrate on semi-simple gauge groups, the inclusion of abelian factors is straightforward when there is no need for additional counterterms and has been carried out in \cite{Riccioni:2001bg} at the two-derivative level.} Being massive it is consistent to disregard them in the low energy effective theory. 

The paper is structured as follows. We first review the structure of (1,0) supergravity and the possible multiplets along with the on-shell duality equations and the anomalies present in the theory. In Section~\ref{sec:HT}, we then perform the Henneaux--Teitelboim analysis to write a non-covariant physical Lagrangian and discuss the global issues appearing in the formalism and a special case where the Henneaux--Teitelboim approach is not needed. In Section~\ref{sec:BdR}, we present the supersymmetric extension to the Green--Schwarz counterterm for the gravitational anomaly. 

\section{Review of matter coupled (1,0) supergravity }
\label{sec:rev}

In this section we review the pseudo-Lagrangian and supersymmetry transformations of six-dimensional chiral  $\,\mathcal{N}=(1,0)$ supergravity coupled  to Yang--Mills, tensor multiplets and hypermultiplets \cite{Ferrara:1996wv,Nishino:1997ff,Riccioni:2001bg}. The model is constructed using the following on-shell multiplets (see for instance~\cite{Romans:1986er}):
\begin{itemize}
\item
a single gravity multiplet containing the vielbein $e_{\mu}{}^a$, the left-handed gravitino $\psi_\mu$ and a 2-form tensor field with on-shell anti-self-dual field strength. 
\item 
an unfixed number $n_T$ of tensor multiplets that will be labelled with an index $r=1,\ldots,n_T$. Each contains on-shell a self-dual tensor field, a right-handed  tensorino and a real scalar field. For $n_T$ tensor multiplets the scalars parametrise the coset $SO(1,n_T)/SO(n_T)$ and the collection of tensorini are denoted by $\chi^r$. The tensor fields, combined with the one from the gravity multiplet, are denoted by $B_{\mu\nu}^I$ with $I=0,1,\ldots, n_T$.
\item
an unfixed number $n_V$ of vector multiplets, each consisting of a vector field $A_\mu$ and a left-handed gaugino $\lambda$. We assume the compact gauge group, in the adjoint of which the vector and gaugino transform, to be semi-simple and exclude abelian factors for simplicity.  The simple factors will be labelled by $z$ and the traces projecting on them will be written as $\Tr_z$, for example the corresponding Yang--Mills kinetic term for one simple factor will be written as $\Tr_z F_{\mu\nu}F^{\mu\nu}$ in this notation that is also employed in~\cite{Ferrara:1996wv}. Here, $F_{\mu\nu}$ denotes the usual non-abelian bosonic field strength of a Yang--Mills field.\footnote{For simplicity we define the Yang--Mills fields as anti-Hermitian, but take nonetheless the trace $\Tr_z$ as positive definite, so equal to minus the matrix representation trace.}
\item 
an unfixed number $n_H$ of hypermultiplets, each consisting of four real scalars, and a symplectic Majorana--Weyl spinor. The $4n_H$ real scalars $\varphi^\alpha$ are coordinates on a quaternionic K\"ahler manifold with structure group $ Sp(n_H) \times Sp(1)_{\rm R}$. One defines the frame $V_\alpha^{X A}$ with $X=1,\ldots,2n_H$  a fundamental index of $Sp(n_H)$ and $A=1,2$ for $Sp(1)_{\rm R}$. The associated torsion-free spin connection splits by construction into $\omega_\alpha{}^{\! X A }{}_{ Y B} = \delta^{A}_B \mathcal{A}_\alpha^X{}_Y + \delta^{X}_Y \mathcal{A}_\alpha^A{}_B$. The hyperini are denoted by $\zeta^X$. 

\end{itemize}
In the next section, where we construct the Henneaux--Teitelboim form of the action and supertransformations, we shall put aside the hypermultiplets, and focus on the tensor-Yang--Mills system coupled to $(1,0)$ supergravity, which captures all subtleties of the construction.  We will re-introduce the hypermultiplets in Section~\ref{sec:BdR} where we describe the higher derivative extension of the model.
We follow mainly the conventions of~\cite{Nishino:1997ff}, thus in particular the space-time signature is $(-+++++)$. Curved six-dimensional indices $\mu$ are split into time and space according to $\mu=(t,i)$ with $i=1,\ldots,5$ and we write a curved time index explicitly as $t$. Flat indices $a=0,\ldots,5$ are split according to $a=(0,\ua)$.  
Our conventions for the Levi--Civita symbol are $\varepsilon^{0\underline{1}\underline{2}
\underline{3}\underline{4}\underline{5}}=+1$ and $\varepsilon^{\ua\ub\uc\ud\ue}=\varepsilon^{0\ua\ub\uc\ud\ue}$. In curved indices $\varepsilon^{t12345}=+1$ and $\varepsilon^{ijklm}= \varepsilon^{tijklm}$. Its indices are lowered with the metric $g_{\mu\nu}$. For further notations and conventions, see Appendix~\ref{app:conv}.

Spinors in six space-time dimensions for $\mathcal{N}=(1,0)$ supersymmetry are symplectic Majorana--Weyl spinors that are defined by the properties that their Majorana conjugate is equal to their charge conjugate (symplectic Majorana) and that they are chiral, i.e.\ eigenspinors of $\gamma_7=-\gamma^0\gamma^{\underline{1}}\cdots\gamma^{\underline{5}}$, where we call a positive eigenvalue left-handed and a negative eigenvalue right-handed. The symplectic condition is defined with respect to the R-symmetry $Sp(1)_{\rm R}\cong SU(2)_{\rm R}$. Further details on spinors and Fierz identities can be found in Appendix~\ref{app:conv}.

The $n_T$ scalar fields contained in the tensor multiplets are known to parametrise the coset space $SO(1,n_T)/SO(n_T)$. We write a coset representative as a block-decomposed matrix  $V\in SO(1,n_T)$ according to 
\begin{align}
V=( v_I ,  v_I{}^r)
\end{align}
where $I=0,1,\ldots, n_T$ is a fundamental index of $SO(1,n_T)$ whose metric $\eta_{IJ}$, used for raising and lowering these indices, we take as $(-++\ldots )$. The conditions for the decomposed matrix $V$ to belong $SO(1,n_T)$ are
\begin{align}
\label{eq:sccond}
v_I v_J \eta^{IJ} = -1\,,\quad
v_I{}^r v_J \eta^{IJ} = 0 \,,\quad
v_I{}^r v_J{}^s \eta^{IJ} = \delta^{rs}\,,\quad 
- v_I v_J + v_I{}^r v_J{}^s \delta_{rs} = \eta_{IJ}\,.
\end{align}
The fields $v_I$ and $v_I{}^r$ will also be referred to as moduli. The indices $r,s$ will be raised and lowered with the Euclidean~$\delta_{rs}$.

We also define the field-dependent coset metric
\begin{align}
M_{IJ} \coloneqq  v_I v_J + v_I{}^r v_J{}^s \delta_{rs} 
\end{align}
and the $SO(1,n_T)$-invariant coset velocity $P_\mu^r$ 
\begin{align}
\partial_\mu v_I = P_\mu^r v_{I r} \,,\quad
D_\mu v_I{}^r = \partial_\mu v_I{}^r + Q_\mu{}^r{}_s v_I{}^s = P_\mu^r v_I \,,
\end{align}
where $Q_\mu{}^r{}_s$ is the composite $SO(n_T)$ connection defined by this equation. 

The Lorentz signature for $\eta_{IJ}$ is related to the different duality conditions for the two-forms in the gravity and tensor multiplets. The $n_T+1$ two-forms will be written collectively as $B_{\mu\nu}^I$. 
In the presence of vector multiplets the field strength of the two-forms is modified by a Chern--Simons term and we define
\begin{align}
\label{eq:HCS}
H_{\mu\nu\rho}^I \coloneqq 3\partial_{[\mu} B_{\nu\rho]}^I - 6 \, b^{Iz} X_{z\,\mu\nu\rho}
\end{align}
with the Chern--Simons three-form for each simple factor of the gauge group given by
\begin{align}
\label{eq:YMCS}
X_{z\mu\nu\rho} := \Tr_z \left( A_{[\mu} \partial_\nu A_{\rho]} +\frac23 A_{[\mu} A_\nu A_{\rho]} \right)\,,
\end{align}
which satisfies the Bianchi identity
\begin{align}
\label{eq:XBianchi}
    4\partial_{[\mu} X_{z\,\nu\rho\sigma]} = \Tr_z \left( F_{[\mu\nu} F_{\rho\sigma]}\right)
\end{align}
with the non-abelian field strength $F_{\mu\nu} = 2\partial_{[\mu} A_{\nu]}+ [A_\mu,A_\nu]$, leading to the following Bianchi identity for the three-form field strength:
\begin{align}
    4 \partial_{[\mu} H_{\nu\rho\sigma]}^I =- 6 b^{Iz} \Tr_z \left( F_{[\mu\nu} F_{\rho\sigma]}\right)\,.
\end{align}

The constants $b^{Iz}$ appearing in~\eqref{eq:HCS} describe the couplings between the tensor and the vector multiplets. From them we can define the following field-dependent quantities
\begin{align}
c^z \coloneqq b^{Iz} v_I \ ,\quad c^{r\, z}\coloneqq b^{Iz} v_I{}^r \ .
\end{align}
The combination $c^z$ will appear for instance in front of the Yang--Mills kinetic term. As the $v_I$ are related to the coset scalar fields, this correspond to the typical scalar-field dependent couplings of vector fields in supergravity.

The bosonic duality equations can be written as 
\begin{align}
\label{eq:bosSD}
M_{IJ} H_{\mu\nu\rho}^J  &= \frac{1} {6 \sqrt{-g}} \eta_{IJ} \varepsilon_{\mu\nu\rho}{}^{\sigma\tau\kappa} 
H_{\sigma\tau\kappa J}\ 
\nn\\
 \Longleftrightarrow \quad\quad \eta_{IJ}\mathcal{E}_{\mu\nu\rho}^J &= \eta_{IJ} H_{\mu\nu\rho}^J -\frac{1} {6 \sqrt{-g}} M_{IJ} \varepsilon_{\mu\nu\rho}{}^{\sigma\tau\kappa} 
H_{\sigma\tau\kappa J} =0 \ ,
\end{align}
where the curved indices have been lowered with $g_{\mu\nu}$. Consistency of the duality equation requires $M_{IJ}\eta^{JK} M_{KL} \eta^{LP}= \delta_I^P$.
For later purposes it will often be useful to consider to consider the following combinations
\begin{align}
H_{\mu\nu\rho} \coloneqq v_I H_{\mu\nu\rho}^I \ ,\quad
H_{\mu\nu\rho}^r \coloneqq v_I{}^r H_{\mu\nu\rho}^I \,.
\end{align}
Below we will also present the supercovariantisations of all these quantities.

\medskip

In this section, and in Section~\ref{sec:BdR}, we shall consider the coupling of $n_H$ hypermultiplets as well. The $4n_H$ scalars  $\varphi^\alpha$ contained in these multiplets parametrise a  quaternionic K\"ahler (QK) manifold of negative scalar curvature \cite{Bagger:1983tt}. Quaternionic K\"{a}hler manifolds have structure group $Sp(n_H)\times Sp(1)_{\rm R}$, and the  vielbein $V_\alpha^{XA}$ and its inverse $V^\alpha_{XA}$ satisfy 
\be
V^\alpha_{XA} V^{\beta  X B} + V^\beta_{XA} V^{\alpha  X B} = g^{\alpha\beta} \delta_A^B\ ,\qquad g_{\alpha\beta} V^\alpha_{XA} V^\beta_{YB} =\Omega_{XY}\varepsilon_{AB}\ ,
\ee
where $g_{\alpha\beta}$ is the metric and $\alpha=1,...,4n_H,\ X=1,...,2n_H,\ A=1,2$. 
A composite $Sp(n_H)\times Sp(1)_{\rm R}$ valued connection is defined through the vanishing torsion condition
\be
\partial_\alpha V_{\beta XA} -\partial_\beta V_{\alpha XA} + {\cal A}_{\alpha X}{}^Y V_{\beta YA}-{\cal A}_{\beta X}{}^Y V_{\alpha YA} +{\cal A}_{\alpha A}{}^B V_{\beta XB}-{\cal A}_{\beta A}{}^B V_{\alpha XB}=0\ .
\ee
For a review of QK manifolds see, for example, \cite{Galicki:1985uwr} and the summary in \cite{Gunaydin:2010fi}.

{\allowdisplaybreaks
The pseudo-Lagrangian is given by
\be
{\cal L}^{\rm cov}= {\cal L}_{\rm B} +{\cal L}_{\rm F}\ ,
\label{pL1}
\ee
where the bosonic part ${\cal L}_{\rm B}$ given in~\cite{Riccioni:2001bg} reads in our conventions 
\bea
e^{-1}\cL_{\rm B} &=& \frac14 R -\frac{1}{48} M_{IJ} H_{\mu\nu\rho}^I H^{\mu\nu\rho J}
-\frac14 P_\mu^r P^\mu_r  -\frac14 c^z  \Tr_z (F_{\mu\nu} F^{\mu\nu}) 
\nn\w2
&& -\frac12 g_{\alpha\beta} \partial_\mu \vp^\alpha \partial^\mu \vp^\beta +\frac{1}{32} e^{-1}\varepsilon^{\mu\nu\rho\sigma\lambda\tau} b^{Iz} B_{\mu\nu}^J \eta_{IJ} \Tr_z \big( F_{\rho\sigma} F_{\lambda\tau}\big)\ ,
\label{bact}
\eea
and the  fermionic part \cite{Riccioni:2001bg}
\bea
e^{-1} \cL_{\rm F} &=& -\frac12 \bar\psi_\mu \gamma^{\mu\nu\rho} D_\nu\Big(\frac{\omega +{\widehat \omega}}{2}\Big) \psi_\rho - \frac12 \bar\chi_r \gamma^\mu D_\mu ({\widehat\omega}) \chi^r -  c^z \Tr_z \Big( {\bar\lambda} \gamma^\mu D_\mu ({\widehat\omega}) \lambda \Big)
\nn\w2
&& -\frac14 \Big( P_\mu^r + {\widehat P}_\nu^r \Big) {\bar\psi}_\mu \gamma^\nu \gamma^\mu \chi_r  +\frac18 \big({H + \widehat H}\big)^{(-)\mu\nu\rho}\bar\psi_\mu \gamma_\nu \psi_\rho -\frac{1}{48} {\widehat H}_{\mu\nu\rho}\, {\bar\chi}_r\gamma^{\mu\nu\rho} \chi^r 
\nn\w2
&& + \frac18 \big({H + \widehat H})_{\mu\nu\rho}^{r(+)}\bar\psi_\mu\gamma_{\nu\rho} \chi^r
+\frac{1}{24} {\widehat H}_{\mu\nu\rho}^r c_r{}^z \Tr_z \big( {\bar\lambda}\gamma^{\mu\nu\rho} \lambda \big)
\nn\w2
&&-\frac14 c^z \Tr_z \Big[ \big(F+{\widehat F}\big)_{\mu\nu} {\bar\psi_\rho} \gamma^{\mu\nu} \gamma^\rho \lambda \Big] 
-\frac12 c^{rz} \Tr_z \big( {\widehat F}_{\mu\nu} {\bar\chi}_r \gamma^{\mu\nu} \lambda \big)
\nn\w2
&& -\frac12{\bar\zeta}^X \gamma^\mu D_\mu (\widehat\omega)\zeta_X +\frac{1}{24}{\widehat H}_{\mu\nu\rho} {\bar\zeta}^X \gamma^{\mu\nu\rho} \zeta_X
\nn\w2
&& +\frac12 
\left(P_\nu^{XA} + \widehat{P}_\nu^{XA}\right)
{\bar\psi}_{\mu A}\gamma^\nu \gamma^\mu \zeta_X + e^{-1} \cL_4\ ,
\label{pL}
\eea
where $\cL_4 $ contains the explicit quartic fermion terms that can be found in~\cite{Riccioni:2001bg}.\footnote{One can add a term quartic in gauge fermions with an arbitrary coefficient without violating the Wess--Zumino consistency conditions that are satisfied by the anomalies \cite{Ferrara:1997gh}.} It is understood that the covariant derivatives of the fermions include the composite $SO(n_T), Sp(1)_{\rm R}$ and $Sp(n_H)$ 
 connections  denoted by $Q_\mu^{rs}, Q_\mu^{AB}=\partial_\mu\vp^\alpha {\cal A}_\alpha^{AB}$ and $Q_\mu^{XY} = \partial_\mu\vp^\alpha {\cal A}_\alpha^{XY}$, respectively. The definitions of the supercovariant curvatures are
\bea
{\widehat\omega}_{\mu ab} &=& \omega_{\mu ab}(e) + {\bar\psi}_\mu \gamma_{[a} \psi_{b]} + \frac12 {\bar\psi}_a \gamma_\mu \psi_b\ ,
\nn\w2
{\widehat H}_{\mu\nu\rho}^I &=& 3\partial_{[\mu} B_{\nu\rho]}^I - 6b^{Iz} X_{z\mu\nu\rho} 
+3 v^I {\bar\psi}_{[\mu} \gamma_\nu \psi_{\rho]} +3 v^{I r} {\bar\chi}_r \gamma_{[\mu\nu} \psi_{\rho]}\ ,
\nn\w2
{\widehat P}_\mu^r &=& P_\mu^r +{\bar\chi}^r \psi_\mu\ ,
\nn\w2
{\widehat F}_{\mu\nu} &=& F_{\mu\nu} - 2{\bar\lambda} \gamma_{[\mu} \psi_{\nu]}\ ,\nn\w2
{\widehat P}_\mu^{XA} & = & V_{\alpha}^{XA} \widehat{\partial_\mu \varphi^\alpha} = P_\mu^{XA} -{\bar\psi}_\mu^A \zeta^X\ .
\label{scf}
\eea
}

The dynamics of the system is described by the Euler--Lagrange equations following from the pseudo-Lagrangian together with the following duality equations that have to be imposed by hand:
\begin{subequations}
\label{eq:decov}
\begin{align}
{\widehat \E}_{\m\n\rh} &\coloneqq 2 {\widehat H}_{\m\n\rh}^{(+)} +\frac12 \bchi^r \gamma_{\m\n\rh} \chi_r - \frac12 {\bar\zeta}^X \gamma_{\mu\nu\rho} \zeta_X=0\ ,
\label{de1}\w2
{\widehat \E}_{\m\n\rh}^r &\coloneqq 2 {\widehat H}_{\m\n\rh}^{r(-)} - c^{rz} \Tr_z \bar\lambda \gamma_{\m\n\rh} \lambda  =0\ ,
\label{Ede}
\end{align}
\end{subequations}
where the projections $(\pm)$ on the (anti-)self-dual parts are defined by
\begin{align}
\widehat{H}^{(\pm)}_{\mu\nu\rho} \coloneqq \frac12 \Bigl( \widehat{H}_{\mu\nu\rho} \pm \frac{1}{6\sqrt{-g} }\varepsilon_{\mu\nu\rho}{}^{\sigma\tau\kappa} \widehat{H}_{\sigma\tau\kappa} \Bigr)\ .
\end{align}
The fact that the different projections appear in~\eqref{eq:decov} for the two parts is due to the different duality properties of the tensor fields in the supermultiplets, expressed by the Lorentzian $\eta_{IJ}$, cf. the bosonic duality equation~\eqref{eq:bosSD}.

{\allowdisplaybreaks
The fermionic field equations in the Einstein frame given in \cite{Riccioni:2001bg}, upon translating to our conventions, read 
\bea
\label{eq:EOMf}
{\cal R}^\mu &=& 
\frac12 \gamma^{\mu\nu\rh} \rho_{\nu\rh}(\ho) -\frac18 {\widehat H}_{\nu ab} \gamma^{\mu\nu\rh}\gamma^{ab} \psi_\rh -\frac{1}{24} {\widehat H}^r_{abc} \gamma^{abc} \gamma^\mu \chi_r
\nn\\
&&+\frac12 {\widehat P}_\nu^r \gamma^\nu\gamma^\mu \chi_r +\frac32 \gamma^{\mu\nu}\chi^r \left({\bar\chi}_r \psi_\nu\right) -\frac14 \gamma^{\mu\nu}\chi^r \left({\bar\chi}_r \gamma_{\nu\rh} \psi^\rh\right) 
\nn\\
&& +\frac14 \gamma_{\nu\rh} \chi^r \left( {\bar\chi}_r \gamma^{\mu\nu}\psi^\rh \right) -\frac12 \chi^r \left( {\bar\chi}_r \gamma^{\mu\nu} \psi_\nu\right) +{\widehat P}_\nu^X \gamma^\nu \gamma^\mu \zeta_X +\frac12 c^z \Tr_z \left(\gamma^{\nu\rho}\gamma^\mu \lambda {\widehat F}_{\nu\rho}\right)
\nn\\
&& + \frac14 c^z \Tr_z \Big[3\gamma^{\mu\nu\rho} \lambda \left({\bar\psi}_\nu \gamma_\rho\lambda\right) -2 \gamma^\mu\lambda \left({\bar\psi}_\nu \gamma^\nu\lambda\right)
+2\gamma^\nu\lambda \left({\bar\psi}_\nu \gamma^\mu \lambda\right) +\gamma_\rho\lambda\left({\bar \psi}_\nu \gamma^\mu{}^{\nu\rho} \lambda\right) \Big]
\nn\\
&& + \frac12 c^{rz} \Tr_z \gamma_\nu \lambda \left( {\bar\chi}_r \gamma^\nu\gamma^\mu \lambda\right)\ ,
\w2
\eta^r &=& \slashed{D} ({\widehat\omega})\chi^r +\frac{1}{24} {\widehat H}_{\m\n\rh} \gamma^{\m\n\rh} \chi^r +\frac{1}{24}{\widehat H}^r_{\m\n\rh} \gamma^\sigma\gamma^{\m\n\rh}\psi_\sigma 
+\frac12 {\widehat P}_\nu \gamma^\mu \gamma^\nu\psi_\mu - \frac12 \gamma^\mu \chi^s {\bar\chi}^s\gamma_\mu \chi^r
\nn\\*
&& +\frac12 c^{rz} \Tr_z \Big[ \gamma^{\mu\nu} {\widehat F}_{\mu\nu} \lambda + \gamma^\mu \gamma^\nu\lambda \left( {\bar\psi}_\mu \gamma_\nu\lambda\right)\Big] +\frac18 c^z \Tr_z \Big[ 3\gamma_{\mu\nu} \lambda {\bar\chi}^r \gamma^{\mu\nu} \lambda 
+2 \lambda {\bar\chi}^r \lambda \Big] 
\nn\\*
&&  +\frac14  \frac{c^{rz}c^{sz}}{c^z} \Tr_z\Big[6\lambda {\bar \chi}_s \lambda - \gamma_{\mu\nu} \lambda {\bar \chi}_s \gamma^{\mu\nu} \lambda \Big] \ ,
\label{EOM1}\w2
\eta^X &=& \gamma^\mu D_\mu ({\widehat\omega})\zeta^X -\frac{1}{12} {\widehat H}_{\mu\nu\rho} \gamma^{\mu\nu\rho} \zeta^X -\gamma^\mu\gamma^\nu \psi_{\mu A} V_\alpha^{XA} {\widehat{\partial_\nu \vp^\alpha}}+\frac{1}{12} \Omega^{XYZW} \gamma^\mu \zeta_Y{\bar\zeta}_Z\gamma_\mu \zeta_W
\nn\\
&& +\frac{1}{48} c^z \Tr_z \left[ \gamma^{\mu\nu\rho}\zeta^X {\bar\lambda}\gamma_{\mu\nu\rho}\lambda \right]\ ,
\eea
where the $Sp(1)_{\rm R}$ doublet index is suppressed in the term ${\widehat P}_\nu^{XA}\gamma^\nu \gamma^\mu \zeta_X$. For the detailed properties of the  quaternionic K\"ahler manifold parametrised by the hypermultiplet scalars, including  the definition of the totally symmetric tensor $\Omega^{XYZW}$, we refer the reader to \cite{Bagger:1983tt} (see also \cite{Galicki:1985uwr,Gunaydin:2010fi,Sezgin:2023hkc}). We have checked that the terms explicitly depending on the gravitino supercovariantise $\rho_{\m\n}(\oh)$ and $D_\mu (\oh)\chi$. 
} 

The supertransformations of the fields are given by \cite{Ferrara:1996wv}\footnote{ The supersymmetry transformation of the gaugino is understood to be for one simple factor $z$ of the gauge group, although we do not write explicitly the label $z$ on $\lambda$ or $F$. There is therefore no sum over $z$ in the bilinear in fermions.}
\bea
\label{eq:susytrm}
\delta_\epsilon e_\mu{}^a &=& {\bar\epsilon} \gamma^a \psi_\mu\ ,
\nn\w2
\delta_\epsilon B_{\mu\nu}^I &=& -2 v^I {\bar\epsilon} \gamma_{[\mu} \psi_{\nu]} +v^I{}_r {\bar\epsilon}\gamma_{\mu\nu} \chi^r -2b^{Iz}\Tr_z \big( A_{[\mu} \delta_\epsilon A_{\nu]} \big)\ ,
\nn\w2
\delta_\epsilon v_I &=& -v_I{}^r {\bar\epsilon} \chi_r\ ,  \hspace{20mm} \delta_\epsilon v_I{}^r = -v_I \eb \chi^r\ ,
\nn\w2
\delta_\epsilon \psi_\mu &=& D_\mu(\widehat\omega) \epsilon -\frac18{\widehat H}_{\mu\nu\rho} \gamma^{\nu\rho}\epsilon -\frac38 \gamma_\mu \chi^r ({\bar \epsilon} \chi_r)-\frac18 \gamma^\nu \chi^r ({\bar\epsilon} \gamma_{\mu\nu} \chi_r) +\frac{1}{16} \gamma_{\mu\nu\rho} \chi^r ({\bar\epsilon} \gamma^{\nu\rho} \chi_r)
\nn\w2
&& -\frac18 c^z \Tr_z \Big( 9\lambda\, {\bar\epsilon} \gamma_\mu \lambda -\gamma_{\mu\nu} \lambda\, {\bar\epsilon}\gamma^\nu \lambda +\frac12  \gamma^{\nu\rho} \lambda\, {\bar\epsilon} \gamma_{\mu\nu\rho} \lambda\Big) -\delta_\epsilon \vp^\alpha {\mathcal A}_\alpha^i \sigma_i \psi_\mu\ ,
\nn\w2
\delta_\epsilon \chi^r &=& -\frac12 {\widehat P}_\mu^r \gamma^\mu\epsilon - \frac{1}{24} {\widehat H}_{\mu\nu\rho}^r \gamma^{\mu\nu\rho} \epsilon -\frac12 c^{r z} \Tr_z \big(\gamma_\mu \lambda\,{\bar\epsilon} \gamma^\mu \lambda\big) -\delta_\epsilon \vp^\alpha {\mathcal A}_\alpha^i\sigma_i \chi^r\ ,
\nn\w2
\delta_\epsilon A_\mu &=&  {\bar\epsilon} \gamma_\mu \lambda\ ,
\nn\w2
\delta_\epsilon \lambda &=& -\frac14 {\widehat F}_{\mu\nu} \gamma^{\mu\nu} \epsilon +  \frac{c^{r z}}{c^z} \Big(\frac14 \lambda \, {\bar\chi}_r \epsilon  +\frac12\epsilon \,{\bar\chi}_r \lambda  -\frac18 \gamma_{\mu\nu} \lambda\,{\bar\chi}_r \gamma^{\mu\nu} \epsilon \Big) -\delta_\epsilon \vp^\alpha {\mathcal A}_\alpha^i \sigma_i \lambda \ ,
\nn\w2
\delta_\epsilon \vp^\alpha &=&  V^\alpha_{XA} \,{\bar\epsilon}^A \zeta^X\ ,
\nn\w2
\delta_\epsilon \zeta^X &=& \gamma^\mu \epsilon_A {\widehat P}_\mu^{XA} -\delta_\epsilon \vp^\alpha {\mathcal A}_\alpha^{XY} \zeta_Y\ .
\label{sr}
\eea

The supersymmetry algebra closes only on-shell and provided that $\eta_{IJ} b^{I z} b^{J z'}= 0$ and besides the fermionic equations of motion one also has to use the duality equations \eqref{eq:decov}. 
When $\eta_{IJ} b^{I z} b^{J z'}\ne  0$, there is a  gauge anomaly for the vector gauge transformations, which act by
\begin{align}
\label{eq:Btrm}
\delta_\Lambda A_\mu &= D_\mu \Lambda = \partial_\mu \Lambda + [A_\mu, \Lambda]\ ,
\nn\\
\delta_\Lambda B_{\mu\nu}^I &=  2 b^{Iz} \Tr_z \big( \Lambda \partial_{[\mu} A_{\nu]}\big)\ .
\end{align}
The gauge variation of the pseudo-Lagrangian~\eqref{pL1} is then anomalous and given by
\begin{align}
\label{eq:consans}
\delta_\Lambda \mathcal{L}^{\text{cov}} = \frac1{16}\eta_{IJ} b^{Iz} b^{Jz'} \varepsilon^{\mu_1\ldots \mu_6} \Tr_z \big(\Lambda \partial_{\mu_1} A_{\mu_2}\big) \Tr_{z'} \big( F_{\mu_3\mu_4} F_{\mu_5\mu_6}\big)\ .
\end{align}
This is the well-known  anomaly that solves the Wess--Zumino consistency condition,  here arising from the variation of a classical Lagrangian according to the Green--Schwarz--Sagnotti mechanism.
This anomaly is referred  to as the consistent anomaly.\footnote{Using field equations one can also write a so-called covariant anomaly, the relation between the consistent and the covariant anomalies is explained in  \cite{Riccioni:1998th}.} 
Because of the Wess--Zumino consistency condition mixing  supersymmetry and gauge invariance, there is also a supersymmetry anomaly and $\delta_\epsilon \mathcal{L}^{\text{cov}} = \mathcal{A}_\epsilon$ for  $\mathcal{A}_\epsilon$ that is explicitly given in~\cite[Eq.~(3.71)]{Ferrara:1997gh}. As a consequence, it also appears that the supersymmetry algebra does not close on the  gaugini whenever $\eta_{IJ} b^{I z} b^{J z'}\ne 0$. This obstruction is a consequence of the supersymmetry anomaly as was explained in detail in \cite{Ferrara:1997gh}. 

It is worth noting that the equations of motion of all the fields with the exception of the two-form potential resulting from the pseudo-Lagrangian \eq{pL1}  transform into each other under the supersymmetry transformations. We also note that writing the Yang--Mills and gravitino field equations as $J^\mu$ and ${\cal R}^\mu$, respectively, one finds that $D_\mu J^\mu \ne 0$ and $D_\mu {\cal R}^{\mu} \ne 0$ on-shell, but rather they are proportional to the gauge and supersymmetry anomalies \cite{Riccioni:1998th}. 

In the next section, we will present a proper Lagrangian that implements the duality equations~\eqref{eq:decov} using the Henneaux--Teitelboim method and breaking manifest Lorentz covariance. This proper Lagrangian will, however, still present the same anomalies under supersymmetry and gauge transformation, a feature that is independent of the self-duality of the tensor fields.

\section{Henneaux--Teitelboim form}
\label{sec:HT}

Henneaux and Teitelboim~\cite{Henneaux:1987hz,Henneaux:1988gg} have proposed a way to write a proper action for self-dual fields coupled to gravity that is invariant under diffeomorphisms, but not manifestly so. The action is defined in the time plus space (ADM) decomposition~\cite{Arnowitt:1962hi} of the metric in which 
\begin{align} 
g_{\mu\nu} dx^\mu dx^\nu = - N^2 dt^2 + h_{ij} (dx^i + N^i dt ) (dx^j + N^j dt ) \; , 
\end{align}
where we introduced the shift $N^i$ and the lapse $N$ together with the spatial metric $h_{ij}$. For supergravity, we also need the generalisation of the formalism to local frames and, using the index conventions explained in the beginning of Section~\ref{sec:rev}, we write the vielbein as
\begin{align}
\label{eq:viel}
 e_\mu{}^0 dx^\mu = N dt \; , \quad e_\mu{}^{\underline{a}} dx^\mu = {e}_i{}^{\underline{a}} ( dx^i + N^i dt ) \; , \quad h_{ij} = {e}_i{}^{\underline{a}}  {e}_{j \underline{a}} 
\end{align}
as well as the inverse vielbein
\begin{align}
\label{eq:invviel}
 e_0{}^\mu \partial_\mu = \frac{1}{N} ( \partial_t - N^i \partial_i ) \; , \qquad e_{\underline{a}}{}^\mu \partial_\mu =  e_{\underline{a}}{}^i \partial_i \,.
 \end{align}

In this section, we will show that by including a Chern--Simons coupling to Yang--Mills fields we can  turn the pseudo-Lagrangian~\eqref{bact} into a proper Lagrangian \`a la Henneaux--Teitelboim that can be used for quantisation. Importantly, the Lagrangian depends only on the spatial components of the fields $B_{ij}^I$, while their time component $B_{ti}^I$ only appears as  an integration constant from the equations of motion. In this section we shall  not consider the coupling to hypermultiplets, since the tensor-Yang--Mills sector already captures all the subtleties of the Henneaux--Teitelboim formalism.  The introduction of  hypermultiplets is straightforward without any complication stemming from the formalism.

\subsection{The bosonic Lagrangian}
\label{sec:HT.1}

Performing the Henneaux--Teitelboim analysis on the tensor-Yang--Mills system one arrives at  the following Lagrangian 
\begin{align}
\label{eq:LHT}
\mathcal{L}
&= \sqrt{-g} \Bigl( \frac14 R-\frac14  v_I b^{Iz} \Tr_z F_{\mu\nu} F^{\mu\nu}  - P_{\mu}^r P^\mu_r \Bigr) \nonumber \\
&\quad  - \frac1{48} \eta_{IJ} \varepsilon^{ijklp} \left( \widecheck{H}_{tij}^I - N^q H_{qij}^I \right) H_{klp}^J  -\frac1{24} N \sqrt{h} h^{il} h^{jp} h^{kq} M_{IJ} H^I_{ijk} H^J_{lpq} \nn\\
&\quad+\frac1{8} \eta_{IJ} b^{Iz} \varepsilon^{ijklp}  B_{ij}^J \Tr_z (F_{tk} F_{lp})
\ ,
\end{align}
where the field strengths $\widecheck{H}^I_{tij}$ and $H^I_{ijk}$ include the Yang--Mills Chern--Simons term~\eqref{eq:YMCS} but, importantly, the time component $B_{ti}^I$ of the $B^I$ field is absent in the electric field strength: 
\be 
\widecheck{H}^I_{tij} = \partial_t B_{ij}^I - 6\, b^{Iz}X_{z tij}\,,\qquad 
H^I_{ijk} = 3\partial_{[i} B^I_{jk]} - 6\, b^{Iz} X_{z ijk} \,. 
\label{Hdefs}
\ee
Since the electric field strength differs from the covariant one in~\eqref{eq:HCS}, we have put a check on it to distinguish it. The relation between this Lagrangian and the pseudo-Lagrangian \eq{bact} will be displayed below; see \eq{eq:Eij}.

The Lagrangian~\eqref{eq:LHT} can be obtained from the Hamiltonian formulation in which $H_{ijk}^I$ is the (dual of the)  momentum conjugate to $B^I_{ij}$~\cite{Henneaux:1987hz,Henneaux:1988gg}. In this way $B_{ij}^I$ must only appear in the Lagrangian through  $H_{ijk}^I$ and the Legendre transform term
\be 
- \frac1{16} \eta_{IJ} \varepsilon^{ijklp}  \partial_t B_{ij}^I  \partial_k B_{lp}^J \; . 
\ee
That this is the case with the final Chern--Simons coupling in the Lagrangian can be seen by writing out the terms in~\eqref{eq:LHT} that are not manifestly of this form and using the Bianchi identity~\eqref{eq:XBianchi}
\begin{align}
&\quad  \ \frac1{8} \eta_{IJ} b^{Jz} \varepsilon^{ijklp} \partial_t B_{ij}^IX_{zklp}  + \frac1{8} \eta_{IJ} b^{Iz} \varepsilon^{ijklp}  B_{ij}^J \Tr_z \bigl[ F_{tk} F_{lp}\bigr]  \\
&=  \frac1{8} \partial_t \Bigl( \eta_{IJ}  b^{Jz} \varepsilon^{ijklp}  B_{ij}^I X_{zklp} \Bigr) - \frac{3}{8} \partial_k \Bigl( \eta_{IJ} b^{Jz} \varepsilon^{ijklp} B_{ij}^I X_{ztlp} \Bigr)  +  \frac{3}{8}  \eta_{IJ} b^{Jz}\varepsilon^{ijklp} \partial_k B_{ij}^I  X_{ztlp}   \,.\nn
\end{align}
 
\subsubsection{Equations of motion and duality}

The Euler--Lagrange equation obtained by varying the Lagrangian~\eqref{eq:LHT} with respect to $B_{ij}^I$ can be written as a total spatial derivative 
\begin{align}
\partial_k \left( \frac12\eta_{IJ} \varepsilon^{ijklp} \bigl( \widecheck{H}_{t lp}^J- N^q H_{lpq}^J\bigr) + M_{IJ}  N \sqrt{h} h^{il} h^{jp} h^{kq} M_{IJ} H^I_{lpq}
 \right) = 0 \,.
\end{align}
Using the Poincar\'e lemma, one obtains that it can be integrated up to the introduction of a total derivative 
\be 
\frac12\eta_{IJ} \varepsilon^{ijklp} \bigl( \widecheck{H}_{t lp}^J- N^q H_{lpq}^J\bigr) + M_{IJ}  N \sqrt{h} h^{il} h^{jp} h^{kq} M_{IJ} H^I_{lpq} = \eta_{IJ} \varepsilon^{ijklp}  \partial_l B_{t p}^J \; , \label{DualityHT} 
\ee
and reproduces in this way the covariant self-duality equation~\eqref{eq:bosSD} for the tensor field, including the Chern--Simons terms.

Varying with respect to the Yang--Mills field gives the following  manifestly diffeomorphism covariant equation in form notation 
\begin{equation}
D \bigl( c^z \star  F_z \bigr)  = - b_I{}^z  H^I \wedge  F_z -\frac14  b_I{}^{z} b^{I z'}\Bigl(  \Tr_z [ F\wedge F] \wedge A_{z'}  +2 \Tr_z [ A d A + \tfrac23 A^3 ]  \wedge  F_{z'} \Bigr)\,,
\end{equation}
where~\eqref{DualityHT} was also used as well as  $b_{I}{}^z=\eta_{IJ} b^{Jz}$. 
In this equation,  the $z$-index is not summed over since this is an equation for each simple factor of the gauge group separately; the $z'$ index is summed over, however. This equation is not gauge invariant: Its covariant differential gives
\begin{align}
D D \bigl( c^z \star  F_z \bigr) =  \frac14 \eta_{IJ} b^{Iz} b^{Jz'} \Tr_z [F\wedge F] \wedge dA_{z'}
\end{align}
as a consequence of the consistent anomaly whenever $\eta_{IJ} b^{Iz} b^{Jz'} \ne 0$~\eqref{eq:consans}. 

\subsubsection{Connection to pseudo-Lagrangian}

The bosonic Lagrangian density~\eqref{eq:LHT} can be rewritten as 
\begin{align}
\label{eq:LcovE}
\mathcal{L} &=  \sqrt{-g} \Bigl( \frac14  R - \frac14  v_I b^{Iz} \Tr_z F_{\mu\nu}  F^{\mu\nu} -\frac1{48} M_{IJ} H^I_{\mu\nu\rho}   H^{\mu\nu\rho J} -P^r_\mu P^\mu_r  \Bigr)  
\nn\\
&\quad + \frac1{32}   \varepsilon^{\mu\nu\rho\sigma\kappa\lambda} b_I{}^z B^I_{\mu\nu}  \Tr_z\bigl[ F_{\rho\sigma} F_{\kappa\lambda} \bigr]     -\frac{N}{16} \sqrt{h} h^{ik} h^{jl} M_{IJ} 
 \mathcal{E}_{0ij}^I 
\mathcal{E}_{0kl}^J  
-\partial_i\left[ \frac1{24} \eta_{IJ} \varepsilon^{ijklp} B_{tj}^I H_{klp}^J \right]
\nn\\
& \equiv \ \mathcal{L}^{\text{cov}} + \mathcal{L}^{\mathcal{E}} -\partial_i\left[ \frac1{24} \eta_{IJ} \varepsilon^{ijklp} B_{tj}^I H_{klp}^J \right]
\,,
\end{align}
where $\mathcal{L}^{\text{cov}}$ is the bosonic pseudo-Lagrangian \eq{bact} in the absence of hypermultiplet scalars. The duality equation ${\cal E}^I_{\mu\nu\rho}$ is defined in \eq{eq:bosSD}, and  ${\cal E}^I_{0ij}$ is obtained by converting one index to a time-like tangent space index by use of the inverse vielbein~\eqref{eq:invviel}
\begin{align}
\label{eq:Eij}
\eta_{IJ}  \mathcal{E}_{0ij}^J = \eta_{IJ} e_0{}^\mu \mathcal{E}_{\mu i j}^J =
N^{-1} \eta_{IJ} \left( \widecheck{H}_{tij}^J + 2 \partial_{[i} B_{j]t}^J  - N^k H_{kij}^J \right) + \frac1{6\sqrt{h}} h_{ik} h_{jl} \varepsilon^{klpqr} M_{IJ} H_{pqr}^J\ .
\end{align}
 Here, we have made the dependence on $B_{ti}^I$ explicit by writing $\widecheck{H}_{tij}^J$. It is crucial that  the field $B_{ti}$ that is introduced  in~\eqref{eq:LcovE}  only appears under a total derivative and has no effect on the dynamics which is still that of the non-covariant true Lagrangian~\eqref{eq:LHT}.

The rewriting~\eqref{eq:LcovE} contains the Lorentz covariant pseudo-Lagrangian~\eqref{bact}, a total derivative and the non-covariant term 
\begin{align}
\mathcal{L}^{\mathcal{E}} = - \frac{N}{16} \sqrt{h} h^{ik} h^{jl} M_{IJ} 
\mathcal{E}_{0ij}^I 
\mathcal{E}_{0kl}^J 
= - \frac{N}{16} \sqrt{h} \delta^{\underline{a}\underline{c}}\delta^{\underline{b}\underline{d}} M_{IJ} \mathcal{E}_{0\underline{a}
\underline{b}}^I 
\mathcal{E}_{0\underline{c}\underline{d}}^J 
\label{defce1}
\end{align}
quadratic in~\eqref{eq:Eij}. In the second step we have converted the spatial indices according to 
\be
\mathcal{E}_{0\underline{a}\underline{b}}^I= e_{\underline{a}}{}^i e_{\underline{b}}{}^j \mathcal{E}_{0ij}\ ,
\label{defce2}
\ee
which is more convenient for some calculations.
The reason for the rewriting in~\eqref{eq:LcovE} is that for the pseudo-Lagrangian we can recycle some of the analysis (in particular supersymmetry) done in~\cite{Riccioni:2001bg}.

\subsubsection{Symmetries of the Lagrangian}

Transitioning to a Henneaux--Teitelboim true Lagrangian also implies that diffeomorphism invariance is not manifest, although still  realised through a modification of the transformations of the two-form fields. To understand this, let us first recall the covariant transformation of the two-form under diffeomorphism, as written for the pseudo-Lagrangian and the equations of motion. It is more convenient to combine it with the appropriate vector and tensor gauge transformations,\footnote{\label{fn:locals}The most general transformation
$\delta B_{\mu\nu}^I = \mathcal{L}_{\xi} B^I_{\mu\nu} + 2\partial_{[\mu} \Lambda^I_{\nu]} + 2 b^{Iz} \Tr_z (\Lambda \partial_{[\mu} A_{\nu]})$
leads to~\eqref{eq:dBcov} for $\Lambda=-\xi^\sigma A_\sigma$ and $\Lambda^I_\mu= \xi^\sigma B^I_{\mu\sigma}$.}
 as
\begin{align}
\label{eq:dBcov}
\delta^{\text{cov}}_\xi B_{\mu\nu}^I = \xi^\sigma H_{\sigma\mu\nu}^I +2 b^{Iz} \Tr_z (A_{[\mu} \xi^\sigma F_{\nu]\sigma}) \,.
\end{align}
However, this cannot be the correct transformation in the Henneaux--Teitelboim formalism since it introduces a term $\xi^t H_{tij}^I$ in the transformation of $B_{ij}^I$ and $H_{tij}^I$ contains $B_{ti}^I$ which is not a dynamical variable of the true Lagrangian. Therefore the transformation must be amended to~\cite{Henneaux:1987hz,Henneaux:1988gg}
\begin{align}
\label{eq:deltacovE}
\delta_\xi  B_{\mu\nu}^I &= \xi^\sigma H_{\sigma\mu\nu}^I +2 b^{Iz} \Tr_z (A_{[\mu} \xi^\sigma F_{\nu]\sigma}) - N \xi^t \mathcal{E}^I_{0\mu\nu} \nn\\
&= \delta^{\text{cov}}_\xi B_{\mu\nu}^I  +  \delta^{\mathcal{E}}_\xi B_{\mu\nu}^I \,.
\end{align}
and where we define  $\mathcal{E}_{0t\mu}^I\coloneqq 0$. 
The role of the `non-covariant' term $\delta^{\mathcal{E}}_\xi B_{ij}^I = -N \xi^t \mathcal{E}_{0ij}^I$ is to remove the  occurrence of $B_{ti}^I$. We note that the redefinition only affects the temporal diffeomorphisms with parameter $\xi^t$; the spatial diffeomorphisms with parameter $\xi^q$ are unaffected.  

For the vector field we find
\begin{align}
\delta_\xi^{\text{cov}} A_\mu = \xi^\sigma F_{\sigma\mu} \,,\hspace{10mm}
\delta_\xi^{\mathcal{E}} A_\mu =0\,,
\end{align}
and the vielbein also transforms only under the covariant transformation
\begin{align}
\delta_\xi^{\text{cov}} N &= \xi^\nu \partial_\nu N + N (\partial_t - N^i \partial_i ) \xi^t \,,\nn\\
\delta_\xi^{\text{cov}}N^i &= \xi^\nu \partial_\nu N^i  +  (\partial_t - N^j \partial_j )\xi^i + N^i  (\partial_t - N^j \partial_j )\xi^t -N^2 h^{ij} \partial_j \xi^t\,,\nn\\
\delta_\xi^{\text{cov}}h_{ij} &= \xi^\nu \partial_\nu h_{ij} + 2 \partial_{(i}\xi^k h_{j)k}  +2N^k  h_{k(i} \partial_{j)} \xi^t \,,
\end{align}
where a compensating Lorentz transformation was included in order to preserve the triangular gauge.\footnote{\label{fn:comp}The full transformation is $\delta e_\mu{}^a = \mathcal{L}_\xi e_\mu{}^a - \Lambda^a{}_b e_\mu{}^b$ and the form~\eqref{eq:viel} requires $\Lambda^0{}_{\underline{b}} = N e_{\underline{b}}{}^i \partial_i \xi^t$.}
The non-covariant transformation is $\delta_\xi^{\mathcal{E}} N= \delta_\xi^{\mathcal{E}} N^i= \delta_\xi^{\mathcal{E}} e_i{}^{\underline{a}} = \delta_\xi^{\mathcal{E}} h_{ij} = 0$. 
The scalar fields similarly transform only under the covariant transformation: $\delta_\xi M_{IJ} =\delta_\xi^{\text{cov}} M_{IJ} = \xi^\sigma \partial_\sigma M_{IJ}$.

From this we deduce
\begin{align}
\delta^{\mathcal{E}}_\xi H_{\mu\nu\rho}^I = - 3 \partial_{[\mu} \left( N\xi^t \mathcal{E}_{|0|\nu\rho]}^I\right)\,.
\end{align}

Using the split of the transformation~\eqref{eq:deltacovE} into covariant and non-covariant piece, we can check invariance of the Lagrangian~\eqref{eq:LcovE} by splitting it into four contributions:
\begin{align}
\label{eq:dL4}
\delta_\xi \Bigl( \mathcal{L} +\partial_i\Bigl[ \frac1{24} \eta_{IJ} \varepsilon^{ijklp} B_{tj}^I H_{klp}^J \Bigr] \Bigr) 
=  \delta_\xi^{\text{cov}} \mathcal{L}^{\text{cov}} + \delta_\xi^{\mathcal{E}} \mathcal{L}^{\text{cov}}  + \delta_\xi^{\text{cov}} \mathcal{L}^{\mathcal{E}} + \delta_\xi^{\mathcal{E}} \mathcal{L}^{\mathcal{E}}\,.
\end{align}

Due to the mixing of the local transformations already mentioned in footnote~\ref{fn:locals} and the anomaly~\eqref{eq:consans} of the covariant Lagrangian under vector gauge transformations, the covariant Lagrangian is not invariant under the covariant transformations, and we find for all the four pieces in turn
\begin{align}
\delta_\xi^{\text{cov}} \mathcal{L}^{\text{cov}} &= \partial_\mu ( \xi^\mu  \mathcal{L}^{\text{cov}} ) - \frac14 \eta_{IJ} b^{Iz} b^{Jz'} \Tr_z (\xi^t A_t dA) \wedge \Tr_{z'} F\wedge F\,,\nn\\
\delta^{\mathcal{E}}_\xi \mathcal{L}^{\text{cov}} &= -\frac1{8} \sqrt{h} M_{IJ} \partial_t\left( N \xi^0 \mathcal{E}^I_{0ij} \right) h^{ik} h^{jl} \mathcal{E}_{0kl}^J + \frac1{16} \varepsilon^{ijklp} \eta_{IJ} \partial_i\left(N \xi^t \mathcal{E}_{0jk}^I\right) N \mathcal{E}_{0lp}^J \nn\\
&\quad + \frac{3}{8} \sqrt{h} M_{IJ} N^p \partial_{[p} \left(N \xi^t \mathcal{E}_{0ij]}^I\right) h^{ik} h^{jl} \mathcal{E}_{0kl}^J\,,\nn\\
\delta^{\mathcal{E}}_\xi  \mathcal{L}^{\mathcal{E}}  &= \partial_\mu ( \xi^\mu \mathcal{L}^{\mathcal{E}} ) -\frac1{16} \eta_{IJ} \varepsilon^{ijklp} \partial_i \xi^t N \mathcal{E}_{0jk}^I N \mathcal{E}_{0lp}^J \,,\nn\\
\delta^{\text{cov}}_\xi \mathcal{L}^{\mathcal{E}} &= \frac1{8} \sqrt{h} h^{ik} h^{jl} M_{IJ} \left[ \partial_t \left(N \xi^t \mathcal{E}_{0ij}^I\right) -3 N^p \partial_{[p} \left(N \xi^t \mathcal{E}_{0ij]}^I\right) \right] \mathcal{E}_{0kl}^J\nn\\
&\quad + \frac1{16} \varepsilon^{ijklp} \eta_{IJ} \partial_i \left(N \xi^t\mathcal{E}_{0jk}^I\right) N \mathcal{E}_{0lp}^J\,.
\end{align}
Their sum gives the expected anomaly
\be
\label{eq:dsusyLt}
    \delta_\xi \mathcal{L} =  - \frac14 \eta_{IJ} b^{Iz} b^{Jz'} \Tr_z \bigl[ \xi^t A_t dA \bigr]  \wedge \Tr_{z'} F\wedge F  + \mbox{total derivative terms}\,.
 \ee
Note that this anomaly could be cancelled by undoing the mixing with the vector gauge transformation of footnote~\ref{fn:locals}. 

When doing the calculation leading to this transformation it is important to keep in mind that coordinate transformation of tangent space fields are accompanied by a compensating Lorentz transformation mentioned in footnote~\ref{fn:comp}. On the component $\mathcal{E}_{0\underline{a}\underline{b}}^I$ this implies for example
\begin{align}
    \delta^{\text{cov}}_\xi \mathcal{E}_{0\underline{a}\underline{b}}^I = \xi^\mu \partial_\mu \mathcal{E}_{0\underline{a}\underline{b}}^I
    -\Lambda^0{}_{\underline{c}}\mathcal{E}_{\underline{c}\underline{a}\underline{b}}^I
    = \xi^\mu \partial_\mu \mathcal{E}_{0\underline{a}\underline{b}}^I 
    +\frac12  Ne_{\underline{c}}^i \partial_i \xi^t \varepsilon_{\underline{a}\underline{b}}{}^{\underline{c}\underline{d}\underline{e}} \mathcal{E}_{0\underline{d}\underline{e}}^I\,,
\end{align}
where the spatial component $\mathcal{E}_{\underline{c}\underline{a}\underline{b}}^I$ was dualised to the meaningful $\mathcal{E}_{0\underline{d}\underline{e}}^I$.

\subsection{Global issues}

In Minkowski signature one generically assumes the spacetime to be globally hyperbolic, ensuring that the 1+5 split can be defined globally. On the contrary, considering the theory in Euclidean signature on a generic spin manifold $\mathcal{M}$ requires us to understand how to define the Euclidean action globally modulo $2\pi i$. This is in particular important for the computation of global anomalies, and imposes specific quantisation of the anomaly coefficients $a^I$ and $b^{Iz}$ \cite{Monnier:2017oqd}. For this purpose it is instructive to keep track of all total derivative terms in writing \eqref{eq:dsusyLt}, to understand how the Henneaux--Teitelboim Lagrangian differs from a true density in Minkowski signature. One computes that
\begin{multline}
    \delta_\xi  \Bigl( \mathcal{L} +\partial_i\Bigl[ \frac1{24} \eta_{IJ} \varepsilon^{ijklp} B_{tj}^I H_{klp}^J \Bigr] \Bigr) = - \frac14 \eta_{IJ} b^{Iz} b^{Jz'} \Tr_z \bigl[ \xi^t A_t dA \bigr]  \wedge \Tr_{z'} F\wedge F  \\
    +\partial_\mu \Bigl( \xi^\mu \mathcal{L} +\xi^\mu \partial_i\Bigl[ \frac1{24} \eta_{IJ} \varepsilon^{ijklp} B_{tj}^I H_{klp}^J \Bigr] \Bigr)+ \partial_i \Bigl(  \xi^t \frac{1}{16} \varepsilon^{ijklp} \eta_{IJ}  N \mathcal{E}_{0jk}^I N \mathcal{E}_{0lp}^J \Bigr)\\
    - \partial_\mu \Bigl( \frac{1}{16} e^{-1}\varepsilon^{\mu\nu\rho\sigma\lambda\tau} b_I{}^z  \xi^\kappa B_{\kappa\nu}^I  \Tr_z \big[ F_{\rho\sigma} F_{\lambda\tau}\big]\Bigr)   \, ,  \label{DiffeoGlobal}
\end{multline}
showing that the Lagrangian 
\be 
\mathcal{L} +\partial_i\Bigl[ \frac1{24} \eta_{IJ} \varepsilon^{ijklp} B_{tj}^I H_{klp}^J \Bigr]  \label{GlobalLagrange}
\ee
transforms as a density up to the anomalous term given in \eqref{eq:dsusyLt}, plus the standard gauge variation of the topological term
\be    -\partial_\mu \Bigl( \frac{1}{16} e^{-1}\varepsilon^{\mu\nu\rho\sigma\lambda\tau} b_I{}^z  \xi^\kappa B_{\kappa\nu}^I  \Tr_z \big[ F_{\rho\sigma} F_{\lambda\tau}\big]\Bigr) \label{CovCS} \ee
expected from the covariant analysis, plus the term
\be
\partial_i \Bigl(  \xi^t \frac{1}{16} \varepsilon^{ijklp} \eta_{IJ}  N \mathcal{E}_{0jk}^I N \mathcal{E}_{0lp}^J \Bigr)\; . \label{DualityContact} 
\ee
We find therefore that the Lagrangian \eqref{GlobalLagrange} transforms in the expected form up to the term \eqref{DualityContact} above. However, because this term is a total spatial derivative of a term quadratic in the duality equation, one may argue that it only produces contact terms in the path integral.  

One can implement consistently the Wick rotation for the two-form gauge fields by passing to Euclidean time $t=  - i t_{\scalebox{0.5}{E}} $ and pure imaginary shift $N^i= i N_{\scalebox{0.5}{E}} ^i$.  
Although perturbation theory based on the Henneaux--Teitelboim Lagrangian does not make use of a quantum field $B_{ti}^I$ locally, we see from~\eqref{GlobalLagrange} that the Euclidean path integral on a generic manifold does require the introduction of $B_{ti}^I$ on intersections of open sets. When the two-forms are defined globally in $\Omega^2(\mathcal{M})$, one can a priori use the duality equation $N \mathcal{E}_{0ij}^I=0$ to solve for $B_{ti}^I$ on these intersections. The term quadratic in $N \mathcal{E}_{0ij}^I$ in \eqref{DualityContact} may not be problematic if there is no local operator inserted at the intersections of open sets such that it would not produce non-covariant contact terms in the path integral. This issue is nevertheless subtle and would require further studies to be addressed. 

Moreover, this does not encompass the general situation in which $d B^I$ are non-trivial in cohomology. Because the selfduality equation is incompatible with the integrality in cohomology classes $H^3(\mathcal{M},\mathbb{Z})$ \cite{Witten:1996hc}, one must consider  two-form fields that are not selfdual in the Euclidean path integral and for these the contribution from 
the term quadratic in $N \mathcal{E}_{0ij}^I=0$ does not vanish and is not a priori well defined. Only if $\mathcal{M} = S^1 \times \mathcal{M}_5$ with $t_{\scalebox{0.5}{E}}$ the coordinate on $S^1$ (or if this is true up to a subset of measure zero in $\mathcal{M}_5$ where the radius of $S^1$ vanishes), one can use the Henneaux--Teitelboim action to compute the globally well-defined action, and indeed in this case the action of free chiral two-forms agrees with the one defined in \cite{Belov:2006jd}. 

\subsection{Fermions and supersymmetry}

We now repeat the analysis of Section~\ref{sec:HT.1} in the presence of fermions. The starting point is the covariant supersymmetric pseudo-Lagrangian $\mathcal{L}^{\rm cov}$ of~\eqref{pL1}. 

Again, the Henneaux--Teitelboim form of the Lagrangian can be written as the covariant pseudo-Lagrangian plus a non-covariant term in the duality equations squared, where now the duality equations~\eqref{eq:decov} including fermionic terms have to be used and we find again the Lagrangian  
\begin{align}
\mathcal{L} = \mathcal{L}^{\text{cov}} + \mathcal{L}^{\mathcal{E}} -\partial_i\left[ \frac1{24} \eta_{IJ} \varepsilon^{ijklp} B_{tj}^I H_{klp}^J \right]\ ,
\label{SHT}
\end{align}   
where $\mathcal{L}^{\text{cov}}$ is the covariant pseudo-Lagrangian given in \eq{pL1} and the non-covariant piece is given by
\be
{\mathcal{L}^{\mathcal{E}} = - \frac{N}{16} \sqrt{h} M_{IJ} \widehat{\mathcal{E}}_{0\underline{a}\underline{b}}^I\widehat{\mathcal{E}}_{0}^{J\underline{a}\underline{b}}}\ ,
\label{LE}
\ee
with $\widehat{\mathcal{E}}_{0\underline{a}\underline{b}}^I$ defined in \eq{eq:Eij} and \eq{defce2}, related to the full duality equation \eq{eq:decov} which can be written in tangent space as 
\begin{subequations}
\begin{align}
{\widehat \E}_{abc} &\coloneqq H_{abc} + \frac16\varepsilon_{abc}{}^{def} H_{def}+ \mathcal{O}_{abc} =0\ ,
\label{de1.2}\w2
{\widehat \E}_{abc}^r &\coloneqq H_{abc}^r - \frac16\varepsilon_{abc}{}^{def} H^r_{def}+ \mathcal{O}^{ r}_{abc}=0\ ,
\label{Ede.2}
\end{align}
\end{subequations}
where $H^I_{abc}$ is as defined in \eq{eq:HCS}, and including the hyperini, we have
\begin{subequations}
\label{fb}
\begin{align}
\mathcal{O}_{abc} &= - 3 \bar \psi_{[a} \gamma_b \psi_{c]} - \frac12 \varepsilon_{abc}{}^{def} \bar \psi_d \gamma_e \psi_f +  \frac12 \bchi^r \gamma_{abc} \chi_r - \frac12 {\bar\zeta}^X \gamma_{abc} \zeta_X \,,\\
\mathcal{O}^{r}_{abc} &= - 3 \bar \psi_{[a} \gamma_{bc]} \chi^r   + \frac12 \varepsilon_{abc}{}^{def}  \bar \psi_{d} \gamma_{ef} \chi^r  - c^{rz} \Tr_z \bar\lambda \gamma_{abc} \lambda \,.
\end{align}
\end{subequations}

For studying the invariance under supersymmetry, we now have to work in vielbein form, where we recall that the triangular gauge~\eqref{eq:viel} requires compensating Lorentz transformations, see footnote~\ref{fn:comp}. In the case of supersymmetry, the transformation~\eqref{eq:susytrm} on the vielbein leads to the compensator
\begin{align}
    \Lambda^0{}_{\underline{b}} = e_{\underline{b}}{}^i \, \bar\epsilon \gamma^0 \psi_i\,,
\end{align}
entering in
\begin{align}
\delta_\epsilon^{\text{cov}} N &= \bar \epsilon \gamma^0 (\psi_t - N^i \psi_i)\nn \\
\delta_\epsilon^{\text{cov}} N^i &= e_{\underline{a}}{}^i  \bar \epsilon \gamma^{\underline{a}} (\psi_t - N^i \psi_i) - N h^{ij} \bar \epsilon \gamma^0 \psi_j \\
\delta_\epsilon^{\text{cov}} e_i{}^{\underline{a}} &= \bar \epsilon \gamma^{\underline{a}} \psi_i \,,\nn
\end{align}
where we have put a superscript `cov' on the transformation to indicate that these are the covariant supersymmetry transformations \eq{eq:susytrm} of the pseudo-Lagrangian.

From the absence of $B_{ti}^I$ in the transformation of all fields in the Henneaux--Teitelboim form, we can again read off the non-covariant modification necessary for the supersymmetry transformations and write
\begin{align}
    \delta_\epsilon = \delta_\epsilon^{\text{cov}}+\delta_\epsilon^{\mathcal{E}}\,.
\end{align}
The non-covariant modification $\delta_\epsilon^{\mathcal{E}}$ is only necessary for 
fields transforming into $H^I_{\mu\nu\rho}$, i.e., the fermions and we find: 
\begin{align}
\delta^{\mathcal{E}}_\epsilon \psi_\mu &= \frac{1}{16}   \gamma^{0\ua\ub} \widehat{\mathcal{E}}_{0\ua\ub} \gamma_\mu \epsilon 
& \Rightarrow \hspace{15mm}   \delta^{\mathcal{E}}_\epsilon \bar\psi_\mu &= -\frac{1}{16} \bar\epsilon \gamma_\mu \gamma^{0\ua\ub} \widehat{\mathcal{E}}_{0\ua\ub}  \ ,
\nn\\
\delta^{\mathcal{E}}_\epsilon \chi^r &= \frac{1}{8}  \gamma^{0\ua\ub} \widehat{\mathcal{E}}_{0\ua\ub}^r  \epsilon
&\Rightarrow \hspace{15mm} \delta^{\mathcal{E}}_\epsilon \bar\chi^r &= \frac{1}{8} \bar\epsilon  \gamma^{0\ua\ub} \widehat{\mathcal{E}}_{0\ua\ub}^r \ . 
\label{susyext}
\end{align}

Equipped with these transformation we can again compute the four terms in analogy with~\eqref{eq:dL4}. The first one follows from the analysis in~\cite{Riccioni:2001bg} and is 
\begin{align}
 \delta^{\text{cov}}_\epsilon  \mathcal{L}^{\text{cov}} = \mathcal{A}_\epsilon + \frac{e}{48} \widehat{\mathcal{E}}^{r\, \mu\nu\rho} \widehat{\mathcal{E}}_{\mu\nu\rho} \bar\chi^r \epsilon + \frac{e}{32} \bar\epsilon \gamma^\nu \psi_\mu \left( \widehat{\mathcal{E}}_{\nu\rho\sigma} \widehat{\mathcal{E}}^{\mu\rho\sigma} + \widehat{\mathcal{E}}_{\nu\rho\sigma}^r \widehat{\mathcal{E}}_r^{\mu\rho\sigma} \right)\,,
\end{align}
where the last term can be rewritten as
\begin{align}
 & \qquad \frac{e}{32}  \bar\epsilon \gamma^\nu \psi_\mu \left( \widehat{\mathcal{E}}_{\nu\rho\sigma} \widehat{\mathcal{E}}^{\mu\rho\sigma} + \widehat{\mathcal{E}}_{\nu\rho\sigma}^r \widehat{\mathcal{E}}_r^{\mu\rho\sigma} \right) \\
&=  -\frac1{32} \bigl( \bar \epsilon \gamma^0 \psi_0 - \bar \epsilon \gamma^{\uc}  \psi_{\uc} \bigr) e ( \widehat{\mathcal{E}}_{0\ua\ub} \widehat{\mathcal{E}}^{0\ua\ub} + \widehat{\mathcal{E}}^r_{0\ua\ub} \widehat{\mathcal{E}}^{0\ua\ub}_r) - \frac1{8} \bar \epsilon \gamma^{\ua} \psi_{\ub}  e ( \widehat{\mathcal{E}}_{0\ua\uc} \widehat{\mathcal{E}}^{0\ub\uc} + \widehat{\mathcal{E}}^r_{0\ua\uc} \widehat{\mathcal{E}}^{0\ub\uc}_r)  \nonumber \\
&\quad  +\frac1{64} \varepsilon^{\ua\ub\uc\ud\ue} \bigl( \bar \epsilon \gamma_0 \psi_{\ue} +\bar \epsilon \gamma_{\ue} \psi_0\bigr) e \bigl( -\widehat{\mathcal{E}}_{0\ua\ub}\widehat{\mathcal{E}}_{0\uc\ud} +\widehat{\mathcal{E}}^r_{0\ua\ub}\widehat{\mathcal{E}}_{0\uc\ud\, r}\bigr) \,.
\end{align} 
The supersymmetry anomaly $\mathcal{A}_\epsilon$ is tied to the gauge anomaly and its explicit form can be found in~\cite{Riccioni:2001bg}.

In order to obtain the other contributions to~\eqref{eq:dL4}, we first record that the duality equations themselves are supercovariant and satisfy
\begin{align}
\delta^{\text{cov}}_\epsilon \widehat{\mathcal{E}}_{\mu\nu\rho} &= - \frac12 \bar\epsilon \gamma^\sigma \gamma_{\mu\nu\rho}\widehat{\mathcal{R}}_\sigma +  \frac32 ( \bar \epsilon \gamma_{\sigma} \psi_{[\mu} +\bar \epsilon \gamma_{[\mu} \psi_{\sigma}) \widehat{\mathcal{E}}_{\nu\rho]}{}^\sigma- \frac12 \bar \epsilon \gamma^\sigma \psi_\sigma  \widehat{\mathcal{E}}_{\mu\nu\rho} \,,
\nn\\
\delta^{\text{cov}}_\epsilon \widehat{\mathcal{E}}_{\mu\nu\rho}^r  &= \bar\epsilon \gamma_{\mu\nu\rho} \widehat{\eta}^r+  \frac32 ( \bar \epsilon \gamma_{\sigma} \psi_{[\mu} +\bar \epsilon \gamma_{[\mu} \psi_{\sigma})  \widehat{\mathcal{E}}^r_{\nu\rho]}{}^\sigma-\frac12 \bar \epsilon \gamma^\sigma \psi_\sigma  \widehat{\mathcal{E}}^r_{\mu\nu\rho} \,,
\end{align}
using the field equations~\eqref{eq:EOMf} (without hypermultiplet contributions).
The second term in both equations can be derived for any variation of the metric using 
\begin{align}
& \delta \Bigl( M_{IJ} H^J_{\mu\nu\rho}{-} \frac{1}{6 \sqrt{g}} \eta_{IJ} g_{\mu\sigma} g_{\nu\kappa} g_{\rho\lambda} \varepsilon^{\sigma\kappa\lambda\varsigma\tau\vartheta} H^J_{\varsigma\tau\vartheta}\Bigr) =- 3 \delta g_{\sigma[\mu} \eta_{IJ} ( \star H)_{\nu\rho]}^J{}^\sigma + \frac{1}{2} g^{\sigma\lambda} \delta g_{\sigma\lambda} \eta_{IJ} ( \star H)_{\mu\nu\rho}^J \nonumber \\
&= \frac{3}{2} \delta g_{\sigma[\mu} M_{IJ} \mathcal{E}_{\nu\rho]}^J{}^\sigma - \frac{1}{4} g^{\sigma\lambda} \delta g_{\sigma\lambda} M_{IJ} \mathcal{E}_{\mu\nu\rho}^J  + \mbox{ same terms in }  H^{(-)}_{\mu\nu\rho}  \mbox{ and } H^{r(+)}_{\mu\nu\rho}  \; . \end{align}

From this one obtains\footnote{To vary $(e \widehat{\mathcal{E}}_{0\ua\ub} \widehat{\mathcal{E}}^{0\ua\ub})$, it is convenient to compute the variation of $\sqrt{e}\widehat{\mathcal{E}}_{0\ua\ub}$ to begin with. The result obtained for it below is consistent  with the duality equation because
\begin{equation*}
\delta_\epsilon^{\text{cov}} \bigl( \sqrt{e} \widehat{\mathcal{E}}_{\ua\ub\uc}\bigr) =  - \frac12 \bar\epsilon \gamma^\sigma \gamma_{\ua\ub\uc}\widehat{\mathcal{R}}_\sigma + \frac{3}{4}  \bigl( \bar \epsilon \gamma^{\ud} \psi_{[\ua} -\bar \epsilon \gamma_{[\ua} \psi^{\ud}\bigr) \sqrt{e} \varepsilon_{\ub\uc]\ud}{}^{\ue\uf} \widehat{\mathcal{E}}_{0\ue\uf} - \frac{3}{2}\bigl( \bar \epsilon \gamma_0 \psi_{[\ua} +\bar \epsilon \gamma_{[\ua} \psi_0\bigr) \sqrt{e} \widehat{\mathcal{E}}_{0\ub\uc]}\ .
\end{equation*}} 
\begin{align}
\delta_\epsilon^{\text{cov}} \bigl( \sqrt{e} \widehat{\mathcal{E}}_{0\ua\ub}\bigr) &= \sqrt{e} e_0{}^\mu e_{\ua}{}^\nu e_{\ub}{}^\rho \delta_\epsilon^{\text{cov}}
\widehat{\mathcal{E}}_{\mu\nu\rho} - \frac12 \bigl( \bar \epsilon \gamma^0 \psi_0 - \bar \epsilon \gamma^{\uc} \psi_{\uc}\bigr) \sqrt{e} \widehat{\mathcal{E}}_{0\ua\ub} \nonumber \\
&\quad    + 2 \bar \epsilon \gamma^{\uc} \psi_{[\ua} \sqrt{e} \widehat{\mathcal{E}}_{0\ub]\uc}+ \frac12 \varepsilon_{\ua\ub}{}^{\uc\ud\ue} \bigl( \bar \epsilon \gamma_0 \psi_{\ue} +\bar \epsilon \gamma_{\ue} \psi_0\bigr) \sqrt{e} \widehat{\mathcal{E}}_{0\uc\ud}\nonumber \\
&=  - \frac12 \bar\epsilon \gamma^\sigma \gamma_{0\ua\ub}\widehat{\mathcal{R}}_\sigma + \bigl( \bar \epsilon \gamma^{\uc} \psi_{[\ua} -\bar \epsilon \gamma_{[\ua} \psi^{\uc}\bigr) \sqrt{e} \widehat{\mathcal{E}}_{0\ub]\uc} + \frac14 \varepsilon_{\ua\ub}{}^{\uc\ud\ue} \bigl( \bar \epsilon \gamma_0 \psi_{\ue} +\bar \epsilon \gamma_{\ue} \psi_0\bigr) \sqrt{e} \widehat{\mathcal{E}}_{0\uc\ud}
\end{align}
and
\begin{align}
\delta_\epsilon^{\text{cov}} \bigl( \sqrt{e} \widehat{\mathcal{E}}^r_{0\ua\ub}\bigr) &= \sqrt{e} e_0{}^\mu e_{\ua}{}^\nu e_{\ub}{}^\rho \delta_\epsilon^{\text{cov}} \widehat{\mathcal{E}}^r_{\mu\nu\rho}  - \frac12 \bigl( \bar \epsilon \gamma^0 \psi_0 - \bar \epsilon \gamma^{\uc} \psi_{\uc}\bigr) \sqrt{e} \widehat{\mathcal{E}}^r_{0\ua\ub} \nonumber \\&\quad    + 2 \bar \epsilon \gamma^{\uc} \psi_{[\ua} \sqrt{e} \widehat{\mathcal{E}}^r_{0\ub]\uc}- \frac12 \varepsilon_{\ua\ub}{}^{\uc\ud\ue} \bigl( \bar \epsilon \gamma_0 \psi_{\ue} +\bar \epsilon \gamma_{\ue} \psi_0\bigr) \sqrt{e} \widehat{\mathcal{E}}^r_{0\uc\ud}\nonumber \\
&=  \bar\epsilon \gamma_{0\ua\ub} \widehat{\eta}^r + \bigl( \bar \epsilon \gamma^{\uc} \psi_{[\ua} -\bar \epsilon \gamma_{[\ua} \psi^{\uc}\bigr) \sqrt{e} \widehat{\mathcal{E}}^r_{0\ub]\uc} - \frac14 \varepsilon_{\ua\ub}{}^{\uc\ud\ue} \bigl( \bar \epsilon \gamma_0 \psi_{\ue} +\bar \epsilon \gamma_{\ue} \psi_0\bigr) \sqrt{e} \widehat{\mathcal{E}}^r_{0\uc\ud}\,.
\end{align}
{\allowdisplaybreaks Using all the above results we obtain 
\begin{align}
\delta^{\mathcal{E}}_\epsilon  \mathcal{L}^{\text{cov}} &=  \frac1{16} \bar\epsilon\gamma^\mu \gamma^{0\ua\ub} \widehat{\mathcal{E}}_{0\ua\ub} \widehat{\mathcal{R}}_\mu - \frac1{8} \bar\epsilon \gamma^{0\ua\ub} \widehat{\mathcal{E}}_{0\ua\ub}^r  \eta^r\,
\nn\\
\delta^{\text{cov}}_\epsilon  \mathcal{L}^{\mathcal{E}} &= -  \frac1{16} \bar\epsilon\gamma^\mu \gamma^{0\ua\ub} \widehat{\mathcal{E}}_{0\ua\ub} \widehat{\mathcal{R}}_\mu + \frac1{8} \bar\epsilon \gamma^{0\ua\ub} \widehat{\mathcal{E}}_{0\ua\ub}^r  \eta^r
\nn\\
&\quad - \frac1{32} \varepsilon^{\ua\ub\uc\ud\ue} \bigl( \bar \epsilon \gamma_0 \psi_{\ue} +\bar \epsilon \gamma_{\ue} \psi_0\bigr) e \bigl( -\widehat{\mathcal{E}}_{0\ua\ub}\widehat{\mathcal{E}}_{0\uc\ud} +\widehat{\mathcal{E}}^r_{0\ua\ub}\widehat{\mathcal{E}}_{0\uc\ud\, r}\bigr) 
\nn\\
\delta^{\mathcal{E}}_\epsilon \mathcal{L}^{\mathcal{E}} &= -\frac e 8 \widehat{\mathcal{E}}_{0\ua\ub} \left( \frac18 \bar\epsilon \gamma^{0\uc\ud} \widehat{\mathcal{E}}_{0\uc\ud}^r \gamma^{0\ua\ub} \chi_r + \frac34 \bar\epsilon \gamma_{[0} \gamma^{0\uc\ud} \gamma_{\ua} \psi_{\ub](+)} \widehat{\mathcal{E}}_{0\uc\ud} \right)\nn\\*
&\quad +
\frac e 8 \widehat{\mathcal{E}}^{0\ua\ub\,r}  \left( \frac38 \bar\epsilon \gamma_{[0} \gamma^{0\uc\ud} \gamma_{\ua\ub](-)} \chi^r \widehat{\mathcal{E}}_{0\uc\ud}  + \frac34 \bar\epsilon\gamma^{0\uc\ud} \gamma_{[0\ua} \psi_{\ub](-)} \widehat{\mathcal{E}}_{0\uc\ud}^r
\right)
\nn\\
&=  -\frac1{32} \bigl( \bar \epsilon \gamma^0 \psi_0 - \bar \epsilon \gamma^{\uc} \psi_{\uc} \bigr) e ( \widehat{\mathcal{E}}_{0\ua\ub} \widehat{\mathcal{E}}^{0\ua\ub} + \widehat{\mathcal{E}}^r_{0\ua\ub} \widehat{\mathcal{E}}^{0\ua\ub}_r) - \frac1{8} \bar \epsilon \gamma^{\ua} \psi_{\ub}  e ( \widehat{\mathcal{E}}_{0\ua\uc} \widehat{\mathcal{E}}^{0\ub\uc} + \widehat{\mathcal{E}}^r_{0\ua\uc} \widehat{\mathcal{E}}^{0\ub\uc}_r)  \nonumber \\*
&\quad  +\frac1{64} \varepsilon^{\ua\ub\uc\ud\ue} \bigl( \bar \epsilon \gamma_0 \psi_{\ue} +\bar \epsilon \gamma_{\ue} \psi_0\bigr) e \bigl( -\widehat{\mathcal{E}}_{0\ua\ub}\widehat{\mathcal{E}}_{0\uc\ud} +\widehat{\mathcal{E}}^r_{0\ua\ub}\widehat{\mathcal{E}}_{0\uc\ud\, r}\bigr)
\end{align}
Summing up these expressions we obtain the expected result
 \begin{equation} \delta_\epsilon \mathcal{L} = \mathcal{A}_\epsilon \; .
\end{equation}
} 

\subsection{\texorpdfstring{The case of $n_{\rm T}=1$}{The case of n(T)=1}}

The case of $n_T=1$ is special since the on-shell non-vanishing three-form field strengths $H^{(-)}$ and $H^{{r=1}(+)}$ (see~\eqref{eq:decov}) can be combined to a single duality-condition-free three-form field strength. A manifestly covariant, and classical gauge invariant and supersymmetric model for $n_T=1$ was constructed long ago in the absence of anomalies~\cite{Nishino:1984gk}. Without anomaly, either the three-form field strength includes the Yang--Mills Chern--Simons term or the Lagrangian includes a topological $B \wedge \Tr( F \wedge F)$ term, but not both. Using the supersymmetric Henneaux--Teitelboim form of the theory we have constructed above,  we shall here show how to write the $n_T=1$ Lagrangian including  both the Chern--Simons and the topological term\footnote{This result cannot be obtained directly from the pseudo-Lagrangian by taking $n_T=1$. Rather, one would need to integrate the field equations to an action. Such a result, apart from the bosonic action in \cite{Duff:1996rs}, has not appeared in the literature so far, to our best knowledge.} 

Let us start with the Henneaux--Teitelboim Lagrangian \eq{SHT}. With only the $B^I$ dependent part of ${\cal L}^{\rm cov}$ kept,\footnote{The other terms will not be affected throughout the computation of the $n_T=1$ case we are considering here and thus we are not displaying them.} it reads 
\bea
\cL_H &=& -\frac{e}{48} M_{IJ} H_{abc}^I H^{abc J}+\frac{1}{32} \varepsilon^{\mu\nu\rho\sigma\lambda\tau} b^{Iz} B_{\mu\nu}^J \eta_{IJ} \Tr_z \big( F_{\rho\sigma} F_{\lambda\tau}\big) \CR
&&  -\frac{e}{24}H_{abc}\mathcal{O}^{abc} -\frac{e}{24} H_{abc}^r \mathcal{O}^{abc}_r +\frac{e}{16}\bigl( \widehat{\mathcal{E}}_{0ab} \widehat{\mathcal{E}}^{0ab}+\widehat{\mathcal{E}}^r_{0ab} \widehat{\mathcal{E}}^{0ab}_r \bigr) \ .
\eea
In order to be able to integrate out the dual field, we expand this Lagrangian in the form~\eqref{eq:LHT} as 
\bea
\cL_H &=&  - \frac1{48} \eta_{IJ} \varepsilon^{ijklp} \left( \widecheck{H}_{tij}^I - N^q H_{qij}^I \right) H_{klp}^J  -\frac1{24} N \sqrt{h} h^{il} h^{jp} h^{kq} M_{IJ} H^I_{ijk} H^J_{lpq} \nn\\
&&\quad+\frac1{8} \eta_{IJ} b^{Iz} \varepsilon^{ijklp}  B_{ij}^J \Tr_z (F_{tk} F_{lp})- \frac{1}{12} N \sqrt{h} H^I_{ijk} e^{\underline{a} i} e^{\underline{b} j} e^{\underline{c} k} \bigl( v_I \mathcal{O}_{\underline{abc}} + v_{I r} \mathcal{O}^r_{\underline{abc}} \bigr) \CR
&& +\frac{e}{16}\bigl( \mathcal{O}_{0ab} \mathcal{O}^{0ab}+\mathcal{O}^r_{0ab} \mathcal{O}^{0ab}_r \bigr) \,.
\label{SHT2}
\eea
For a single tensor multiplet, i.e. $n_T=1$, the index $I=(+,-)$ in light-cone basis 
takes two values and our aim will be to integrate out one of the spatial field strengths $H_{ijk}^-$ in a light-cone basis from the Henneaux--Teitelboim Lagrangian above to obtain a covariant Lagrangian for the other field.
To this end
we introduce some notation adapted to breaking the $I=(+,-)$ index via
\bea
b^{Iz} &=& ( b^{+z}, {b}^{-z} )\ ,\qquad v^I= (v^+, v^-)\ , \qquad v^{I r} = (v^+{}_1,  v^-{}_1)\ , 
\nn\\
B^I_{ij} &=& (B^+_{ij},  B^-_{ij})\ ,
\label{single}
\eea
with, by convention, 
$\eta_{IJ} v^J = ( v^- , v^+)$.

Up to the action of $O(1)\times O(1)$ we can always choose a convention in which \eq{eq:sccond} is solved by\footnote{We can use the local action of $O(1)\times O(1)$ to change independently $v^\pm \rightarrow - v^\pm$ or $v^\pm{}_1 \rightarrow - v^{\pm}{}_1$.}
\be 
v^+ = \frac{1}{\sqrt{2} \, y} \; , \quad v^+{}_1 = \frac{1}{\sqrt{2} \, y} \; , \quad v^- = - \frac{y}{\sqrt{2}} \; , \quad v^-{}_1 = \frac{y}{\sqrt{2}} \; . 
\ee
Using the above variables and relations, the Lagrangian \eq{SHT2} for $n_T=1$ can be expressed as 
\bea
\cL_H &=&- \frac1{48} \varepsilon^{ijklp} \left( \widecheck{{H}}_{tij}^- - N^q {H}_{qij}^- \right) H^+_{klp} - \frac1{48} \varepsilon^{ijklp} \left( \widecheck{{H}}_{tij}^+ - N^q {H}_{qij}^+ \right) {H}^-_{klp}  
\CR
&& -\frac1{24} N \sqrt{h} h^{il} h^{jp} h^{kq} \bigl( y^2 H^+_{ijk} H^+_{lpq}+y^{-2} {H}^-_{ijk} {H}^-_{lpq} \bigr) 
\nn\\
&& +\frac1{8} b^{+z} \varepsilon^{ijklp}  {B}^-_{ij} \Tr_z (F_{tk} F_{lp})+\frac1{8} {b}^{-z} \varepsilon^{ijklp}  {B}^+_{ij} \Tr_z (F_{tk} F_{lp})
\CR
&& - \frac{1}{12} N \sqrt{h} {H}^-_{ijk} e^{\underline{a} i} e^{\underline{b} j} e^{\underline{c} k} \frac{y^{-1}}{\sqrt{2}} \bigl( \mathcal{O}_{\underline{abc}} {+}  \mathcal{O}^1_{\underline{abc}} \bigr)- \frac{1}{12} N \sqrt{h} {H}^+_{ijk} e^{\underline{a} i} e^{\underline{b} j} e^{\underline{c} k} \frac{y}{\sqrt{2}} \bigl( - \mathcal{O}_{\underline{abc}} {+} \mathcal{O}^1_{\underline{abc}} \bigr)
\CR
&& +\frac{e}{16}\bigl( \mathcal{O}_{0ab} \mathcal{O}^{0ab}+\mathcal{O}^r_{0ab} \mathcal{O}^{0ab}_r \bigr)\ .
\label{SHT3}
\eea
The terms in the penultimate line can be put in the form
\bea
&&  - \frac{1}{12} N \sqrt{h} {H}^-_{ijk} e^{\underline{a} i} e^{\underline{b} j} e^{\underline{c} k} \frac{y^{-1}}{\sqrt{2}} \bigl( \mathcal{O}_{\underline{abc}} +  \mathcal{O}^1_{\underline{abc}} \bigr)- \frac{1}{12} N \sqrt{h} {H}^+_{ijk} e^{\underline{a} i} e^{\underline{b} j} e^{\underline{c} k} \frac{y}{\sqrt{2}} \bigl( - \mathcal{O}_{\underline{abc}} + \mathcal{O}^1_{\underline{abc}} \bigr)\CR
&=& - \frac{1}{12} N \sqrt{h} G^{\underline{abc}}\frac{y^{-1}}{\sqrt{2}} \bigl( \mathcal{O}_{\underline{abc}} +  \mathcal{O}^1_{\underline{abc}} \bigr)
- \frac{1}{12} e H^{+ abc}  \frac{y}{\sqrt{2}} \bigl( -\mathcal{O}_{{abc}} +  \mathcal{O}^1_{{abc}} \bigr)\ , 
\label{lemma1}
\eea 
where we have used the  self-duality of $\mathcal{O}_{abc}$ and the anti-self-duality of $\mathcal{O}_{abc}^1$ manifest in \eqref{fb}, and defined
\be
G^{\underline{abc}} \coloneqq  e^{\underline{a} i} e^{\underline{b} j} e^{\underline{c} k} \Bigl(  {H}^-_{ijk} + \frac{y^2}{2N \sqrt{h}} h_{is} h_{jt} h_{ku} \varepsilon^{stuwv} \bigl( H^+_{twv}- N^z H^+_{wvz} \bigr) \Bigr) \ .
\ee
Using this result in the Lagrangian \eq{SHT3}, upon  
adding a Lagrange multiplier $B^+_{ti}$ for the Bianchi identity of ${H}^-_{ijk}$  and up to total derivatives, it can be written
as 
\bea 
&& \cL_H +\frac{1}{4} \varepsilon^{ijklp} \partial_i B^+_{t j} \bigl( {H}^-_{klp} + 6 {b}^{-z} X_{z klp} \bigr) 
\\
&=&-e \frac{y^2}{24} H^+_{\mu\nu\rho} H^{+ \mu\nu\rho} + \frac{1}{16} {b}^{-z} \varepsilon^{\mu\nu\rho\sigma\kappa\lambda} B^+_{\mu\nu} \Tr_z F_{\rho\sigma} F_{\kappa\lambda} - \frac{1}{12} e H^{+abc} \frac{y}{\sqrt{2}} \bigl( -\mathcal{O}_{{abc}} +  \mathcal{O}^1_{{abc}} \bigr)
\CR
&& + \frac14 \varepsilon^{\mu\nu\rho\sigma\kappa\lambda} b^{+z} X_{z \mu\nu\rho} b^{-z'} X_{z' \sigma\kappa\lambda}
+\frac{e}{48}  \mathcal{O}^{{abc}}\mathcal{O}^1_{{abc}} \CR
&& - \frac{y^{-2}}{24} N \sqrt{h}\Bigl( G_{\underline{abc}} + \frac{y}{\sqrt{2}} \bigl( \mathcal{O}_{\underline{abc}} +  \mathcal{O}^1_{\underline{abc}} \bigr) \Bigr) \Bigl( G^{\underline{abc}} + \frac{y}{\sqrt{2}} \bigl( \mathcal{O}^{\underline{abc}} +  \mathcal{O}^{1\, \underline{abc}} \bigr) \Bigr) \ ,
\label{res} \nonumber
\eea
where we have used the self-duality of $\mathcal{O}_{abc}$ and the anti-self-duality of $\mathcal{O}^1_{abc}$ that give 
\bea 
&&  -\frac{e}{48}\bigl( \mathcal{O}_{\underline{abc}} \mathcal{O}^{\underline{abc}}+\mathcal{O}^1_{\underline{abc}} \mathcal{O}^{\underline{abc}}_1 \bigr) + \frac{e}{48}    \bigl(  \mathcal{O}_{\underline{abc}} + \mathcal{O}^1_{\underline{abc}} \bigr)  \bigl(  \mathcal{O}^{\underline{abc}} +  \mathcal{O}^{\underline{abc}}_1 \bigr)
\CR
&=& 
\frac{e}{24}   \mathcal{O}^{\underline{abc}}\mathcal{O}^1_{\underline{abc}}= \frac{e}{48}   \mathcal{O}^{{abc}}\mathcal{O}^1_{{abc}}\ .
\eea
Note that notwithstanding the $\pm$ labels, $H^+$ and $H^-$ are not subject to (anti) self-duality conditions. Thanks to the Lagrange multiplier, we can now treat  ${ H}^-_{ijk}$ as an independent field and integrate it out, making the Lagrange multiplier $B^+_{ti}$ a dynamical field. This gives the proper and manifestly covariant Lagrangian for the case of $n_T=1$ from which all field equations can be derived. The bosonic part of this Lagrangian, upon defining $y^2\coloneqq  e^{2\phi}$,  and re-introducing the $B$-independent part, is given by 
\bea
e^{-1}\cL_{\rm B} &=& \frac14 R -\frac14 \partial_\mu\phi \partial^\mu\phi - \frac{1}{24} e^{2\phi} H^+_{\mu\nu\rho} H^{+\mu\nu\rho} 
-\frac{1}{4\sqrt2} \left(- b^{+z} e^\phi +{ b}^{-z} e^{-\phi} \right)\Tr_z (F_{\mu\nu} F^{\mu\nu})  \nn\w2
&& + \frac{1}{16} e^{-1} {b}^{-z} \varepsilon^{\mu\nu\rho\sigma\kappa\lambda} B^+_{\mu\nu} \Tr_z F_{\rho\sigma} F_{\kappa\lambda} + \frac14 e^{-1}\varepsilon^{\mu\nu\rho\sigma\kappa\lambda} b^{+z} X_{z \mu\nu\rho} b^{-z'} X_{z' \sigma\kappa\lambda}\ ,
\label{bact2}
\eea
where $H^+_{\mu\nu\rho} = 3\partial_{[\mu} B^+_{\nu\rho]} - 6b^{+z} X_{z\mu\nu\rho}$. This is in agreement with the action discussed in \cite{Duff:1996rs}.

{\allowdisplaybreaks 
The remaining part of the full Lagrangian is given by the sum of  all quartic fermion terms in \eq{pL} that are independent of $H$ (with the hypermultiplets suppressed), and the last term in \eq{res}. As for the supertransformations, they are obtained from \eq{sr} and \eq{susyext}, where the duality equation $H_{abc}^-= \frac16 \varepsilon_{abc}{}^{def} e^{2\phi}  H^{+}_{def}$ is to be used to remove $H_{abc}^-$ in favour of $H_{abc}^+$, and up to cubic fermion terms they take the form~\cite{Nishino:1984gk}
\begin{align}
\delta_\e e_\mu{}^a &= {\bar\epsilon}\gamma^a \psi_\mu\ ,
\nn\\
\delta_\e B_{\mu\nu}^+ &= -\frac{1}{\sqrt2} e^{-\phi} \left(2{\bar\psi}_{[\mu}\gamma_{\nu]}- {\bar\epsilon}\gamma_{\mu\nu} \chi\right) \ ,
\nn\\
\delta_\e \phi &= {\bar \e}\chi\ ,
\\
\delta_\e \psi_\mu &= D_\mu \epsilon + 
\frac{\sqrt{2}}{48} e^\phi H_{\nu \rho \sigma}\gamma^{\nu \rho \sigma}\gamma_\mu \epsilon \ ,
\nn\\
\delta_\e \chi &= \frac12 \gamma^\mu \partial_\mu \phi\, \epsilon - \frac{\sqrt{2}}{24} e^\phi \gamma^{\mu\nu\rho} H_{\mu\nu\rho} \epsilon \ ,\nn
\end{align}
where we have set $\chi^1 \equiv \chi$.\footnote{Note that relative to \cite{Nishino:1984gk} the gaugino has been rescaled by a dilaton-dependent factor.}
} 

Integrating out ${H}^-_{ijk}$ was a choice and we could have equivalently integrated out $H^+_{ijk}$ and obtained a dual Lagrangian for the covariant ${B}^-_{\mu\nu}$. Doing so amounts to the replacements $\phi\to -\phi$, $B^+\to B^-$, $b^{+z} \leftrightarrow b^{-z}$ and change the overall sign of the Yang--Mills kinetic term in the results above.

Finally, we note that setting $b^{-z}=0$ gives the Lagrangian which was constructed long ago in \cite{Nishino:1984gk}, which is classically gauge invariant and supersymmetric. Setting $b^{z+}=0$ instead gives the dual formulation \cite{Nishino:1985xp} which is also gauge invariant.

\section{Higher-derivative extension of the model}
\label{sec:BdR}
In order to construct the $R^2$ type corrections it is convenient to use the Bergshoeff--de Roo trick  which is based on finding a Poincar\'e to Yang--Mills map in the heterotic string frame in ten dimensions \cite{Bergshoeff:1989de}.  If one distinguishes the two-derivative Yang--Mills Lagrangian as multiplied by $\beta$ and the $R^2$ correction as multiplied by $\alpha$, the Bergshoeff--de Roo supersymmetric Lagrangian takes the schematic form 
\be 
\cL = R - H^2   - \alpha \Tr R(\omega_-)^2 - \beta \Tr F^2 + \alpha\,  t_8  \bigl( \alpha \Tr R(\omega_-)^2 + \beta \Tr F^2 \bigr)^2 + \dots  
\ee
and the order $\alpha^2$ and $\alpha \beta$ terms are all comprised in the definition of three-form field strength $H$ and $\omega_- = \omega - \tfrac12 H$.  The two-derivative Lagrangian described in Section~\ref{sec:rev} corresponds to the truncation at $\alpha = 0$. In this case the supersymmetry transformations are known exactly as we have reviewed. Because $\alpha$ has the dimension of a length squared, the corrections in $\alpha$ to the action are higher derivative. Only when one benefits from an off-shell formulation one can hope to get supersymmetric higher-derivative invariants that do not require a modification of the (off-shell) supersymmetry transformations. For $n_T>1$ there is no such a formulation, and one can only hope to solve the problem perturbatively as a formal expansion in $\alpha$. In this paper we shall only consider the leading correction linear in $\alpha$ and to all orders in $\beta$. 

Before starting this section, let us quickly review general facts about the low-energy expansion in $\alpha$. The perturbative expansion of the Lagrangian and the supersymmetry transformations expand as 
\be S = \sum_{n=0}^\infty \alpha^n S^{\scalebox{0.6}{$(n)$}} \; , \qquad \delta_\epsilon = \sum_{n=0}^\infty \alpha^n \delta_\epsilon^{\scalebox{0.6}{$(n)$}}  \; . \ee
In the low-energy effective action one can consider $\alpha$ as a small parameter, and one is allowed to use field redefinitions. More formally this is well described within the Batalin--Vilkovisky formalism \cite{Henneaux:1994lbw}. The application of the Noether procedure can then be formulated as the cohomology problem of finding a cohomology class of ghost number 0 of the Batalin--Vilkovisky  BRST operator in the local functionals of the fields defined modulo total derivatives. This cohomology is isomorphic to the cohomology of $\delta_\epsilon^{\scalebox{0.6}{$(0)$}}$
inside the Koszul--Tate cohomology, i.e. in the set of local functionals of the fields satisfying the first order equations of motion of $S^{\scalebox{0.6}{$(0)$}}$ \cite{Henneaux:1994lbw,Barnich:1994db}. It is therefore enough to find that $  \delta_\epsilon^{\scalebox{0.6}{$(0)$}} S^{\scalebox{0.6}{$(1)$}} \approx0 $ modulo the equations of motion to ensure the existence of $\delta_\epsilon^{\scalebox{0.6}{$(1)$}}$ such that 
\be \delta_\epsilon^{\scalebox{0.6}{$(0)$}} S^{\scalebox{0.6}{$(1)$}}+\delta_\epsilon^{\scalebox{0.6}{$(1)$}} S^{\scalebox{0.6}{$(0)$}} = 0 \; . \ee
Let us note nonetheless that there are two complications that do not allow us to apply directly the theorem of \cite{Henneaux:1994lbw,Barnich:1994db} in (1,0) supergravity. The first is that we shall use the duality equation for the three-form that is not strictly speaking an Euler--Lagrange equation  for $S^{\scalebox{0.6}{$(0)$}}$, only its spatial curl is in the Henneaux--Teitelboim formulation. 
The second is related to the anomaly and comes from the Green--Schwarz--Sagnotti mechanism, since $ \delta_\epsilon^{\scalebox{0.6}{$(0)$}} S^{\scalebox{0.6}{$(0)$}} \ne 0$ and $(\delta_\epsilon^{\scalebox{0.6}{$(0)$}})^2 \not\approx 0$ modulo the equations of motion. Here we shall ignore possible difficulties associated to these two complications  and will not discuss in detail the solution to the Wess--Zumino consistency condition at order $\alpha$.  

In order to find a solution $S^{\scalebox{0.6}{$(1)$}} $ for an $R^2$ type supersymmetry invariant we will check that that the fields of the theory can be mapped to a given off-shell formulation for which one can write an off-shell  supersymmetry invariant (which would then give the complete $\alpha$ expansion after integrating out the auxiliary fields perturbatively). Because the map is only valid modulo the equations of motion of $S^{\scalebox{0.6}{$(0)$}} $, in our case we only obtain instead the first order correction to the action in $\alpha$.

\subsection{Off-shell Poincar\'e multiplet from tensor calculus}

The Bergshoeff--de Roo trick was applied  in six-dimensional minimal supergravity coupled to a single tensor multiplet using the off-shell Poincar\'e multiplet \cite{Bergshoeff:2012ax}. The latter contains the dilaton scalar $L$ and a single Kalb--Ramond field $\mathcal{B}_{\mu\nu}$~\cite{Bergshoeff:1985mz}.
The $R^2$ type correction can then be identified with a one-loop $R^2$ correction in type IIA string frame, with the identification $L = \scalebox{0.9}{Vol(K3)} e^{- 2 \phi_{\scalebox{0.4}{IIA}}}$ for a reduction on a K3 surface.\footnote{ The truncation to (1,0) supergravity of the complete one-loop correction in type IIA requires also the inclusion of another supersymmetry invariant \cite{Novak:2017wqc}, but this will play no role in our discussion.}  Poincar\'e supergravity in string frame can be obtained as a specific gauge fixing of the dilaton-Weyl multiplet coupled to a linear multiplet with fields \cite{Bergshoeff:1985mz} (writing out the symplectic indices)
\be
\{ e_\mu{}^a, \psi_\mu^A, \mathcal{B}_{\m\n}, V_\m^{AB}, b_\mu, \psi^A, \sigma\}\ ,\qquad \{E_{\m\n\rh\sigma}, \varphi^A, L^{AB} \} \ ,
\label{wL}
\ee
where $e_\mu{}^a$, $\psi_\mu^A$, $V_\mu^{AB}$ and $b_\mu$ are the gauge fields for the superconformal transformations, $E_{\m\n\rh\sigma}$ is a totally antisymmetric gauge field, $\psi^A$ and $\varphi^A$ are anti-chiral symplectic-Majorana fields, and $\sigma$ and $L^{AB}$ are real scalar fields. We use the convention that $L^{AB}= L^{BA} = L^i \sigma_i^{AB}$, where $i=1,2,3$ is the $Sp(1)_{\rm R}$ triplet index and $\sigma^i_A{}^B$ are the Pauli matrices. The off-shell supertransformations of these multiplets are given in \cite{Bergshoeff:1985mz}. To obtain the off-shell Poincar\'e supergravity in string frame, a convenient set of gauge fixing conditions are \cite{Bergshoeff:2012ax}\footnote{Note that $\sigma_2^{AB} = \delta^{AB}$ in our conventions.}
\be
\sigma =1\ , \qquad L^{AB} = \frac{L}{\sqrt2}\delta^{AB} \ ,\qquad \psi^A=0\ ,\qquad b_\mu=0\ .
\label{cg}
\ee
This gauge choice breaks the R-symmetry group $Sp(1)_{\rm R}$ down to $U(1)_{\rm R}$. The compensating local $Sp(1)_{\rm R}$ transformation 
\be 
\Lambda^i = \bar \epsilon \sigma^i \chi
\label{ChooseSU2}
\ee
is determined up to a local $U(1)_{\rm R}$ transformation along the $i=2$ component. In \cite{Bergshoeff:2012ax}, the component $\Lambda^2$ was chosen to vanish, but we find it to be more convenient to use \eqref{ChooseSU2} such that almost all supersymmetry transformations are $Sp(1)_{\rm R}$ covariant. Defining $\chi_A=\delta_{AB}\varphi^B/(2L)$, in the gauge \eq{cg} we find\footnote{To compare these results with those of \cite{Bergshoeff:2012ax}, we need to send the fields there to ours as follows 
\bea
&& \psi_\mu \to \sqrt{2} \psi_\mu\ , \qquad \epsilon\to\sqrt{2} \epsilon\ , \qquad V_{\mu AB}\to V_\mu^i (\sigma^i)_{AB}\ ,
\nn\\
&& \varphi \to -2i\, L \sigma^2 \chi \ , \qquad E_\mu \to \sqrt2 \widehat{E}_\mu\ ,\qquad E_{\mu\nu\rho\sigma} \to \sqrt2 E_{\m\n\rho\sigma}\ .
\nn
\eea
In \cite{Bergshoeff:1985mz} the hat notation is not used for the supercovariant $E_\mu$, and in \cite{Bergshoeff:2012ax} $E_\mu$ is purely bosonic, defined as $(1/4!)\varepsilon^{\m\n_1...\n_5} \partial_{\n_1} E_{\n_2...\n_5}$. In going from \cite{Bergshoeff:1985mz} to \cite{Bergshoeff:2012ax}, one needs to also send $V_\m^{AB} \to -2 V_\m^{AB}$. } 
\bea
\delta_\epsilon \chi&=& \frac1{4L} \gamma^\mu  \partial_\mu L\,  \epsilon - \frac{1}{2}  (\bar \psi_\mu  \sigma^i \sigma_2 \chi) \sigma_2 \sigma_i  \gamma^\mu \epsilon  - \frac12 ( V_a^{ i} \sigma_i  - V_a^2 \sigma_2)  \gamma^a \epsilon - \frac{1}{24} \widehat{\mathcal{H}}_{abc} \gamma^{abc} \epsilon 
\CR
&& -2  \chi \bar \epsilon \chi   - \frac{i}{4 L}   \gamma^a \widehat{E}_a  \sigma_2 \epsilon + \Lambda_i \sigma_2 \sigma^i \sigma_2 \chi \ ,
\label{dchi}
\eea
where $\Lambda^i$ is given in~\eqref{ChooseSU2}. 
{\allowdisplaybreaks
Altogether, the resulting supersymmetry transformations of the off-shell Poincar\'e multiplet
\be
\{ e_\mu{}^a, \psi_\mu^A, \mathcal{B}_{\m\n}, V_\mu^i, E_{\m\n\rh\sigma},\chi^A, L \} \ ,
\label{pm}
\ee
upon taking into account the compensating symmetry transformations needed to stay in the gauge \eq{cg} as detailed in \cite{Bergshoeff:2012ax}, are given by \cite{Bergshoeff:2012ax}
\bea
\delta_\epsilon e_\mu{}^a &=&  \eb \gamma^a \psi_\mu\ ,
\nn\w2
\delta_\epsilon \psi_\mu &=& D_\mu (\oh_+,V) \e +  \Lambda^i \sigma_i  \psi_\mu\ ,
\nn\w2
\delta_\epsilon \mathcal{B}_{\mu\nu} &=& -2 \eb\gamma_{[\mu} \psi_{\nu]}\ ,
\nn\w2
\delta_\epsilon \chi &=& \frac1{4L} \gamma^\mu  \widehat{\partial_\mu L}\,  \epsilon- \frac12 ( V_a^{ i} -{\bar\chi}\sigma^i\psi_a  ) \sigma_i  \gamma^a \epsilon - \frac{1}{24} \widehat{\mathcal{H}}_{abc} \gamma^{abc} \epsilon  - \chi \bar \epsilon \chi \CR
 &&   + \frac{1}{2} \Bigl( - \frac{i}{2 L} \widehat{E}_a  + V_a^2 -{\bar\chi}\sigma_2 \psi_a + \bar \chi \sigma_2 \gamma_a \chi  \Bigr) \gamma^a \sigma_2  \epsilon  \ , 
\nn\w2
\delta_\epsilon L  &=& 2 L \eb \chi\ ,
\nn\w2
\delta_\epsilon E_{\m\n\rho\sigma} &=& -2 i L \eb \sigma_2 \big( 2 \gamma_{[\m\n\rho}\psi_{\sigma]} + \gamma_{\m\n\rho\sigma} \chi \big) \ ,
\nn\w2
\delta_\epsilon V_\mu^i  &=&  - \frac12 e_\mu{}^a\bar\epsilon \sigma^i  \gamma^b {\widehat \rho}_{ab +} - \frac1{12} \widehat{\mathcal{H}}_{abc} \bar \epsilon \sigma^i \gamma^{abc} \psi_\mu - \partial_\mu \Lambda^i  +2i\ve^{i}{}_{jk} \Lambda^j V_\mu^k\ . 
\label{DilWeylSusy}
\eea
where we have added the cubic terms in fermions using \cite{Bergshoeff:1985mz}. The result for $\delta_\epsilon \chi$ is a rewriting of~\eqref{dchi} by following the following steps. First we supercovariantise $\partial_\mu L$, and observe that
\be
\frac12 \left({\bar\psi}_\mu \chi\right) \gamma^\mu\e - \frac{1}{2}  (\bar \psi_\mu  \sigma^i \sigma_2 \chi) \sigma_2 \sigma_i  \gamma^\mu \epsilon= \frac12 \left({\bar\chi} \sigma^i\psi_\mu\right) \sigma_i\e -\frac12 \left({\bar\chi}\sigma_2 \psi_\mu\right) \sigma_2 \e\ .
\ee
Next, we find by Fierz rearrangement that 
\be
\chi{\bar\chi}\e +\sigma_2\sigma^i\sigma_2 \chi \left({\bar\chi}\sigma_i\right) =  -\frac12 \left( {\bar\chi}\sigma_2 \gamma^a \chi\right) \gamma_a\sigma_2\e\ .
\ee
We have also used $V_\mu^i = - \frac12 \sigma^i_{AB} V_\mu^{AB}$ and 
\be
\widehat{\cal H}_{\mu\nu\rho} = 3\,\partial_{[\mu} {\cal B}_{\nu\rho]} +3{\bar\psi}_{[\mu} \gamma_\nu \psi_{\rho]}\ .
\label{pureH}
\ee
Here we write $\mathcal{B}_{\mu\nu}$ for the off-shell Poincar\'e multiplet two-form, in order not to confuse it with the set of two-forms $B_{\mu\nu}^I$ of the theory coupled to $n_{\rm T}$ tensor multiplets. 
} 

In the presence of several tensor multiplets the three-form field strength acquires Chern--Simons couplings of the form
\be 
H^I = d B^I + b^{I z} \Tr_z\Bigl[  A d A + \frac{2}{3} A^3\Bigr] - a^I \Tr\Bigl[ \omega d \omega  + \frac{2}{3} \omega^3\Bigr] \ ,  
\ee
where the constants $a^I$ defining the Lorentz--Chern--Simons term in the definition of $H^I$ determine the corresponding $R^2$ type correction to the effective action. The gravitational Chern--Simons term is higher order in derivatives and for this reason did not appear in the previous sections. We can write a covariant Weyl rescaling with respect to the moduli dependent scalar $y = v_I \ag^I$. For a single tensor multiplet and when $\ag^I$ is light-like, i.e. $\eta_{IJ} \ag^I \ag^J = 0 $, one can identify $L = y^{-2}$ with the effective type IIA dilaton in six dimensions. The Weyl rescaling to ``type IIA string frame'' can in this way be generalised to an arbitrary number of tensor multiplet and a non-light-like vector $\ag^I$. Note however, that type IIA string constructions of (1,0) supergravity in six dimensions generally give a single tensor multiplets, and only in type IIB one can get multiple tensors. The tensor multiplet scalar fields include generally the K\"{a}hler structure moduli of the four-dimensional base in F-theory compactifications~\cite{Vafa:1996xn}, in particular the volume of K3 and Kalb--Ramond fields over K3 two-cycles in perturbative orbifold constructions \cite{Gimon:1996ay,Dabholkar:1996ka}.  The type IIB axio-dilaton is always in the hypermultiplet sector. Nevertheless, we shall refer to the frame obtained by Weyl rescaling with respect to $y$ as the  ``type IIA string frame'', or simply string frame for short.

We will show that there is a map from the field content of (1,0) supergravity coupled to $n_{\rm T}$ tensor multiplet in this frame to the off-shell Poincar\'e supermultiplet introduced above. In this way we will be able to use the results of \cite{Bergshoeff:2012ax} that gives an explicit map from the off-shell Poincar\'e spin connection $\widehat{\omega}_-$ and the Rarita--Schwinger field strength $\widehat{\rho}_+$  to the off-shell Yang--Mills multiplet and derive the full $R^2$-type supersymmetry invariant to order $\alpha$, including the octic fermion terms. 

\subsection{The string frame and the embedding}
\label{StringFrameSection}

As a first step towards finding the Poincar\'e to Yang--Mills map, starting from the supertransformation \eq{sr},  we go to string frame by redefining the fields as follows  
\bea
e_\mu{}^a  &=& y^{-1/2} e'_\mu{}^{ a}\ ,
\nn\\
\psi_\mu &=& y^{-\frac14} \Big( \psi_\mu' 
+ \frac12 y^{-1} y_r e'_\mu{}^{ a} \gamma_a \chi'^r \Big)\ ,
\nn\\
\chi^r &=& y^{\frac14} \chi^{\prime r}\ , 
\nn\\
\zeta^X &=& y^{\frac14} \zeta^{\prime X}\ ,
\nn\\
\delta \lambda &=& y^{\frac34} \lambda'\ ,
\label{defs}
\eea
where the primed fields are in  string frame, $y'=y$, $\vp'=\vp$ and 
\be
y:= a_I v^I\ ,\qquad y_r := a_I v^I_r\ .
\ee
We also redefine the supersymmetry parameter and transformation by 
\begin{equation}
    \epsilon = y^{-\frac14} \epsilon'\ ,\qquad  
    \delta_\e + \delta_\Lambda = \delta_{\e'}\ ,  
\end{equation}
where $\Lambda_{ab} = \frac12 y^{-1}y^r\,\eps\gamma_{ab} \chi_r \in \mathfrak{so}(1,5)$ is the Lorentz rotation that is required to put the supertransformation of the vielbein into canonical form.

For example, to obtain the supertransformation of the vielbein in  string frame, we proceed as follows: 
\be 
\delta_{\e'} \big(y^{-1/2} e_\mu^{\prime}{}^a \big)=  (\delta_\e  + \delta_\Lambda)  e_\mu{}^a =  \Big( \eps \gamma^a \psi_\mu -\Lambda^{a}{}_{b} e_{\mu}{}^b \Big)_{\Phi \to \Phi'}\ ,
\ee
where the notation $()_{\Phi \to \Phi'}$ indicates that we express all the fields in terms of the string frame fields according to the map defined above. In this example this gives 
\be
 e^\prime_\mu{}^a \delta_{\e'} y^{-\frac12} + y^{-\frac12} \delta_{\e'}  e^\prime_\mu{}^a = y^{-\frac12} \eps^\prime \gamma^a \Big(\psi^\prime_\mu + \frac12 y^{-1}y^r \gamma_\mu \chi^\prime_r \Big) -\frac12 y^{-\frac32} y^r \eps\gamma^{a}{}_b\chi^\prime_r e^\prime_{\mu}{}^b\; . 
 \ee
From this formula, and using $\delta_{\e'} y= - y^r \eps^\prime \chi^\prime_r$, we readily get
\be
\delta_{\e'} e^\prime_\mu{}^a = \eps^\prime \gamma^a \psi^\prime_\mu\ . 
\ee
For short we shall \mbox{\it drop all the primes} in the following and all the fields in this section are from now on understood to be in the dual string frame unless we specify otherwise. Using the procedure described above, we find that the supertransformations \eq{sr} in the string frame take the form: 
\bea
\delta_\epsilon e_\mu{}^a &=& {\bar\epsilon} \gamma^a \psi_\mu\ ,
\nn\w2
\delta_\epsilon B_{\mu\nu}^I &=& -2 y^{-1}\, v^I {\bar \epsilon} \gamma_{[\mu} \psi_{\nu]} +y^{-1}\big( v^{Ir} - y^{-1} y^r v^I \big)\, {\bar\epsilon} \gamma_{\mu\nu} \chi_r \ ,
\nn\w2
\delta_\epsilon v_I &=& -v^r{}_I {\bar\epsilon} \chi_r\ ,  \hspace{20mm} \delta_\epsilon v^r{}_I = -v_I \eps \chi^r\ , 
\nn\w2
\delta_\epsilon \psi_\mu &=& D_\mu ({\widehat \omega}_+, V) \e +(\eb\sigma^i\chi) \sigma_i\psi_\m  \ ,
\nn\w2
\delta_\epsilon \chi^r &=& -\frac12 {\widehat P}_\mu^r \gamma^\mu \e -\frac{1}{24} y{\widehat H}^r_{\mu\nu\rho} \gamma^{\mu\nu\rho} \e + \frac14\chi^r  (\eps\chi) -\frac18 \gamma^{ab}\chi^r (\eps\gamma_{ab}\chi) 
\nn\w2
&& +\frac14 \gamma^a\e ({\bar\chi}\gamma_a \chi^r)  -\frac{1}{16} \gamma^{abc} \e ({\bar\chi}\gamma_{abc} \chi^r )\ ,
\label{susyprime}
\eea
where 
\begin{subequations}
\be
V_\mu^i = X_\mu^i+{\bar\chi} \sigma^i \psi_\mu
\label{key1}
\ee
\be
\chi = y^{-1} y_r \chi^r\ ,
\label{defCV}
\ee
\end{subequations}
and 
\begin{subequations}
\begin{align}
X_\mu^i &= \frac14 \big( {\bar\chi}\gamma_\mu \sigma^i \chi - {\bar\chi}^r \gamma_\mu \sigma^i \chi_r \big) \ ,
\w2
D_\mu (\widehat\omega_+, V) \e &= D_\mu(\widehat\omega_+)\e + V_\mu^i \sigma_i \epsilon\ ,
\label{cd1}
\end{align}
\be
{\widehat\omega}_{\mu\pm}{}^{ab} = {\widehat\omega}_\mu{}^{ab} \pm \frac12 \ag_I {\widehat H}^I_\mu{}^{ab}\ .
\label{defoh}
\ee
\end{subequations}
In obtaining the transformation rule for the gravitino, we have used the duality equation \eq{Ede}. The occurrence of $\ag_I H^I$ in $\delta_\epsilon \psi_\mu$ is obtained thanks to the identity
\be
y H = -\ag_I H^I + y_r H^r\ ,
\label{id}
\ee
and the second term above can be replaced by a bilinear in fermion using the duality equation \eqref{Ede}. We will write explicitly the contraction $\ag_I H^I$ so that it is not confused with $H:= v_I H^I$. The supercovariant fields ${\widehat\omega}_{\mu ab}$ and  ${\widehat P}_\mu^r$  have the same form as in \eq{scf}, with all fields understood to be in the dual string frame. However, the covariant field strength ${\widehat H}_{\mu\nu\rho}^I$  is given in the dual string frame by\footnote{Recall that we are not considering the coupling to Yang--Mills multiplets in this section.}
\be
{\widehat H}_{\mu\nu\rho}^I = 3\partial_{[\mu} B_{\nu\rho]}^I +3y^{-1} v^I {\bar\psi}_{[\mu} \gamma_\nu \psi_{\rho]} 
 +3y^{-1}\big( v^{r I} - y^{-1} y^r v^I  \big){\bar\chi}_r \gamma_{[\mu\nu} \psi_{\rho]}\ ,
\label{defsprime}
\ee
and the duality equations are modified to 
\bea
{\widehat\E}_{\mu\nu\rho} &=& 2 {\widehat H}_{\mu\nu\rho}^{(+)} +\frac12 y^{-1}{\bar\chi^r}\gamma_{\m\n\rh}\chi_r +\frac32 y^{-1}{\bar\chi} \gamma_{\m\n\rh} \chi\ ,
\nn\w2
{\widehat\E}_{\mu\nu\rho}^r &=& 2 {\widehat H}_{\mu\nu\rho}^{r(-)} \ .
\eea
The fields $V_\mu^i$ and $\chi$ defined in \eq{defCV} turn out to transform as they should in the off-shell Poincar\'e multiplet \eq{pm}, as we shall see below. The vielbein and the gravitino field are identified without modification. For the remaining members of the off-shell Poincar\'e multiplet, we find that the following identifications are appropriate:
\begin{subequations}
\bea
L &=& y^{-2}\ ,
\label{defL}\w2
\mathcal{B}_{\mu\nu} &=& \ag_I B^{I}_{\mu\nu} \ ,
\w2
\widehat{E}^\mu &=& \frac{1}{24} \varepsilon^{\mu\nu\rho\sigma\kappa\lambda} \widehat{\partial_\nu E_{\rho\sigma\kappa\lambda}}=  -\frac{i}{2 y^2} \Bigl(5  \bar \chi \gamma^\mu \sigma^2 \chi -  \bar\chi_r \gamma^\mu  \sigma^2 \chi^r \Bigr)\ .
\label{def2}
\eea
\end{subequations}
To see this, to begin with we note that $L$ as defined above transforms as in \eq{DilWeylSusy}. Next, contracting $\delta_\epsilon B_{\mu\nu}^I$ in \eq{susyprime} with $a_I$, we readily obtain the formula for $\delta_\epsilon {\mathcal B}_{\mu\nu}$ as in \eq{DilWeylSusy}.  Turning to the supertransformation of the dilaton defined in \eq{defCV}, we find
\bea 
\delta_\epsilon \chi
&=& -\frac12 {\widehat P}_\mu \gamma^\mu \e -\frac{1}{24} \ag_I{\widehat H}^I_{\mu\nu\rho} \gamma^{\mu\nu\rho} \e  -\frac12 X_\mu^i \gamma^\mu \sigma_i \e -(\eps\chi) \chi- \frac{1}{48} y \widehat{ \mathcal{E}}_{\mu\nu\rho} \gamma^{\mu\nu\rho} \e \ .
\label{cdc}
\eea
We find that this result agrees with the supertransformation of the dilatino in \eq{DilWeylSusy}, where we use our ansatz for ${\widehat E}^\mu$. Note that even though $\delta_\epsilon \chi$ is not $Sp(1)_{\rm R}$ covariant in \eq{DilWeylSusy}, the elimination of ${\widehat E}^\mu$ using \eqref{def2} in \eq{DilWeylSusy} gives rise to the $Sp(1)_{\rm R}$ covariant result \eq{cdc}. Note also that $E^\mu$ must be a conserved current, and it can indeed be identified as the $i=2$ component of the $Sp(1)_{\rm R}$ current in the dual string frame,
\be 
J^{\mu\, i}_{\rm R} = - \frac{i}{2y^{2}} \Bigl(  \bar\psi_\nu  \gamma^{\mu\nu\rho} \sigma^i \psi_\rho  + 4 \bar \psi_\nu \gamma^{\mu\nu} \sigma^i \chi   + 5 \bar \chi \gamma^\mu \sigma^i \chi - \bar\chi_r \gamma^\mu \sigma^i \chi^r \Bigr)\ .  
\label{JR} 
\ee
We are left with the most involved part of the computation, namely checking check that the supersymmetry transformation of $V_\mu^i$ defined in \eq{defCV} in the dual string frame indeed matches the supersymmetry transformation of the auxiliary field $V_\mu^i$ in \eq{DilWeylSusy}, modulo equations of motion. To this end we first compute the supersymmetry variation of $X_a^i $ and find 
\be 
\delta_\epsilon X^i_a = - \frac12 \bar\epsilon \sigma^i \Bigl[ \mathcal{O}_a +2 X_a^j \sigma_j \chi  \Bigr]\ ,
\label{sx}
\ee
where
\bea
\cO_\mu \hspace{-2mm}&\coloneqq&\hspace{-2mm}  -\frac12 \widehat{P}_\nu \gamma^\nu  \gamma_\mu \chi  +\frac12 \widehat{P}^r_\nu \gamma^\nu \gamma_\mu \chi_r  
- \frac14 \gamma^{ab} \chi_r y  \widehat{H}^r_{\m ab} + \frac{1}{4} \gamma^{ab} \chi y_r \widehat{H}^r_{\m ab}+ 2 X^i_\mu \sigma_i \chi \CR
&& + \frac{1}{48} \widehat{\mathcal{E}}^r_{abc} \gamma_\mu \gamma^{abc} ( y \chi_r - y_r \chi)  \ .
\label{geom3}
\eea
In order to match the correct off-shell transformation we need to use the fermion field equations $\R^\mu=0$ and $\eta=0$, written in terms of the string frame fields. In the remainder of this subsection, we shall write them in the absence of the hyper and Yang-Mills  multiplets, but we will keep the terms that are proportional to the equations of motion ${\mathcal E}_{abc}$ and ${\mathcal E}_{abc}^r$ because in Sections~\ref{sec:4.3} and~\ref{sec:4.5} they will play a role when we include the couplings of the hyper and Yang-Mills multiplets. Thus, the fermionic field equations in string frame are given by  
\bea
\R^\mu \hspace{-2mm}&=&\hspace{-2mm} \frac12 \gamma^{\m\n\rh} {\widehat \rho}^{\, +}_{\n\rh} +2 \gamma^{\m\n}\widehat{D_\nu(\widehat{\omega}_+)\chi}  -\frac52 {\widehat P}^\mu \chi  -\frac32 \gamma^{\m\n} {\widehat P}_\n \chi 
+\frac12 \widehat{P}_\nu^r \gamma^\nu \gamma^\mu \chi_r  + y{\widehat H}^{\mu ab} \gamma_{ab} \chi 
\nn\w2
&&  -\frac{1}4 y {\widehat H}_r^{ \mu  ab}  \gamma_{ab} \chi^r +\frac14 y^r {\widehat H}{}_r^{\mu ab} \gamma_{ab} \chi  + \frac78 \gamma_{ab}\chi \bchi\gamma^{\mu ab} \chi + \frac18 \gamma_{ab} \chi^r \bchi_r\gamma^{\mu ab} \chi
\label{eom1}\w2
&& -\frac34 \gamma^{\mu\nu} \chi^r \bchi_r\gamma_\nu \chi +\frac{11}{4}  \chi^r \bchi_r\gamma^\mu \chi - \frac{1}{8} y \widehat{\mathcal{E}}_{abc} \gamma^{abc}  \gamma^\mu \chi +\frac1{48} \widehat{\mathcal E}^r_{abc} \gamma^\mu \gamma^{abc}( y \chi_r - 6 y_r \chi ) \ ,
\nn\w2
\eta \hspace{-2mm} &=& \hspace{-2mm} \widehat{\slashed{D} (\widehat{\omega}_+)\chi} -\widehat{P}_\mu \gamma^\mu \chi -\widehat{P}^r_\mu \gamma^\mu  \chi_r +\frac16 y{\widehat H}_{abc} \gamma^{abc}\chi -\frac12\gamma^a \chi^r \bchi_r\gamma_a \chi -\frac{y_r}{16}  \widehat{\mathcal E}^r_{abc} \gamma^{abc}  \chi\ , 
\label{eom2}
\eea
where $\eta:= y^{-1}y^r \eta_r$ and 
\be \label{gc2}
\widehat{\rho}^{\, +}_{\mu\nu}  = D( {\widehat\omega}_+,V)_\mu \psi_\nu -D({\widehat\omega}_+,V)_\nu \psi_\mu\ ,
\ee
not to be confused with the expression for it in Einstein frame given in \eq{EOM1}. These equations are not the Euler--Lagrange equation for the string frame fields, but the Euler--Lagrange equation for the Einstein frame fields (rescaled by the appropriate power of $y$) written in terms of the string frame fields. Note also that the covariant derivative of the dilatino does not have a $V$-term in its definition. 

We can now write the term in $\mathcal{O}_\mu$ in \eq{sx} as the term $ \gamma^\nu {\widehat \rho}^{\, +}_{\mu\nu}$ appearing in the off-shell Poincar\'e supersymmetry transformation~\eqref{DilWeylSusy} of $V_\mu^i$ using the fact that 
\be    \gamma^\nu {\widehat \rho}^{\, +}_{\mu\nu} + 2 \widehat{D_\mu(\oh_+) \chi}- \mathcal{O}_\mu  = \mathcal{E}_\mu  \ee
vanishes on-shell with\footnote{Note that to derive this equation one gets  duality equation terms from \eq{geom3}, \eq{eom1} and \eq{eom2}, and from the simplification of terms involving the 3-form field strengths.}
\be \mathcal{E}_\mu \coloneqq \frac{1}{4}\gamma_\mu \gamma_\nu \mathcal{R}^\nu - \mathcal{R}_\mu - \frac12 \gamma_\mu \eta  -  \frac{1}{24} y \widehat{\mathcal{E}}_{abc}  \gamma^{abc} \gamma_\mu \chi+\frac{1}{48} y_r \widehat{\mathcal E}^r_{abc} \gamma_\mu \gamma^{abc}  \chi \; . \label{EmuCombination}\ee
Solving for ${\cal O}_\mu$ and substituting into the expression for the supersymmetry variation of $X_\mu^i$ given in \eq{sx} yields 
\be
\delta_\epsilon X^i_\mu =- \frac12 \bar\epsilon \sigma^i \Bigl[\gamma^\nu {\widehat \rho}_{\mu\nu +} +  2\widehat{D_\mu(\oh_+) \chi}  +  2X_\mu^j \sigma_j \chi  \Bigr] + \bar \epsilon \gamma^a \psi_\mu X^i_a +\frac12 \bar\epsilon \sigma^i \mathcal{E}_\mu \;  . 
\label{sxBetter}
\ee
Next, we compute 
\bea 
\delta_\epsilon  \big( \bar \chi \sigma^i \psi_\mu \big)\hspace{-2mm} &=&   \hspace{-2mm}
\bar \chi \sigma^i \bigl[ D_\mu ({\widehat \omega}_+) \e + \bigl( X_\mu^j +  \bar \chi \sigma^j \psi_\mu \bigr) \sigma_j \epsilon  + ( \bar \epsilon \sigma^j \chi) \sigma_j \psi_\mu  - \frac{1}{96} y_r \widehat{\mathcal{E}}^r_{abc} \gamma^{abc} \gamma_\mu  \epsilon \bigr]
\w2
&&\!  +  \bar \psi_\mu \sigma^i \Big[\frac12 \gamma^a \epsilon \widehat{P}_a + \frac{1}{24} a_I  \widehat{H}^I_{abc} \gamma^{abc} \epsilon   + \chi \bar \epsilon  \chi +\frac12  X_a^j \gamma^a \sigma_j \epsilon + \frac{1}{48} y  \widehat{\mathcal{E}}_{abc} \gamma^{abc} \epsilon  \Big]\ .
\label{sv2} \nn
\eea
Thus, the complete transformation of $V_\mu^i = \bar \chi \sigma^i \psi_\mu + X_\mu^i$ takes the form
\bea  
\label{dV}
\delta_\epsilon V_\mu^i &=&  - \frac12 e_\mu{}^a\bar\epsilon \sigma^i  \gamma^b {\widehat \rho}_{ab +} - \frac1{12} \ag_I \widehat{H}^I_{abc} \bar \epsilon \sigma^i \gamma^{abc} \psi_\mu -  \partial_\mu \Lambda^i - 2 i \varepsilon^{i}{}_{jk} V_\mu^j \Lambda^k 
\w2
 && +\frac12 \bar\epsilon \sigma^i \Bigl[\frac{1}{4}\gamma_\mu \gamma_\nu \mathcal{R}^\nu -\mathcal{R}_\mu - \frac12 \gamma_\mu \eta - \frac{1}{12} y \gamma^{abc} \widehat{\mathcal{E}}_{abc} \bigl( \psi_\mu + \frac12 \gamma_\mu \chi \bigr)  \Bigr]\ ,\nonumber 
\eea
where
\be
\Lambda^i = \bar \epsilon \sigma^i  \chi\ ,
\label{L}
\ee
in agreement with the off-shell transformation of $V_\mu^i$ in the off-shell Poincar\'e multiplet given in \eqref{DilWeylSusy}, upon using ${\mathcal R}_\mu=0$ and $\eta=0$, in accordance with the fact that the supertransformations \eq{susyprime} only close on-shell. 
We have not  checked explicitly that the $Sp(1)_{\rm R}$ currents $J^{\mu\, 2}_{\rm R}$ in \eqref{JR} transforms as $E^\mu$, but it must by closure of the supersymmetry algebra. 

Having identified the fields of the off-shell Poincar\'e multiplet in terms of the fields of the model summarised in Section~\ref{sec:rev}, using these identifications in \eq{DilWeylSusy} we find that the supertransformations of $\{e_\mu{}^a, \psi_\mu, {\mathcal B}_{\mu\nu}, \chi, L\}$ agree with those given in \eq{sr}, with \eq{defCV} and \eq{defL} understood. It is worth noting that in \cite{Bergshoeff:2012ax} expressions for the auxiliary fields $V_\mu^i$ and $E_\mu$ are obtained from the field equations of an {\it off-shell} two-derivative supergravity Lagrangian for the case of $n_T=1$. Here we do not have an off-shell two-derivative supergravity action in presence of multi-tensor multiples, but rather we have the pseudo-Lagrangian \eq{pL1}. Nonetheless, in what follows we will be able to use the results of this section to find the Poincar\'e-Yang--Mills map that will enable us to construct the four-derivative extension of the model whose two-derivative sector is the one given in \eq{pL1}. 

\vskip 2mm

To summarise, the key result here is that the identifications described above allow us to use all the formulas computed in \cite{Bergshoeff:2012ax} for the off-shell Poincar\'e multiplet and to identify the Poincar\'e to Yang--Mills map in the next section. 

\subsection{Inclusion of vector multiplets}
\label{sec:4.3}

To include the vector multiplets, we need to extend the definition of the auxiliary field given in \eq{key1} by taking into account the gaugino contributions to it. Because the supersymmetry transformation of the $B$ field gets a correction 
\be \delta_\epsilon (a_I B^I_{\mu\nu}) = - 2 \bar \epsilon \gamma_{[\mu} \psi_{\nu]} - 2 a_I b^{Iz} \Tr_z\bigl[ A_{[\mu} \bar \epsilon \gamma_{\nu]} \lambda \bigr] \; , \ee
one cannot get a map to the off-shell Poincar\'e  multiplet whenever $a_I b^{I z}\ne 0$. This is of course due to the fact that the Wess--Zumino consistency condition implies then that the $R^2$ type invariant cannot be fully supersymmetric in the presence of a mixed anomaly. In this section we prove that the map to the  off-shell Poincar\'e  multiplet exists when there is no anomaly, and all required identities are satisfied in general up to terms proportional to $a_I b^{I z} $. 

For the modification of the 4-form auxiliary field we observe that 
\be \frac{1}{24} \varepsilon^{\mu\nu\rho\sigma\kappa\lambda} {\partial_\nu E_{\rho\sigma\kappa\lambda}} =  J^{\mu\, 2}_{\rm R}\; , \ee
gets a correction because the $Sp(1)_{\rm R}$ current in the presence of vector multiplet is 
\be  J^{\mu\, i}_{\rm R} = - \frac{i}{2y^{2}} \Bigl(  \bar\psi_\nu  \gamma^{\mu\nu\rho} \sigma^i \psi_\rho  + 4 \bar \psi_\nu \gamma^{\mu\nu} \sigma^i \chi   + 5 \bar \chi \gamma^\mu \sigma^i \chi - \bar\chi_r \gamma^\mu \sigma^i \chi^r-2 y c^{z}  \Tr_z \bigl[  \bar \lambda \sigma^i \gamma^\mu \lambda \bigr] \Bigr)\  . \ee
This suggests that one must add $- \frac{1}{2} y c^{z}  \Tr_z [  \bar \lambda \sigma^i \gamma_\mu \lambda ]$ to the definition of the auxiliary field $V_\mu^i$, but the transformation of the dilatino 
\bea \delta_\epsilon \chi^r &=& -\frac12 {\widehat P}_\mu^r \gamma^\mu \e -\frac{1}{24} y{\widehat H}^r_{\mu\nu\rho} \gamma^{\mu\nu\rho} \e + \frac14\chi^r  (\eps\chi) -\frac18 \gamma^{ab}\chi^r (\eps\gamma_{ab}\chi) 
\nn\w2
&& +\frac14 \gamma^a\e ({\bar\chi}\gamma_a \chi^r)  -\frac{1}{16} \gamma^{abc} \e ({\bar\chi}\gamma_{abc} \chi^r )+ \frac14 y c^{r z} \Tr_z \bigl[ \bar \lambda \sigma^i \gamma^\mu \lambda \bigr] \gamma_\mu \sigma_i \epsilon \ ,
\label{fin3}\eea
requires instead the definition 
\be 
{V_\mu^i = X_\mu^i+{\bar\chi} \sigma^i \psi_\mu - \frac{1}{2} y_r c^{r z}  \Tr_z \bigl[  \bar \lambda \sigma^i \gamma_\mu \lambda \bigr]} \ ,
\label{fin2}
\ee
in order to reproduce the first line of $\delta_\epsilon \chi$ in \eqref{DilWeylSusy}. 
The difference is proportional to the mixed anomaly coefficient
\be a_I b^{I z} = - y c^{z} + y_r c^{r z}  \; . \ee
With the definition \eqref{fin2} we get indeed 
\bea 
\delta_\epsilon \psi_\mu \hspace{-2mm}&=&\hspace{-2mm} D_\mu(\widehat{\omega}^+,V) \epsilon + a_I  b^{Iz}  Z_{\mu z} \e\ ,  \label{Z} \\
\delta_\epsilon \chi
\hspace{-2mm}&=&\hspace{-2mm} -\frac12 {\widehat P}_\mu \gamma^\mu \e -\frac{1}{24} \ag_I{\widehat H}^I_{\mu\nu\rho} \gamma^{\mu\nu\rho} \e  -\frac12 ( V_\mu^i -  {\bar\chi} \sigma^i \psi_\mu ) \gamma^\mu \sigma_i \e -(\eps\chi) \chi- \frac{1}{48} y \widehat{ \mathcal{E}}_{\mu\nu\rho} \gamma^{\mu\nu\rho} \e \; ,  \nonumber
\eea
where $Z_{\mu z}$ is the Clifford algebra valued 1-form 
\be
Z_{\mu z} \coloneqq \Tr_z \Bigl[ -\lambda {\bar\lambda}\gamma_\mu  +\frac14 ( {\bar\lambda}\sigma^i \gamma_\mu \lambda) \sigma_i \Bigr]  \  .
\label{fin1}
\ee
Accordingly, the gravitino field strength gets a super-torsion term in the presence of an anomaly 
\be \label{gc2Vector}
\widehat{\rho}^{\, +}_{\mu\nu}  = D( {\widehat\omega}_+,V)_\mu \psi_\nu -D({\widehat\omega}_+,V)_\nu \psi_\mu  + a_I b^{Iz} \bigl( Z_{\mu z} \psi_\nu - Z_{\nu z} \psi_\mu \bigr) \ .
\ee
With this definition one obtains that the supersymmetry transformation of the supercovariant 3-form field strength is supercovariant 
\be \delta_\epsilon ( a_I \widehat{H}^I_{abc}) = 3 \bar \epsilon \gamma_{[a} \widehat{\rho}^{\, +}_{bc]}- 6 a_I b^{Iz} \Tr_z\bigl[ \bar\epsilon \gamma_{[a} \lambda \widehat{F}_{bc]} \bigr] \; . \ee 
For the vector multiplet, we get
\bea
\label{SusySFonShell} 
\delta_\epsilon A_\mu \hspace{-2mm}&=&\hspace{-2mm}  {\bar\epsilon} \gamma_\mu \lambda\ ,
\\
\delta_\epsilon \lambda \hspace{-2mm}&=&\hspace{-2mm}  - \frac{1}{4} \gamma^{\mu\nu} \,  \widehat{F}_{\mu\nu}   \epsilon+\Bigl( \frac{c^{rz}}{c^z} - \frac{y^r}{y} \Bigr) \left( \frac14 \lambda {\bar\chi}_r\e +\frac12 \e {\bar\chi}_r \lambda -\frac18 \gamma_{ab}\lambda{\bar\chi}_r\gamma^{ab}\e\right) + ( \bar \epsilon \sigma^i \chi) \sigma_i \lambda \ . \nonumber
\eea
Note that we have the convention that $z$ takes the value associated to the gauge algebra component of $\lambda$, even if we do not write the label $z$ on the gauge multiplet fields themselves. The second term vanish if $a^I = b^{Iz}$, giving then the standard string frame supersymmetry transformation of the gaugini on-shell. 

As in the preceding section, the most complicated step is the computation of $\delta_\epsilon V_\mu^i$. It is convenient to decompose this calculation in steps. First we need to take into account the corrections to \eqref{EmuCombination} that depend on the vector multiplets which were disregarded in the preceding section. They read 
\bea 
\mathcal{E}_\mu \Big|_{\rm YM} \hspace{-2mm}&\coloneqq&\hspace{-2mm}  - \frac12 y c^z \gamma^{\nu\rho} \gamma_\mu \Tr_z\bigl[ \lambda \widehat{F}_{\nu\rho}\bigr] - \frac14 a_I b^{Iz} \gamma_\mu \gamma^{\nu\rho}   \Tr_z\bigl[ \lambda \widehat{F}_{\nu\rho}\bigr]  \CR 
&&  + \frac12  y c^z \Tr \Bigl[ \Bigl( \frac34 \gamma_\mu \gamma_{ab} \lambda \bar \lambda \gamma^{ab} +\gamma^a \lambda \bar \lambda \gamma_a \gamma_\mu  + \frac32 \gamma_\mu \lambda \bar \lambda \Bigr) \chi \Bigr]  + \frac12 y c^{rz} \Tr_z \Bigl[ - \gamma^\nu \lambda \bar \lambda \gamma_\mu \gamma_\nu \chi_r \Bigr]  \CR
&& +\frac14 a_I b^{I z} \Bigl( \frac{y^r}{y} - \frac{c^{rz}}{c^z}\Bigr)  \Tr_z \Bigl[ \frac12 \gamma_\mu \gamma_{ab} \lambda \bar \lambda \gamma^{ab} \chi_r + 3 \gamma_\mu \lambda \bar \lambda \chi_r  \Bigr]\CR
&& - \frac1{48}( a_I b^{I z}  + y c^{z} ) \gamma_\mu \gamma^{abc} \chi \Tr_z \bigl[ \bar \lambda \gamma_{abc} \lambda \bigr]  \ , \eea
where we did not include the terms involving a naked gravitino field, which are understood to be absorbed in the supercovariant derivatives. In total we obtain 
\bea 
\delta_\e V_\mu^i &=&- \frac12 e_\mu{}^a\bar\epsilon \sigma^i  \gamma^b {\widehat \rho}_{ab +} - \frac1{12} \ag_I \widehat{H}^I_{abc} \bar \epsilon \sigma^i \gamma^{abc} \psi_\mu -  \partial_\mu \Lambda^i -  2 i \varepsilon^{i}{}_{jk} \Bigl( X_\mu^i+{\bar\chi} \sigma^i \psi_\mu  \Bigr) \Lambda^k 
\CR
 && +\frac12 \bar\epsilon \sigma^i \Bigl[\frac{1}{4}\gamma_\mu \gamma_\nu \mathcal{R}^\nu -\mathcal{R}_\mu - \frac12 \gamma_\mu \eta - \frac{1}{12} y \gamma^{abc} \widehat{\mathcal{E}}_{abc} \bigl( \psi_\mu + \frac12 \gamma_\mu \chi \bigr)  -\mathcal{E}_\mu \Big|_{\rm YM} \Bigr]\CR
&& - \frac14 ( y c^z + a_I b^{Iz}) \bar \epsilon \sigma^i \gamma^{\nu\rho} \gamma_\mu \Tr_z\bigl[ \lambda \widehat{F}_{\nu\rho}\bigr] + \frac12 y \bigl( c^{rz} ( \bar \epsilon \chi_r) + c^z ( \bar \epsilon \chi) \bigr) \Tr_z \bigl[ \bar \lambda \sigma^i \gamma_\mu \lambda \bigr] \CR
&& + ( y c^z + a_I b^{I z}) \Bigl( \frac{y^r}{y} - \frac{c^{rz}}{c^z}\Bigr)  \Tr_z \Bigl[ \bar \lambda \sigma^i \gamma_\mu \Bigl( \frac14 \lambda \bar \chi_r \epsilon + \frac12 \epsilon \bar \chi_r \lambda - \frac18 \gamma_{ab} \lambda \bar \chi_r \gamma^{ab} \epsilon \Bigr) \Bigr] \CR
&& + i \varepsilon^{i}{}_{jk}  y_r c^{r z}  \Tr_z \bigl[  \bar \lambda \sigma^j \gamma_\mu \lambda \bigr]  \Lambda^k  \CR
&& + \frac18 \Bigl( y^r c^z - y c^{rz} + a_I b^{I z} \frac{y^r}{y} \Bigr) \bar \chi_r \gamma_\mu \gamma_\nu \sigma^i \sigma^j \epsilon \Tr_z \bigl[ \bar \lambda \gamma^\nu \sigma_j \lambda \bigr] \CR
&& - \frac12 ( y c^z + a_I b^{Iz})  \bar \chi \sigma^i \sigma^j \epsilon \Tr_z\bigl[ \bar \lambda \gamma_\mu \sigma_j \lambda \bigr]  + a_I  b^{Iz} \bar \chi \sigma^i  Z_{\mu z} \e \,,
\eea
where the third, fourth and fifth lines come from the variation of the $\lambda$-dependent extra term in \eqref{fin2}, the sixth line comes from the extra terms in the variation of $X_\mu^i$ and the seventh line from the $\lambda$-dependent extra terms in the variation of $\psi_\mu$. The expression above simplifies to
\bea
\delta_\e V_\mu^i &=&   - \frac12 e_\mu{}^a\bar\epsilon \sigma^i  \gamma^b {\widehat \rho}_{ab +} - \frac1{12} \ag_I \widehat{H}^I_{abc} \bar \epsilon \sigma^i \gamma^{abc} \psi_\mu -  \partial_\mu \Lambda^i - 2 i \varepsilon^{i}{}_{jk} V_\mu^j \Lambda^k 
\w2
 && +\frac12 \bar\epsilon \sigma^i \Bigl[\frac{1}{4}\gamma_\mu \gamma_\nu \mathcal{R}^\nu -\mathcal{R}_\mu - \frac12 \gamma_\mu \eta - \frac{1}{12} y \gamma^{abc} \widehat{\mathcal{E}}_{abc} \bigl( \psi_\mu + \frac12 \gamma_\mu \chi \bigr)   \Bigr]\CR
&& +\frac18 a_I b^{I z} \bar\epsilon \sigma^i \bigl( \gamma_\mu \gamma^{\nu\rho} - 2 \gamma^{\nu\rho} \gamma_\mu\bigr)  \Tr_z\bigl[ \widehat{F}_{\nu\rho} \lambda \bigr] \CR
&& + \frac18 a_I b^{I z}  \Bigl( \frac{y^r}{y} - \frac{c^{rz}}{c^z}\Bigr) \Tr_z\Bigl[ - 2\bar \epsilon \sigma^i \gamma_\mu \lambda \bar \lambda \chi_r +\bar \epsilon \sigma^j \sigma^i \gamma_\mu \gamma_\nu \chi_r \bar \lambda \gamma^\nu \sigma_j \lambda  \Bigr]\; . \nonumber \eea

We have therefore obtained that the map to the off-shell Poincar\'e  multiplet is defined on-shell in the presence of vector multiplets and provided we assume the vanishing of the mixed anomaly $a_I b^{Iz} = 0$. The explicit modifications proportional to $a_I b^{Iz} $ permit in principle to compute the solution to the Wess--Zumino consistency condition when there is a mixed anomaly, but we shall not do it in this paper. 

\subsection{\texorpdfstring{The $R^2$ correction via Poincar\'e to Yang--Mills map}{The R2 correction via Poincar\'e to Yang--Mills map}}

To construct the curvature-squared extension of the model, we seek a map between the Poincar\'e supermultiplet and the off-shell Yang--Mills multiplet \cite{Bergshoeff:1987rb,Bergshoeff:2012ax} in the dual string frame. It is defined as the identification 
\be
\Big( {\widehat\omega}_{-\mu}{}^{ ab} , -\hrho_{+}^{\, ab} , {\widehat F}^{ab \, i}(V) \Big) \quad \longrightarrow \quad \Big( \mathcal{A}_\mu , \uplambda , \mathcal{Y}^i \Big)\ ,
\label{map}
\ee
where the torsionful supercovariant connection is defined in \eq{defoh}, the supercovariant Rarita--Schwinger field strength in \eqref{gc2} and
\be
{\widehat F}_{\mu\nu}^i(V) = 2\partial_{[\mu} V^i_{\nu]} + i \varepsilon^{i}{}_{jk} V_\mu^j V_\nu^k - e_{[\mu}{}^a\bar\psi_{\nu]} \sigma^i  \gamma^b {\widehat \rho}_{ab +}  + \frac1{12} \ag_I \widehat{H}^I_{abc} \bar \psi_\mu  \sigma^i \gamma^{abc} \psi_\nu  \ .
\label{defFV}
\ee
For this we first assume that $a_I b^{Iz}=0$ everywhere, and will discuss the case in which there is a mixed anomaly at the end.

The off-shell Yang--Mills supermultiplet, and its coupling to the off-shell $(1,0)$ Poincar\'e multiplet described in the last sections has been determined in \cite{Bergshoeff:1985mz}. In that case, the Yang--Mills multiplet fields transform as \cite{Bergshoeff:1985mz,Bergshoeff:2012ax}\footnote{To compare with \cite{Bergshoeff:2012ax}, we send the  field there to ours as $\uplambda \to- \frac{1}{\sqrt{2}}\uplambda$ and $\mathcal{Y}_{AB} \to \mathcal{Y}^i (\sigma^i)_{AB}$.}
\bea
\delta_\epsilon \mathcal{A}_\mu &=& \bar\e \gamma_\mu \uplambda\ ,
\label{v1}\w2
\delta_\epsilon \uplambda  &=& -\frac14 \gamma^{ab} \widehat{\mathcal{F}}_{ab} \e+ \mathcal{Y}_i  \sigma^i \e + \Lambda_i \sigma^i \uplambda \ ,
\label{v2}\w2
\delta_\epsilon \mathcal{Y}^i &=& \frac12 \bar\e\sigma^i \gamma^\mu \widehat{D_\mu \uplambda} -\frac{1}{48}  \bar\e\sigma^i \gamma^{abc}\widehat{\mathcal{H}}_{abc} \uplambda + 2 i \varepsilon^{i}{}_{jk} \Lambda^j \mathcal{Y}^k \ ,
\label{v3}
\eea
where $\Lambda^i$ takes the same value as in \eqref{ChooseSU2}. We stress that the off-shell vector multiplet described above should not be confused with the on-shell vector multiplet of the model, and this is why we use a different font to denote them. Only when $a^I = b^{I r}$ one finds that \eqref{SusySFonShell} can be put in the form \eqref{v3} with $\mathcal{Y}^i=0$. 

The off-shell superconformal Lagrangian is given in \cite{Bergshoeff:1985mz}, and gauge-fixing the super-conformal invariance using \cite[Eq. (3.1)]{Bergshoeff:2012ax}, one obtains the off-shell Yang--Mills Lagrangian in the dual string frame 
\bea
e^{-1} \cL &=& \Tr \Big[ -\frac14 \mathcal{F}_{\m\n} \mathcal{F}^{\m\n}   -\bar\uplambda \slashed D(\widehat{\omega},V) \uplambda -\mathcal{Y}^i \mathcal{Y}_i-\frac{1}{16 e} \varepsilon^{\mu\nu\rho\sigma\lambda\tau} \mathcal{B}_{\mu\nu} \mathcal{F}_{\rho\sigma}  \mathcal{F}_{\lambda\tau} 
\nn\w2
&& -\frac14 \bar\uplambda \gamma^\mu \gamma^{\nu\rho} \left(\mathcal{F}_{\nu\rho}  +\widehat{\mathcal{F}}_{\nu\rho} \right) \psi_\mu
+\frac{1}{24} \bar\uplambda \gamma^{abc} \widehat{\mathcal{H}}_{abc} \uplambda \Big]\ .
\label{result}
\eea

Using the map to the off-shell Poincar\'e multiplet described in the preceding section, it follows from the computations of \cite{Bergshoeff:2012ax} that on-shell
\bea 
\delta_\epsilon \widehat{\omega}_{\mu ab}^{-} &=& - \bar \epsilon \gamma_\mu  \widehat{\rho}^{\; +}_{ab} \; , \CR
\delta_\epsilon \widehat{\rho}_{ab+} &=& \frac{1}{4} \widehat{R}(\widehat{\omega}_-)_{cdab} \gamma^{cd} \epsilon+ \widehat{F}^i_{ab}(V) \sigma_i \epsilon  + ( \bar \epsilon \sigma^i \chi) \sigma_i \widehat{\rho}^{\, +}_{ab}\; , \CR
\delta_\epsilon \widehat{F}^i_{ab}(V) &=& \frac12 \bar \epsilon \sigma^i  \gamma^c \widehat{D_c \widehat{\rho}_{ab+}} - \frac1{48} \ag_I \widehat{H}^I_{bcd} \bar\epsilon \sigma^i \gamma^{bcd} \widehat{\rho}_{ab+} + 2 i \varepsilon^i{}_{jk} (\bar \epsilon \sigma^j \chi) \widehat{F}^k_{ab}(V)\; , \label{GravityYM}
\eea
where\footnote{Note that we can as well define the covariant derivative of $\widehat{\rho}_{ab+}$ with respect to the spin connection $\widehat{\omega}_{ -}$, by modifying the coefficient of the three-form field strength coupling. But to exhibit the map to the off-shell Yang--Mills multiplet we want to distinguish the Yang--Mills $\mathfrak{so}(1,5)$ connection $\widehat{\omega}_{ -}$  from the spin connection $\widehat{\omega}$.} 
\begin{align}
\widehat{D \widehat{\rho}_{ab+}}  &= d \widehat{\rho}_{ab+} + \tfrac14 \widehat{\omega}_{cd} \gamma^{cd} \widehat{\rho}_{ab+} - 2 \widehat{\omega}^{\, -}_{[a}{}^c \widehat{\rho}_{b]c+} {+} V^i \sigma_i  \widehat{\rho}_{ab+}{-}\tfrac{1}{4} \widehat{R}(\widehat{\omega}_-)_{cdab} \gamma^{cd} \psi- \widehat{F}^i_{ab}(V) \sigma_i \psi \ , \CR
\widehat{R}(\widehat{\omega}_-)_{\mu\nu ab} &=R_{\mu\nu ab}(\widehat{\omega}_-) +2 \bar \psi_{[\mu} \gamma_{\nu]} \widehat{\rho}_{ab+} \; . 
\end{align}
Comparing \eqref{GravityYM} with \eqref{v3} shows that we have indeed the map \eqref{map}.
Here we have defined a map from the on-shell supergravity multiplet to the off-shell Yang--Mills multiplet coupled to the off-shell Poincar\'e  multiplet. We can therefore use the Lagrangian \eqref{result} to write a Lagrangian that is supersymmetric modulo the two-derivative field equations. 

It is important to note that in the computations of this section we have never used the property that $a^I$ is lightlike.  Due to the anomaly, if $\eta_{IJ} a^I a^J\ne 0$, there will be an obstruction at the next order to obtain a supersymmetric Lagrangian as there is for Yang--Mills. For $a_I b^{Iz}\ne 0$ there is already an obstruction at first order in $a_I$ and we cannot rely directly on the map to the off-shell Poincar\'e  multiplet since there are corrections proportional to $a_I b^{Iz}$.   For instance, the variation of the torsionful spin connection gives in this case
\bea 
\delta_\epsilon \widehat{\omega}_{\mu ab}^{-} \hspace{-2mm}&=&\hspace{-2mm} - \bar \epsilon \gamma_\mu  \widehat{\rho}^{\; +}_{ab}  + a_I b^{Iz}  \Tr_z \bigl[  3  e_a{}^\nu e_b{}^\rho \bar \epsilon \gamma_{[\mu} \lambda \widehat{F}_{\nu\rho]}+2 \bar \epsilon \gamma_{[a} \lambda \bar \lambda \gamma_{b]} \psi_\mu \bigr]\ , \\
\delta_\epsilon \widehat{\rho}_{ab+} \hspace{-2mm}&=&\hspace{-2mm} \frac{1}{4} \widehat{R}(\widehat{\omega}_-)_{cdab} \gamma^{cd} \epsilon+ \widehat{F}^i_{ab}(V) \sigma_i \epsilon  + ( \bar \epsilon \sigma^i \chi) \sigma_i \widehat{\rho}^{\, +}_{ab} - \frac{3}{4} a_I b^{Iz} \Tr_z\bigl[ \widehat{F}_{[ab} \widehat{F}_{cd]}\bigr] \gamma^{cd} \epsilon \CR
&& +2 a_I b^{Iz} \widehat{ D_{[a} Z_{b]z}} \epsilon  + 2 a_I b^{I z} a_J b^{J z'} Z_{[a z} Z_{b] z'} \epsilon \; , \nonumber
\eea
where $\widehat{\rho}^{\; +}$ is defined in \eqref{gc2Vector} and $Z_{\mu z} $ in \eqref{fin1}.  One finds that the additional anomalous terms  are very similar to  \cite[Eq.~(2.14)]{Bergshoeff:1989de} in ten dimensions, suggesting that there should be a correction of the type
\be 
e\,  t^{abcdefgh} a_I b^{Iz} \Tr_z \bigl[ F_{ab} F_{cd}   \bigr]  y c^{z'} \Tr_{z'} \bigl[ F_{ef} F_{gh}   \bigr] 
\ee
in the effective action in presence of mixed anomaly. We have used the $t_8$-tensor familiar from higher-derivative corrections \cite{Bergshoeff:1989de}.

When there is no anomaly, one may hope in principle to obtain a complete supersymmetry invariant using the off-shell map, i.e. to all order in the expansion parameter $a_I$ and therefore arbitrary high order derivative terms. For this purpose one would need to write the two-derivative Lagrangian in a partly off-shell formulation such that the expression \eqref{defCV} of $V_\mu^i$ would be obtained from its equation of motion. To obtain such a formulation one needs to split the tensor fields into $a_I B^I$ and the $n_T-1$ extra tensor multiplets, such that $a_I B^I$ would appear in the Lagrangian as in the theory with one-tensor multiplet. Such a formulation might exist but it will necessarily break the manifest $SO(1,n_T)$ symmetry and we shall not attempt to define it in this paper.

To summarise, we have established the map \eq{map}, with key definitions given in \eq{key1}, \eq{gc2} and \eq{defFV}. Using this map in \eq{result} yields the higher derivative extension of $(1,0)$ supergravity coupled to tensor multiplets, with the Lagrangian  
\bea
e^{-1} \cL_{ R^2} &=& -\frac14 {R}_{abcd}(\widehat{\omega}_-)  {R}^{abcd}(\widehat{\omega}_-)  -\frac{1}{16 e} \ag_I \varepsilon^{\mu\nu\rho\sigma\lambda\tau} {B}^I_{\mu\nu} {R}_{\rho\sigma}{}^{ab}(\widehat{\omega}_-){R}_{\lambda\tau ab}(\widehat{\omega}_-)
\nn\w2
&& -\overline{\widehat{\rho}^{\, +}_{ab}} \slashed D(\widehat{\omega}{}_{-} ,V) \widehat{\rho}^{\, ab+} -{\widehat F}^{ab \, i}(V) {\widehat F}_{ab \, i}(V) 
\label{R2result}\w2
&& +\frac14 \overline{\widehat{\rho}^{\, +}_{ab}}  \gamma^\mu \gamma^{\nu\rho} \left({R}_{\nu\rho}{}^{ab}(\widehat{\omega}_-) {+}\widehat{R}_{\nu\rho}{}^{ab}(\widehat{\omega}_-) \right) \psi_\mu
-\frac{1}{12} \ag_I \overline{\widehat{\rho}^{\, +}_{ab}}  \gamma^{\mu\nu\rho} \widehat{{H}}^I_{\mu\nu\rho} \widehat{\rho}^{\, ab+}  \ .
\nn
\eea

\subsection{Inclusion of hypermultiplets}
\label{sec:4.5}

In the presence of hypermultiplets one can modify the definition of the auxiliary field defined in \eq{key1}, which we will now denote by $V_\mu^{i0}$, as follows
\be
V_\mu^i= V_\mu^{i0} + Q_\mu^i\ ,
\label{nV}
\ee
where we recall that $Q_\mu^i= \partial_\mu \varphi^{\alpha} \mathcal{A}_\alpha^i $.  In computing the supertransformations of the newly defined $V_\mu^i$ we need to use the supertransformations \eq{sr} in dual string frame. We begin by the supertransformation of $V_\mu^{i0}$ for which we find 
\bea  
\delta_\epsilon V_\mu^{i0} \hspace{-2mm}&=&\hspace{-2mm}  - \frac12 e_\mu{}^a\bar\epsilon \sigma^i  \gamma^b {\widehat \rho}_{ab +} - \frac1{12} \ag_I \widehat{H}^I_{abc} \bar \epsilon \sigma^i \gamma^{abc} \psi_\mu -  \partial_\mu \Lambda^{i0} - 2 i \varepsilon^{i}{}_{jk} V_\mu^j \Lambda^{k0} +  2i \varepsilon^i{}_{jk} V_{\mu}^{j 0} \delta \varphi^{\alpha} \mathcal{A}_\alpha^k
\nn\w2
 && -\frac12 \bar\epsilon \sigma^i \Bigl[\mathcal{R}_\mu^0 -\frac{1}{4}\gamma_\mu \gamma^\nu \mathcal{R}_\nu^0 + \frac12 \gamma_\mu \eta + \frac{1}{12} y \gamma^{abc} \widehat{\mathcal{E}}^{\, 0}_{abc} \bigl( \psi_\mu + \tfrac12 \gamma_\mu \chi \bigr)\Bigr]\ ,
\label{dV0}
\eea
where $\Lambda^{i0}$ is as defined in \eq{L}, and $ \left(  {\mathcal R}_\mu^0, \eta, {\widehat\rho_{ab_+}}, \right)$ are as defined in \eq{eom1}, \eq{eom2} and \eq{gc2}, but covariantised by the inclusion of the composite connection $Q_\mu^i$. 
The last term in the first line comes from the variation of the term ${\bar\chi}\sigma^i\psi_\mu$, and the derivative of $\Lambda^{i0}$ has been covariantised by employing $V_\mu^i$.  
We deduce from the Lagrangian \eq{pL} that the full gravitino equation, which we will denote by ${\mathcal R}_\mu$ is  given by
\be 
\mathcal{R}_\mu^A = \mathcal{R}_\mu^{A 0} + \gamma^\nu \gamma_\mu  \zeta_X\,\bigl[ { P}_\nu^{XA} + \bar \zeta^X ( \psi_\mu^A  + \tfrac12 \gamma_\mu \chi^A ) \bigr]\ .  
\label{gcurve}
\ee
Using Fierz identities one obtains that\footnote{We use $\zeta_X \bar \zeta^X = \frac{1}{48} \gamma^{abc} \bar \zeta^X \gamma_{abc} \zeta_X$.}
\be 
\mathcal{R}^{0\, A}_\mu -\frac{1}{4}\gamma_\mu \gamma^\nu \mathcal{R}^{0\, A}_\nu = \mathcal{R}^A_\mu -\frac{1}{4}\gamma_\mu \gamma^\nu \mathcal{R}^A_\nu-2  \zeta_X { P}_\mu^{XA} - \frac{1}{24} \gamma^{abc} ( \psi_\mu^A + \tfrac12 \gamma_\mu \chi^A)  \bar \zeta^X \gamma_{abc} \zeta_X \; , 
\ee
where the cubic term in fermions compensates the one from the duality equation. 
Using this equation as well as the variation 
\be
\delta_\epsilon Q_\mu^{AB}= D_\mu \left( \delta_\epsilon \varphi^\alpha \mathcal{A}_\alpha^{AB} \right)  +2{ P}_{\mu X}^{(A} {\bar\epsilon}^{B)} \zeta^X\ , 
\ee
where $D_\mu (\delta_\epsilon \vp^\alpha {\mathcal A}^i_\alpha) = \partial_\mu (\delta_\epsilon\vp^\alpha {\mathcal A}^i_\alpha) + 2i \varepsilon^i{}_{jk} Q_\mu^j (\delta_\epsilon\vp^\alpha {\mathcal A}_\alpha^k)$,  we find 
\bea
\delta_\epsilon V_\mu^i &=& \delta_\epsilon V_{\mu}^{i 0}   + \delta_\epsilon Q_\mu^i
\nn\w2
&=&  - \frac12 e_\mu{}^a\bar\epsilon \sigma^i  \gamma^b {\widehat \rho}_{ab +} - \frac1{12} \ag_I \widehat{H}^I_{abc} \bar \epsilon \sigma^i \gamma^{abc} \psi_\mu -  \partial_\mu \Lambda^{i 0} - 2 i \varepsilon^{i}{}_{jk} V_\mu^j   \Lambda^{k0} 
\nn\w2
&& - \frac12 \bar \epsilon \sigma^i \bigl( \mathcal{R}_\mu - \tfrac14 \gamma_\mu \gamma^\nu \mathcal{R}_\nu  + \tfrac12 \gamma_\mu \eta + \tfrac{1}{12} y \gamma^{abc} \widehat{\mathcal{E}}_{abc} ( \psi_\mu + \tfrac12 \gamma_\mu \chi )\bigr)  - \bar \epsilon^A  \zeta_X { P}_\nu^{XB} \sigma^i_{AB}
\nn\w2
&& + 2i \varepsilon^i{}_{jk} V_{\mu}^{j 0} \delta_\epsilon \varphi^{\alpha} \mathcal{A}_\alpha^k + D_\mu ( \delta_\epsilon \varphi^\alpha  \mathcal{A}_\alpha^i  )+ \bar \epsilon^A  \zeta_X P_\mu^{XB}\sigma^i_{AB} 
\nn\w2
&& +\frac{1}{48}  \widehat{\mathcal E}^r_{abc} {\bar\epsilon} \sigma^i \gamma_\mu \gamma^{abc}( y \chi_r{-}y_r \chi) \ .
\eea
Defining $\Lambda^i= \Lambda^{i0}- \delta_\epsilon \vp^\alpha {\mathcal A}_\alpha^i$ and recalling \eq{nV}, it follows that
\bea
\delta_\epsilon V_\mu^i  &=&  - \frac12 e_\mu{}^a\bar\epsilon \sigma^i  \gamma^b {\widehat \rho}_{ab +} - \frac1{12} \ag_I \widehat{H}^I_{abc} \bar \epsilon \sigma^i \gamma^{abc} \psi_\mu -  \partial_\mu \Lambda^{i } - 2 i \varepsilon^{i}{}_{jk} V_\mu^{j }   \Lambda^{k } 
\nn\w2
&& - \frac12 \bar \epsilon \sigma^i \bigl( \mathcal{R}_\mu - \tfrac14 \gamma_\mu \gamma^\nu \mathcal{R}_\nu  + \tfrac12 \gamma_\mu \eta + \tfrac{1}{12} y \gamma^{abc} \widehat{\mathcal{E}}_{abc} ( \psi_\mu + \tfrac12 \gamma_\mu \chi )\bigr )
\nn\w2
&& +\frac{1}{48}  \widehat{\mathcal E}^r_{abc} {\bar\epsilon} \sigma^i\gamma_\mu \gamma^{abc}( y \chi_r{-}y_r \chi) \ .
\label{deltaV}
\eea 
This result agrees with \eq{dV}, and therefore we can use the vector field $V_\mu^i$ now defined as in \eq{nV} in the Poincar\'e-Yang--Mills map, and therefore in the formula \eq{result}. This gives the previous result for $\cL_{\rm R^2}$ given in \eq{R2result} plus a new term given by
\be
{\cL}_{R^2}^{\rm extra} = -F_{\mu\nu}^i (Q) F^{\mu\nu}_i (Q)\ ,
\label{LQ}
\ee
where
\be
F_{\mu\nu}^i (Q) = 2\partial_{[\mu} Q_{\nu]}^i +i \ve^i{}_{jk} Q_\mu^j Q_\nu^k\ .
\ee

\subsection{Back to Einstein frame}
We have written the supersymmetry invariant in string frame in \eqref{R2result}, but the two-derivative Lagrangian (\ref{bact}) and (\ref{pL}) is written in Einstein frame. Recall moreover that what we call ``type IIA string frame'' is not an actual string frame, but corresponds to the Weyl rescaling with respect to the K\"{a}hler modulus $y = a_I v^I$ in type IIB whenever $n_T>1$. Writing the two-derivative in this frame would break manifest $SO(1,n_T)$ invariance and does not seem particularly helpful. We will rather choose to rewrite the higher derivative correction \eqref{R2result} in Einstein frame. 

Starting from \eq{defoh} where $\oh$ is given in string frame, we can perform the inverse of the transformations \eq{defs} to go back to Einstein frame. In particular we obtain, the torsionful spin connection in Einstein frame
\be  
\widehat{\omega}_{-\, ab} =  \widehat{\omega}_{ab} + {\widehat T}_{ab} \ ,
\ee
where $ \widehat{\omega}_{ab}$ is the torsion-free spin connection and the torsion ${\widehat T}_{ab}=e^c {\widehat T}_{c,ab}$ is defined in flat indices as 
\be 
\widehat{T}_{c,ab} = \frac12 y^{-1}\bigl( 2y_r \eta_{c[a} \widehat{P}_{b]}^r - {\bar\psi}_c\gamma_{ab}\chi^r- \ag_I \widehat{H}^I_{abc}\bigr)+\frac14 y^{-2} y_r y_s \bar \chi^r \gamma_{abc} \chi^s  \ .
\ee
One can straightforwardly write $\widehat{\rho}^+_{ab}$ in Einstein frame as well, but we shall concentrate here on the bosonic part of the higher derivative correction. The bosonic part of the higher derivative extension that contains the Riemann squared term reads 
\be 
\cL_{{\rm B},R^2} = - \frac14 e  y  R({\omega}_-)_{abcd} R({\omega}_-)^{abcd} - \frac{1}{16} \varepsilon^{\mu\nu\rho\sigma\kappa\lambda} \ag_I B^I_{\mu\nu} R(\omega_-)_{\rho\sigma}{}^{ab} R(\omega_-)_{\kappa\lambda ab}\; , 
\label{r2A}
\ee
where $\omega_{-ab} = \omega_{ab}+ T_{ab}$ and $T_{ab}$ is the bosonic part of ${\widehat T}_{ab}$. Note that this Lagrangian correction would give rise to ghost degrees of freedom, and one needs to carry out field redefinitions in order to ensure that the effective Lagrangian is well defined. To display the dependence on $H$ explicitly, we note that 
\bea 
&& - \frac{1}{2} \ag_I B^I \wedge R_{ab}(\omega_-)  \wedge R^{ab}(\omega_-) 
\CR
&=& - \frac{1}{4} \ag_I B^I \wedge R_{ab} \wedge R^{ab}  - \frac14  \Bigl( \omega_{ab} d \omega^{ab}- \frac{2}{3} \omega^a{}_b \wedge \omega^b{}_c  \wedge \omega^{c}{}_a \Bigr) \wedge a^I M_{IJ} \star H^J 
\CR
&&-  \Bigl( \frac12 T_{ab} D T^{ab} + T_{ab} \wedge R^{ab} - \frac13 T^a{}_b \wedge T^b{}_c \wedge T^c{}_a \Bigr) \wedge  a^I M_{IJ} \star H^J  \; , 
\eea
up to a total derivative and modulo the  duality equation for the three-form field strengths, and where $T_{ab}$ is the bosonic part of ${\widehat T}_{ab}$. Using this result in \eq{r2A} and combining it with the bosonic part of the  two-derivative action in Einstein frame given in~\eq{bact} gives 
\bea
e^{-1}\cL_{\rm B} \hspace{-2mm} &=& \hspace{-2mm}  \frac14 R -\frac{1}{48} M_{IJ} H_{\mu\nu\rho}^I H^{\mu\nu\rho J}
-\frac14 P_\mu^r P^\mu_r  -\frac12 
g_{\alpha\beta} \partial_\mu\varphi^\alpha \partial^\mu \varphi^\beta-\frac14 c^z  \Tr_z \bigl[ F_{\mu\nu} F^{\mu\nu} \bigr] 
\nn\w2
&& +\frac{1}{32} e^{-1}\varepsilon^{\mu\nu\rho\sigma\lambda\tau} B_{\mu\nu}^I \Bigl( b_I^z  \Tr_z \big[ F_{\rho\sigma} F_{\lambda\tau}\big]  - \ag_I R_{\rho\sigma ab} R_{\lambda\tau}{}^{ab} \Bigr) \\
&&  \hspace{-5mm}- \frac14  y  \bigl( R_{abcd}  + 2 D_{[a} T_{b],cd} + T_{a,c}{}^e T_{b,ed}-T_{b,c}{}^e T_{a,ed} \bigr) \bigl( R^{abcd} + 2 D^{[a} T^{b],cd} + 2T^{a,c}{}_f T^{b,fd}\bigr)\CR
&&  \hspace{-5mm} - \frac12 \Bigl(T_{\mu,ab} D_\nu T_{\rho,}{}^{ab} + T_{\mu,ab}  R_{\nu\rho}{}^{ab} - \frac23 T_{\mu,a}{}^b \ T_{\nu,b}{}^c  T_{\rho,c}{}^a \Bigr) \ag^I M_{IJ} H^{J\, \mu\nu\rho}-{\widehat F}^{ab \, i}(Q) {\widehat F}_{ab \, i}(Q)  \ ,\nonumber
\label{ba}
\eea
where 
\be 
d H^I = -b^{I z}  \Tr_z \big( F \wedge  F\big)  + \ag^I R_{ab} \wedge R^{ab}\; . 
\ee
One can analyse the complete Lagrangian writing 
\be 
\cL = -\frac{1}{48} e  M_{IJ} H_{\mu\nu\rho}^I H^{\mu\nu\rho J}+\frac{1}{32} \varepsilon^{\mu\nu\rho\sigma\lambda\tau} B_{\mu\nu}^I \Bigl( b_I^z  \Tr_z \big( F_{\rho\sigma} F_{\lambda\tau}\big)  - \ag_I R_{\rho\sigma ab} R_{\lambda\tau}{}^{ab} \Bigr) + \cL_{\rm extra} \ , 
\ee
where we separated the kinetic term and the (generalised) topological term from the term $\cL_{\rm extra}$ that is defined to only depend on the field strengths $H^{(-)}$ and $H^{(+) r}$. This complete Lagrangian is obtained by combining two-derivative Lagrangian of Section~\ref{sec:rev} with the Riemann squared invariant \eqref{R2result} in Einstein frame. Then the duality equation at first order in $\alpha$ can be written as
\bea 
\widehat{\mathcal{E}}_{\mu\nu\rho} \hspace{-2mm} &=&\hspace{-2mm}  2 H^{(+)}_{\mu\nu\rho} - 24\frac{ \delta S_{\rm extra}}{\delta H^{(-) \mu\nu\rho}} \,,\CR
\widehat{\mathcal{E}}^r_{\mu\nu\rho} \hspace{-2mm} &=&\hspace{-2mm}  2 H^{r (-)}_{\mu\nu\rho} - 24\frac{ \delta S_{\rm extra}}{\delta H^{(+) \mu\nu\rho}_r} \; . 
\label{newE}
\eea
In summary, the total proper Lagrangian in Einstein frame is given by
\be
\label{eq:L2}
{\cal L}= {\cal L}^{\rm cov} +{\cal L}^{\cal E} +{\cal L}_{R^2}\ ,
\ee
where ${\cal L}^{\rm cov}$ is given in  \eq{pL1}, \eq{bact}, \eq{pL}, ${\cal L}^{\cal E}$ is given in \eq{LE} with $\widehat{\mathcal{E}}_{\mu\nu\rho}$ and $\widehat{\mathcal{E}}_{\mu\nu\rho}^r$ from \eq{newE}, and ${\cal L}_{R^2}$ is given in \eq{R2result} with $V_\mu^i$ from \eq{nV}, and going to Einstein frame straightforwardly using \eq{defs}. The bosonic part of the resulting ${\cal L}_{R^2}$ is given in \eq{r2A}. The supersymmetry transformations are given in  \eq{eq:susytrm}.\footnote{We recall that the corrections linear in $a^I$ in the supersymmetry transformations necessarily exist and we have not computed them explicitly in this paper.}

\subsection{String theory low energy effective action}
\label{String4}
There are several four-derivative supersymmetry invariants one can write in (1,0) supergravity. One finds two types of $R^2$ type corrections in off-shell (1,0) supergravity coupled to one tensor multiplet, the  Riemann squared type discussed in this paper \cite{Bergshoeff:2012ax} and the Gauss--Bonnet type \cite{Novak:2017wqc,Butter:2018wss}. Their sum gives the $R^2$ correction in the (1,0) truncation of (1-loop) type IIA on K3 \cite{Liu:2013dna}. Their difference instead only depends on the Riemann tensor  through terms that can be eliminated by field redefinitions \cite{Chang:2022urm}, such that its bosonic component can be written in Einstein frame as
\be \mathcal{L}_{H^4,{\rm B}} = e  y  \Bigl(\frac1{24} H^1_{\mu\rho\sigma} H^1_\nu{}^{\rho\sigma} H^{1 \mu\kappa\lambda} H^{1 \nu}{}_{\kappa\lambda} + H^{1\mu\rho\sigma} H^{1 \nu}{}_{\rho\sigma} P_\mu^1 P_\nu^1 - \frac12 P_\mu^1 P^{1\mu} P_\nu^1 P^{1 \nu} \Bigr) \; . \label{H4Nt1} \ee
In the low energy effective action, one finds therefore that there is a unique $R^2$ type correction associated to the gravitational anomaly, while the other correction mentioned above is understood as a  matter multiplet $H^4$ type higher derivative invariant. Note that this second kind of supersymmetry invariant is not protected and can be written at leading order in $\alpha'$ as the (on-shell) full superspace integral of an arbitrary function of the tensor multiplet scalar fields. 
The correction to the action of the type above generalises then  to $n_T$ tensor multiplets  (but neglecting vector and hyper multiplets) as\footnote{This formula can be derived using the on-shell harmonic superspace formalism as the integral $\int d^4\theta d^2u \mathcal{E} F_{rstu}(v)  \bar \chi^r \sigma^{++} \gamma^a \chi^s \bar \chi^t \sigma^{++} \gamma_a \chi^u  $, with a tensor function of the scalar field $v^I$ satsifying $D_{[v} F_{r]stu}(v) = 0$ and $\chi^{r +} = D_\alpha^+ F^r$ in \cite{Sokatchev:1988aa} with first component $\chi^{r A}$ for $A=+$. 
This is compatible with  the truncation of the $(2,0)$ supersymmetry invariant of the same type for $f(y)=y$ \cite{Bossard:2018rlt} and one recovers (\ref{H4Nt1}) for $n_T=1$. Note that $\chi^{r +}$ is not a G-analytic superfield in the presence of vector or hyper multiplet, so including them requires corrections.}
\begin{multline} \mathcal{L}_{H^4,{\rm B}} = \frac{1}{8} e  \Bigl( 3 f(y) \delta_{(rs} \delta_{tu)} + \frac{6f'(y)}{y} \delta_{(rs} y_{t} y_{u)} + \frac{yf''(y)-f'(y)}{y^3} y_r y_s y_t y_u\Bigr)  \\
 \Bigl(\frac1{24} H^r_{\mu\rho\sigma} H^s_\nu{}^{\rho\sigma} H^{t \mu\kappa\lambda} H^{u \nu}{}_{\kappa\lambda} + H^{r\mu\rho\sigma} H^{s \nu}{}_{\rho\sigma} P_\mu^t P_\nu^u - \frac12 P_\mu^r P^{s\mu} P_\nu^t P^{u \nu} \Bigr) \; . \end{multline}
We conclude that our supersymmetry analysis exhibits the expected result that the $R^2$ term is uniquely determined by the anomaly coefficient vector $a^I$. There are also Yang--Mills $F^4$ type and hypermultiplet $(\partial \varphi)^4$ type corrections to the effective action at the same order in derivatives. The tensor multiplets and the hypermultiplet corrections are not protected by supersymmetry, so one does not know much about them in string theory. 

\vskip 2mm

Let us now discuss the $R^2$ type term obtained in this paper in relation to string theory compactifications. The known supersymmetry vacua with (1,0) supersymmetry in six dimensions can be understood as F-theory compactifications  \cite{Kumar:2009us,Kumar:2009ae,Kumar:2009ac,Kumar:2010ru}. In quantum gravity the coefficients $2a^I$ and $b^{Iz}$ are quantised in the self-dual lattice $L_{1,n_T}$ of BPS string states \cite{Kumar:2009ac,Seiberg:2011dr,Monnier:2017oqd}, with the definition $\Tr_z = \frac{2}{h_z^{\vee}} \Tr_{{\rm adj}\, z}$.\footnote{Where $\Tr_{\rm adj}$ is the trace in the adjoint representation and $h_z^{\vee}$ the dual Coxeter number of the simple group $G_z$.} In F-theory, $L_{1,n_T} = H_2(B,\mathbb{Z})$, the second homology group of the K\"{a}hler base for the elliptically fibered Calabi--Yau manifold. The vectors $b^{Iz}$ can be interpreted as the homology cycles on which the elliptic fibre is degenerate, and $v_I b^{Iz}\ge 0$ is their volume. The vector $-2a^I$ is the canonical divisor cycle. In the co-dimension one locus where $v_I b^{Iz} = 0$, the BPS string of charge $b^I \in L_{1,n_T}$ becomes tensionless and the low-energy effective theory breaks down. One finds indeed that the kinetic terms of the Yang--Mills Lagrangian goes to zero, indicating a strong coupling \cite{Sagnotti:1992qw,Duff:1996rs,Seiberg:1996vs}. This is easier to interpret after a Weyl rescaling by $c^z$, so that the Lagrangian becomes 
\be 
e^{-1} \mathcal{L} = - \frac{1}{4(c^z)^2} R  - \frac14 \Tr_z F_{\mu\nu} F^{\mu\nu} - \frac14 \sum_{z\ne z'} \frac{c^{z'}}{c^z} \Tr_{z'}  F_{\mu\nu} F^{\mu\nu} + \dots 
\ee
and one understands that the Weyl rescaled Planck length is going to zero, so that gravity decouples. 

There is a priori a similar interpretation if $y = v_I a^I$ goes to zero. With the appropriate normalisation, one gets that 
\be 
4 \eta_{IJ} a^I a^J = n_T-9 \; , 
\ee
so that $y = v_I a^I$ cannot vanish for $n_T\le 9$. One may then wonder for $n_T\ge 10$ if it is consistent to reach a singularity at $y=0$.   Note that because of the anomaly constraint \cite{Randjbar-Daemi:1985tdc}
\be {\rm dim}(G) = n_H + 29 n_T-273 \; , \ee
the dimension of the gauge group is always positive for $n_T\ge 10$. In F-theory compactifications, it is only possible to reach $y=0$ if all the gauge couplings are going to zero simultaneously \cite{Kumar:2009ae}, because 
\be 
-24 v_I a^I \ge n_z v_I b^{I z}  
\ee
for positive integers $n_z \ge 1$ determined by the simple groups $G_z$, and each $v_I b^{Iz}\ge 0$ for the Yang--Mills kinetic terms to be well defined. One may wonder if it is a condition from F-theory or if it is a more general consequence of quantum gravity that $-v_I a^I\ge 0$. It does not a priori follow from a unitarity bound on the $R^2$ coefficient, since it is allowed to get a small negative value  at weak gravity coupling \cite{Caron-Huot:2022jli}.

At the level of the effective action, it is natural to consider the limit $y\rightarrow 0$ in the ``string frame'' described in Section~\ref{StringFrameSection}. The only singularities in the supersymmetry transformations involve then either $P_\mu$ or terms in $(\frac{c^{r z}}{c^z} {-} \frac{y^r}{y}) \Tr_z \lambda \lambda \chi_r$, similarly as for the locus $v_I b^{I z}= 0$ where the corresponding gauge coupling diverges. In this frame we find therefore that gravity decouples at $y\rightarrow 0$ with 
\be 
e^{-1} \mathcal{L} = - \frac{1}{4y^2} R  - \frac1{4 y} c^z \Tr_z F_{\mu\nu} F^{\mu\nu}  -\frac14 \widehat{R}_{abcd}(\widehat{\omega}_-)  \widehat{R}^{abcd}(\widehat{\omega}_-) + \dots 
\ee

Let us end this section with a simple explicit example. A  perturbative type~I theory with $n_T=10$ tensor multiplets can be obtained by the orientifold of type IIB on the $\mathbb{Z}_3$ orbifold locus in the K3 moduli space. The orientifold includes a K3 automorphism that exchanges the two twisted sectors \cite{Polchinski:1996ry}, such that the orientifold is only defined for the K\"ahler moduli in $O(3,11) / (O(3)\times O(11))$, giving eleven neutral hypermultiplets in $O(4,11) / (O(4)\times O(11))$ at tree-level. To fix conventions we define the modulus $V = \frac{\rm Vol( K3)}{(2\pi)^2\alpha'}$ and we denote the nine  axions that are in the twisted sectors collectively by $B$. Together $V$ and the nine $B$ give the ten scalars of the tensor multiplets. We define $Q^I$ the vector of string charges in $L_{1,10}$, that we decompose into the D1 charge $m$, $q$ the vector of charges of the nine D3 branes wrapping the 2-cycles odd under the orientifold K3 automorphism and $n$ the charge of the D5 brane wrapping K3. Such a BPS string has mass $v_I Q^I / \sqrt{\alpha'}$ with
\be 
v_I Q^I = \frac{1}{\sqrt{2V}} \bigl( m + (  B , q) + \bigl( \tfrac12 (B,B) + V \bigr) n \bigr) \; . \label{vItypeI} 
\ee
There are two inequivalent orientifold actions one can define, the standard one $\Omega$ and  $\Omega g$ including the $\mathbb{Z}_6$ generator $g$ \cite{Gimon:1996ay}. They are called the $\mathbb{Z}_3^{\rm A}$ and the $\mathbb{Z}_6^{\rm B}$ orientifold in \cite{Gimon:1996ay} and they are T-dual to each other. The low energy effective theory includes 10 tensor multiplets, gauge group $U(8) \times SO(16)$ with charged hypermultiplets in the $({\bf 28},{\bf 1}) \oplus ( {\bf 8},{\bf 16})$ plus eleven neutral hypermultiplets. One straightforwardly computes the anomaly polynomial of the model  \cite{Alvarez-Gaume:1983ihn}\footnote{We define $\Tr_{\scalebox{0.6}{$SO(2n)$}}$ in the vector representation and $\Tr_{\scalebox{0.6}{$U(n)$}}$ in the fundamental representation.}
\be 
\Bigl( \frac12 \Tr R^2 + 2 \Tr_{\scalebox{0.6}{$SO(16)$}} F^2 - 2 \Tr_{\scalebox{0.6}{$U(8)$}} F^2\bigr)^2 
 - \Tr_{\scalebox{0.6}{$U(8)$}}F  \Bigl(\Tr_{\scalebox{0.6}{$U(8)$}}F \, \Tr R^2   +16  \Tr_{\scalebox{0.6}{$U(8)$}}F^3  \Bigr)\; . 
 \label{FtrF3}
\ee
The first term is taken care of by the Green--Schwarz--Sagnotti mechanism, while the second is resolved through the gauging of a neutral hypermultiplet axion in the twisted sector \cite{Berkooz:1996iz}. The form of the Chan--Paton representation matrix \cite[Eq.~(5.5)]{Gimon:1996ay} implies that the gauged axion is the sum of the nine twisted RR scalars, that we write as the scalar product $(u,C)$ for $u = (\tfrac13,\tfrac13,\tfrac13,\tfrac13,\tfrac13,\tfrac13,\tfrac13,\tfrac13,\tfrac13)$ of unit norm. We must have the gauge transformation 
\be \delta_\Lambda A = d \Lambda  + [ A , \Lambda ] \; , \qquad \delta_\Lambda C =  u \Tr_{\scalebox{0.6}{$U(8)$}} \Lambda \; , \label{AxionGauging} \ee
and the second anomalous term is canceled by a Green--Schwarz counterterm of the form 
\be  ( u, C)   \Bigl(\Tr_{\scalebox{0.6}{$U(8)$}}F \, \Tr R^2   +16  \Tr_{\scalebox{0.6}{$U(8)$}}F^3  \Bigr)\; . \label{AbelianGS}\ee
The gauging \eqref{AxionGauging} is more easily defined by dualising the axion to a four-form $(u,C_4)$, in which case the hypermultiplet is dualised to a linear multiplet and the gauging is realised through the term  
\be (u,C_4) \wedge \Tr_{\scalebox{0.6}{$U(8)$}}F \ee
that appears in the supersymmetric Lagrangian \cite[Eq. 4.15]{Bergshoeff:1985mz} coupling the linear multiplet to an abelian vector multiplet.  It would be interesting to supersymmetrise the Green--Schwarz counterterm \eqref{AbelianGS}. We expect that the $F^4$ and $F^2 R^2$  supersymmetry invariants will give the correct Green--Schwarz counterterm \eqref{AbelianGS} in presence of the gauging. 

As it was explained in \cite{Berkooz:1996iz}, the gauging implies that the abelian vector multiplet combines with the hypermultiplet including the scalar field $(u,C)$ to define a massive vector multiplet. Indeed, integrating out the abelian vector multiplet auxiliary field in the Lagrangian \cite[Eq. 4.15]{Bergshoeff:1985mz} gives a mass to the three linear multiplet scalar fields while the axion $(u,C)$ is absorbed in the massive vector. The low energy effective theory for massless fields then only includes the unbroken gauge group $SU(8) \times SO(16)$ and ten massless neutral hypermultiplets.

The gauge coupling $v_I b^{Iz}$ can be computed using the method introduced in \cite{Antoniadis:1999ge}, showing that the coupling to the nine twisted scalar fields $B$ are all equal, with a $-1/2$ factor between $SU(8)$ and $SO(16)$.\footnote{The trace over the Chan--Paton representation matrix of the orbifold action needed in \cite[Eq.~(3.20)]{Antoniadis:1999ge} can be computed using \cite[Eq.~(5.5)]{Gimon:1996ay}.} The anomaly factorises as
\be 
\Bigl( \frac12 \Tr R^2 + 2 \Tr_{\scalebox{0.6}{$SO(16)$}} F^2 - 2 \Tr_{\scalebox{0.6}{$SU(8)$}} F^2\bigr)^2 \; .\label{6Danomaly} 
\ee
For the $\mathbb{Z}_3^{\rm A}$ orientifold the vector multiplets come from D9 branes and one must get consistency with the type I Chern--Simons coupling in ten dimensions in the large volume  limit $V\gg1 $
\be 
d H^{\scalebox{0.6}{10D}} = \Tr R^2 +  \Tr_{\scalebox{0.6}{$SO(32)$}} F^2 \; . 
\ee
Writing the anomaly coefficients as $a^I = (m_a,q_a,n_a)$ and $b^{Iz} = (m_z,q_z,n_z)$, the consistency with type I in ten dimensions fixes $n_a=-1$ and $n_z=1$ for both gauge groups using $\Tr_{\scalebox{0.6}{$SO(32)$}} F^2= \Tr_{\scalebox{0.6}{$SO(16)$}} F^2+2\Tr_{\scalebox{0.6}{$SU(8)$}} F^2$, while \eqref{6Danomaly} fixes $q_a,q_z$ and sets $m_a=m_z=0$ using $q_{\scalebox{0.6}{$SU(8)$}}= - \frac12 q _{\scalebox{0.6}{$SO(16)$}}\propto u$ from the computation of \cite{Antoniadis:1999ge}, i.e. 
\be 
a^I = (  0 ,- \tfrac12 u,- 1 ) \; , \quad b^I_{\scalebox{0.6}{$SO(16)$}} = ( 0 , 2 u , 1) \; , \quad  b^I_{\scalebox{0.6}{$SU(8)$}} = ( 0 , -u , 1) \; , 
\ee
where the unit vector $u$ is defined as above with all components equal to $1/3$. 

One gets the couplings in the $\mathbb{Z}_6^{\rm B}$ orientifold by T-duality. Then the gauge fields come from D5 branes and 
\be 
a^I = (  -1 ,- \tfrac12 u, 0 ) \; , \quad b^I_{\scalebox{0.6}{$SO(16)$}} = ( 1 , 2 u , 0) \; , \quad  b^I_{\scalebox{0.6}{$SU(8)$}} = ( 1 , -u , 0) \; . 
\ee
The positivity of the gauge couplings can be computed from \eqref{vItypeI} and one obtains the constraint 
\be 
- \frac12<  (u,B) <1\;  
\ee
in the $\mathbb{Z}_6^{\rm B}$ orientifold. One finds therefore that one reaches the strong coupling regime for the $SO(16)$ gauge group before  reaching the point $y=0$ at $(u,B)= - 2$, consistently with the general F-theory inequality. The case of the $\mathbb{Z}_3^{\rm A}$ orientifold is identical by T-duality.

\bigskip

\subsubsection*{Acknowledgments}

We thank Emilian Dudas and Yoshiaki Tanii for useful discussions. GB and AK are grateful to Texas A\&M for the generous hospitality during the early stages of this work.  The work of ES is supported in part by the NSF grants PHYS-2112859  and PHYS-2413006.

\appendix

\section{Conventions and Fierz identities}
\label{app:conv}

As stated in Section~\ref{sec:rev}, our space-time signature is $(-+++++)$. Curved six-dimensional indices $\mu$ are split into time and space according to $\mu=(t,i)$ with $i=1,\ldots,5$ and we write a curved time index explicitly as $t$. Flat indices $a=0,\ldots,5$ are split according to $a=(0,\ua)$.   Our conventions for the Levi--Civita symbol are 
$\varepsilon^{0\underline{1}\underline{2}
\underline{3}\underline{4}\underline{5}}=+1$ and $\varepsilon^{\ua\ub\uc\ud\ue}=\varepsilon^{0\ua\ub\uc\ud\ue}$. In curved indices $\varepsilon^{t12345}=+1$ and $\varepsilon^{ijklm}= \varepsilon^{tijklm}$. 

\subsection*{Relationship between different conventions}

We convert the expressions in~\cite{Riccioni:2001bg} to the ones in this paper by using the following substitutions:
\begin{align}
 \eta_{rs} &\to  - \eta_{IJ}\ ,&\quad \eta_{ab} & \to -\eta_{ab}\ ,&\quad \varepsilon^{\mu_1...\mu_6} &\to -\varepsilon^{\mu_1...\mu_6}\ ,
\nn\w2
 B_{\mu\nu}^r &\to \frac12 B^I\ ,& \quad c^{rz} &\to b^{Iz}\ ,&\quad \tr &\to -\frac12 \Tr\ ,
\nn\w2
v_r &\to v_I\ ,&\quad x_r^M &\to v_I^r\ ,& \quad {\mathcal A}_\alpha &\to -{\mathcal A}_\alpha\ ,&\quad \omega_\mu{}^{mn} &\to -\omega_\mu{}^{ab}\ ,
\nn\w2
 \gamma_\m &\to i\gamma_\m\ ,& \quad \chi^M &\to \chi^r\ ,& \quad \lambda& \to -\sqrt2 \lambda\ .
\label{cc2}
\end{align}
Note that $A_\mu$ is anti-Hermitian in our conventions, but we define the trace with a minus sign such that it is positive definite. In the appropriate basis one has $\Tr \left(F_{\mu\nu} F^{\mu\nu}\right) = \delta_{PQ}\,F_{\mu\nu}^{P} F^{\mu\nu Q} $. Moreover we use $\bar \epsilon \chi = \bar \epsilon^A \chi_A$ whereas~\cite{Riccioni:2001bg} uses $\bar \epsilon \chi = \bar \epsilon_A \chi^A$ so all fermion bilinears get an extra minus sign.

\subsection*{Our conventions and notations}
%
In our conventions 
\begin{align}
\eta_{ab} &= {\rm diag} (-+++++)\ ,\qquad \eta_{IJ}= (-+++...+)\ , \qquad \gamma^{\mu_1\ldots \mu_6} = - \varepsilon^{\mu_1\ldots \mu_6} \gamma_7\ ,
\nn\w2
\gamma_7 \epsilon &= \epsilon \ ,\quad \gamma_7\chi = -\chi \ , \quad \delta_\lambda \epsilon = -\frac14 \lambda_{ab}\gamma^{ab} \epsilon\ ,\quad  (\sigma^i \epsilon)_A = (\sigma^i)_A{}^B \epsilon_B\ , \quad \bar\chi\lambda = {\bar\chi}^A\lambda_A\ ,
\nn\w2
 y & = a^I v_I \ ,\qquad y^r = a^I v_I^r\ ,\qquad c^z = b^{Iz} v_I \ ,\qquad c^{rz} = b^{Iz} v_I^r\ ,
\nn\w2
\partial_\mu v_I & = P_\mu^r v_{Ir}\ ,\qquad \partial_\mu v_I^r = P_\mu^r v_I\ ,\qquad v_I v^I =-1\ , \qquad
\nn\w2
P_\mu & = y^{-1} y^r P_\mu^r\ ,\qquad \chi = y^{-1} y^r \chi^r\ .
\end{align}
The Hodge dual of a $p$-form $\alpha$ in six dimensions is defined by 
\begin{align}
    (\star\alpha)_{\mu_1\ldots \mu_{6{-}p}} = \frac{1}{p! \sqrt{-g}} g_{\mu_1 \nu_1} g_{\mu_2\nu_2} \dots g_{\mu_{6{-}p} \nu_{6{-}p}} \varepsilon^{\nu_1\nu_2 \dots \nu_{6{-}p} \sigma_1 \dots  \sigma_{p}}  \alpha_{\sigma_1\ldots \sigma_p}
\end{align}
where $\varepsilon^{\mu_1\mu_2 \dots \mu_6}$ is constant and satisfies $\varepsilon^{t12345}=1$ and its indices are lowered by the metric $g_{\mu\nu}$. Note also that
\be
\gamma^{a_1...a_n} = \frac{(-1)^{\lfloor n/2\rfloor}}{(6{-}n)!} \varepsilon^{a_1...a_n b_1...b_{6{-}n}} \gamma_{b_1...\gamma_{6{-}n}}\gamma_7\ .
\ee
In the Henneaux--Teitelboim for of the model we split the worldline and tangent space indices as follows
\be
\mu=(t, i)\ ,\qquad a=(0,\underline{a})\ ,\qquad i,\underline a= 1,...,5\ .
\ee

\subsection*{Spinors and Fierz rearrangement formulae}

We use the same convention as in \cite{Nishino:1997ff} except that we do not write explicitly Sp(1) indices. So it is always understood that Sp(1) indices are contracted as the Lorentz indices so that %
\be   
[ \lambda \bar \chi \epsilon ]_A = [(\bar \chi \epsilon)  \lambda  ]_A = \lambda_A \bar \chi^B \epsilon_B \; . \ee
We also use the Pauli matrices $\sigma^{i AB}$. 
For the purpose of the appendix we define 
\be P_\pm = \frac{1\pm \gamma_7}{2} \; . 
\ee
In our conventions $\lambda,\epsilon , \psi =dx^\mu \psi_\mu$ are chiral and $\chi$ anti-chiral. Moreover $\psi$ commutes with itself because it is a $1$-form. 
In this way the elementary Fierz rearrangements can be written as\footnote{By definition, the Sp(1) indices are not contracted for $\psi \bar \epsilon$ and they are for $\bar \epsilon \gamma^a \psi$, etc.} 
\bea 
\psi \bar \epsilon  &=& \Bigl( - \frac18 \bar \epsilon \gamma^a \psi \gamma_a+ \frac{1}{96} \bar \epsilon  \gamma^{abc} \psi \gamma_{abc} - \frac{1}{8} \bar \epsilon \sigma^i \gamma^a \psi \sigma_i \gamma_a + \frac{1}{96} \bar \epsilon \sigma^i \gamma^{abc} \psi \sigma_i \gamma_{abc} \Bigr) P_-\ ,
\nn\w2
\psi \bar \chi &=&\Bigl(  - \frac18 \bar \chi \psi +\frac1{16} \bar \chi \gamma^{ab} \psi \gamma_{ab} - \frac18 \bar \chi \sigma^i \psi \sigma_i +\frac1{16} \bar \chi \sigma^i \gamma^{ab} \psi \sigma_i \gamma_{ab} \Bigr) P_+ \ .   
\eea
Similarly, we have the Fierz identity
\be
\chi\, \bar\chi \e = -\frac18 \gamma^a\sigma^i\e\, \bar\chi \gamma_a\sigma_i \chi +\frac{1}{96}\gamma^{abc}\e\, \bar\chi \gamma_{abc}\chi\ .
\label{fchi}
\ee
The symplectic Majorana--Weyl reality condition implies that 
\be 
\bar \lambda \gamma_{a_1\dots a_{2n}} \chi = \bar \chi \gamma_{a_{2n} \dots a_1} \lambda \ , \qquad \bar \epsilon \gamma_{a_1\dots a_{2n+1}} \psi =-  \bar \psi \gamma_{a_{2n+1} \dots a_1} \epsilon \ .
\ee
The more general condition is
\begin{align}
\bar\epsilon \gamma_{[n]} \sigma^{[m]} \psi = (-1)^{m+n} \bar\psi \gamma_{[n]^T}\sigma^{[m]^T} \epsilon\,.
\end{align}

We write $\psi$ for the gravitino 1-form, which is commuting with itself. In this case the Fierz identity is symmetric so that 
\be 
\psi \bar \psi  =  \Bigl( \frac18 \bar \psi \gamma^a \psi \gamma_a - \frac{1}{96} \bar \psi \sigma^i \gamma^{abc} \psi \sigma_i \gamma_{abc}  \Bigr) P_- \ee 
Using this one obtains 
\be 6 \psi \bar \psi - \gamma^{ab} \psi \bar \psi \gamma_{ab} = 2  \bar \psi \gamma^a \psi  \gamma_a P_- \; . 
\ee
Written for two independent spinors this identity becomes 
\be 
3 \lambda \bar \epsilon - 3 \epsilon \bar \lambda - \frac12 \gamma^{ab} \lambda \bar \epsilon \gamma_{ab} + \frac12 \gamma^{ab} \epsilon \bar \lambda \gamma_{ab} = - 2 \bar \epsilon \gamma^a \lambda \gamma_a P_-\ .
\label{G1}
\ee
Some useful Fierz identities are
\bea 
\gamma^{bcd} \gamma_a \chi \bar \chi \gamma_{bcd} &=& - 12 \gamma^b \gamma_a \chi \bar \chi \gamma_b - 48 \chi \bar \chi \gamma_a + (\bar\chi \gamma^{bcd} \chi) \gamma_{bcd} \gamma_a\ ,
\label{g1}\w2
\chi_r \bar \chi^r \sigma^i \gamma_a \chi + \frac14 \gamma^b \gamma_a \chi \bar \chi_r \sigma^i \gamma_b \chi^r &=& \frac1{8} \sigma^i \gamma^{bc} \chi \bar \chi_r \gamma_{abc} \chi^r - \sigma^i \chi_r \bar \chi^r \gamma_a \chi\ , 
\label{g2}\w2
2\sigma^i \psi_{[\mu} \bar\chi \sigma^i \psi_{\nu]} &=& 2\psi_{[\mu} \bar\chi \psi_{\nu]} -\gamma^a\chi \bar\psi_\mu \gamma_a \psi_\nu =0\ ,
\label{g3}\w2
\sigma_i \chi ( \bar \chi_r \sigma^i \gamma_a \chi^r ) &=& \frac14 \gamma^{bc} \chi ( \bar \chi_r \gamma_{abc} \chi^r) - \gamma^b \gamma_a \chi_r ( \bar \chi^r \gamma_b \chi) - 2 \chi_r ( \bar \chi^r \gamma_a \chi) \CR
&=& \frac14 \gamma^{bc} \chi_r ( \bar \chi^r \gamma_{abc} \chi) - \frac12 \gamma^b \gamma_a \chi_r ( \bar \chi^r \gamma_b \chi)  \  . 
\label{ge}
\eea

\bibliographystyle{utphys}
\providecommand{\href}[2]{#2}\begingroup\raggedright\endgroup

\end{document}